\documentstyle[12pt,amssymb,latexsym,amsmath%,srcltx
]{article}
\newcommand{\beq}{\begin{equation}}
\newcommand{\eeq}{\end{equation}}
\newcommand{\ba}{\begin{array}}
\newcommand{\ea}{\end{array}}
\begin{document}

\title{New multidimensional partially integrable 
generalization of $S$-integrable $N$-wave equation}

\author{
A.I. Zenchuk\\
Center of Nonlinear Studies of L.D.Landau Institute
for Theoretical Physics  \\
(International Institute of Nonlinear Science)\\
Kosygina 2, Moscow, Russia 119334\\
E-mail: zenchuk@itp.ac.ru
}

\maketitle
\begin{abstract}
This paper develops a modification of the dressing method based on  inhomogeneous linear integral equation with integral operator having nonempty kernel. Method allows one to construct
the  systems of 
multidimensional Partial Differential Equations (PDEs) having  differential polynomial structure  in any dimension $n$. Associated   solution space is not full, although it is parametrized by  certain number of arbitrary functions of $(n-1)$ variables. 
We consider $4$-dimensional generalization of the classical (2+1)-dimensional $S$-integrable $N$-wave equation 
as an example.
\end{abstract}

%%%%%%%%%%%%%%%
\section{Introduction}
%%%%%%%%%%%%%%%

Completely integrable nonlinear Partial Differential Equations (PDEs) became an attractive field of research after the 
  discovery of  complete integrability of 
the Korteweg-de Vries equation \cite{GGKM}. $S$-integrable \cite{ZSh1,ZSh2,ZM,BM,AKNS,ZMNP,K} and $C$-integrable \cite{Calogero,C_int1,C_int2,C_int3,C_int4,C_int5} systems are mostly remarkable among multidimensional nonlinear PDEs. The dressing method is one of the promoted methods for constructing and solving $S$-integrable PDEs. After the original version of the dressing method \cite{ZSh1} several important modifications have been developed
 \cite{ZSh2,ZM,BM,Zak,Zakharov1,Zakharov2} (see also \cite{ZMNP,K}). All of them are based either  on a Riemann-Hilbert or on a $\bar\partial$-problem \cite{BC,ABF}, which are uniquely solvable linear integral equations for some matrix function $U(\lambda;x)$
  depending on spectral parameter $\lambda$, where independent variables of nonlinear PDEs appear as a  set of additional parameters $x=(x_1,x_2,\dots)$.

This paper is based on the modification of the $\bar\partial$-problem
introduced in \cite{SAF,Z}.
It was shown \cite{Z}, that classical $S$-integrable PDEs, some types of $C$-integrable PDEs and some types of ''mixed'' PDEs (i.e. equations having $S$- and $C$-integrable systems as the particular cases) can be
studied by the dressing method based on the following integral
equation, which is equivalent to the
$\bar\partial$-problem \cite{ZM,BM}:
\begin{eqnarray} \label{Sec1:U}
\Phi(\lambda;x)=\int \Psi(\lambda,\mu;x)U(\mu;x)d\Omega(\mu)
\equiv \Psi(\lambda,\mu;x) * U(\mu;x)\equiv \hat \Psi U.
\end{eqnarray}
Here  $\lambda$ and $\mu$ are spectral variables,  
$U$ is the unknown matrix function. The matrix functions $\Phi$ and $\Psi$ are defined by
some extra conditions, which fix their dependence on the additional  parameters $x_i$. $\Omega$ is some largely arbitrary 
scalar measure in the $\mu$-space. Apart from $\Omega$, all the functions appearing in this paper 
are $Q\times Q$ matrix functions.

To describe the classical completely integrable equations, kernel of the integral operator $ \Psi$ must be trivial, i.e.  $\dim{\mbox{ker}}\;  \hat\Psi=0$. It was shown in \cite{ZS} that this  requirement  is too restrictive. More general dressing algorithm must be based on the integral operator with $\dim{\mbox{ker}}  \hat\Psi> 0$. So, some examples of nonlinear PDEs  corresponding to  $\dim{\mbox{ker}}  \hat\Psi =1$ have been derived in \cite{ZS}. Generalization $\dim{\mbox{ker}} \hat \Psi>0 $
removes any formal restriction on  dimensionality of the constructed nonlinear PDEs, although derived systems have  restricted solution spaces. Namely, $n$-dimensional nonlinear system admit at most $(n-2)$-dimensional solution space, $n\ge 1$. The  increase of dimensionality of nonlinear PDEs in comparison with classical $S$-integrable equations  happens due to the presence of  {\it the external dressing function} in the algorithm, whose dimensionality has no restriction.  Remember that the
classical dressing algorithm uses only  {\it the internal dressing functions} whose dimensionalities are strongly restrictive and usually equal 1. 
However, the new
version of the dressing algorithm  has 
an important  common feature with the classical algorithm. Namely, derivatives  $\Psi_{x_j}(\lambda,\mu;x)$ are separated functions of the spectral parameters $\lambda$ and $\mu$. This fact fixes dimensionalities of the internal dressing functions mentioned above. This property  has been removed in \cite{Z2} (i.e.  $\Psi_{x_j}(\lambda,\mu;x)$ are not separated function of spectral parameters  there) keeping $\dim{\mbox{ker}} \hat\Psi=0$. But there is  serious obstacles for  construction of the explicit solutions, which have not been found in \cite{Z2}.
 
 Another problem appeared in \cite{ZS} is a complicated "block" structure of the derived nonlinear PDEs, which is not typical for the equations of mathematical physics. 
 
Combination of basic ideas of both refs. \cite{ZS} and \cite{Z2}  enriches the solution space  and  improves structure of the system of nonlinear PDEs derived in this paper.
 The dimensionality $n$  of the resulting system of nonlinear PDEs has no restriction  and its solution space is parametrized by  certain number of arbitrary functions of $(n-1)$ variables, although the full integrability is not achieved yet. Nevertheless, this is a progress in comparison with the results of \cite{ZS}, where the solution space of the derived system of nonlinear PDEs is parametrized  by  arbitrary functions of $(n-2)$ variables. 
Structure of  nonlinear PDEs is also improved: system of equations has differential polynomial structure.

Algorithm represented in this paper  is based on the following modification of the eq.(\ref{Sec1:U}):
\begin{eqnarray}\label{U0}
\Phi^{(sk)}(\lambda;x)&=& \sum_{n=1}^{Q} 
\int {\cal{T}}^{(n)}
\Psi^{(s)}(\lambda,\mu;x)
U^{(nk)}(\mu;x) d\Omega(\mu) = \\\nonumber
&& \sum_{n=1}^{Q}
{\cal{T}}^{(n)}\Psi^{(s)}(\lambda,\mu;x) *
U^{(nk)}(\mu;x)= \\\nonumber
&&
 \sum_{n=1}^{Q}
 \Psi^{(sn)}(\lambda,\mu;x)*
U^{(nk)}(\mu;x)\equiv 
 \sum_{n=1}^{Q}
\hat\Psi^{(sn)}
U^{(nk)} ,\\\nonumber
&&
 \Psi^{(sn)}(\lambda,\mu;x)  = {\cal{T}}^{(n)} \Psi^{(s)}(\lambda,\mu;x)
,\;\;s=1,\dots, Q,
\end{eqnarray}
which must be solved for  {\it the matrix spectral functions} $U^{(nm)}$. Here
${\cal{T}}^{(n)}$ are constant diagonal matrices,  functions
$\Phi^{(s)}$ and $\Psi^{(s)}$ are called {\it the internal  dressing
functions} and will be specified below.
We write superscripts inside of parenthesis in order to distinguish them from the power notations.

It was observed that  the integral equation (\ref{U0}) may be simplified taking
\begin{eqnarray}
{\cal{T}}^{(n)}_\alpha=\delta^{(n)}_\alpha=\left\{
\begin{array}{ll}
1,& n=\alpha\cr
0, & n\neq \alpha
\end{array}
\right.
\end{eqnarray}
and the diagonal function $\Phi^{(sm)}$  in all examples studied in this paper. So that the integral equation reads:
\begin{eqnarray}\label{U}
\Phi^{(sk)}_{\beta}\delta_{\alpha\beta} (\lambda;x)= 
\sum_{n,\gamma=1}^Q \delta^{(n)}_\alpha \Psi^{(s)}_{\alpha\gamma}(\lambda,\mu;x)*U^{(n k)}_{\gamma\beta}\;\;\Rightarrow\\\label{U3}
 \left\{
\begin{array}{ll}
\sum_{\gamma=1}^Q \Psi^{(s)}_{\beta\gamma}(\lambda,\mu;x)*U^{(\beta k)}_{\gamma\beta}(\mu;x)=\Phi^{(sk)}_{\beta} (\lambda;x) , &\alpha=\beta \cr
\sum_{\gamma=1}^Q \Psi^{(s)}_{\alpha\gamma}(\lambda,\mu;x)*U^{(\alpha k)}_{\gamma\beta}(\mu;x)=0,&\alpha\neq\beta
\end{array}\right..
\end{eqnarray}
Hereafter (except  Sec.\ref{Solutions}) double subscript (usually Greek) means the element of the appropriate matrix and  single subscript denotes the nonzero element of the appropriate diagonal matrix unless different is specified. 

At first glans, following the philosophy of the dressing method based on the integral equation, integral equations for $U^{(\beta k)}_{\gamma\beta}(\lambda;x)$ and $U^{(\alpha k)}_{\gamma\beta}(\lambda;x)$, $\alpha\neq\beta$ are decoupled. This might result in appropriate  decoupling of the system of nonlinear PDEs. Moreover, the integral eqs. (\ref{U3}a) and (\ref{U3}b)  with different values of the  indexes $\alpha$  and $\beta$  might be  decoupled as well.  However, we will show that   general  situation is different and the above statements are   partially correct. Namely, we will obtain an example of the complete system
 of nonlinear  PDEs  written for the fields produced by 
 $U^{(1 k)}_{\gamma 1}(\lambda;x)$ and $U^{(1 k)}_{\gamma Q}(\lambda;x)$, $k,\gamma=1,\dots,Q$.  However, more general PDEs may couple  fields produced by all functions $U^{(\alpha k)}_{\gamma \beta}(\lambda;x)$, $k,\alpha,\beta,\gamma=1,\dots,Q$ (see Sec.\ref{Proposition}, item (d.3) of  Proposition).

In the next section (Sec.\ref{Section:Nw}) we 
represent a  generalization of the dressing method for the classical (2+1)-dimensional $N$-wave hierarchy. In  Sec. \ref{dimker0} we discuss the case $\dim{\mbox{ker}}\hat \Psi =0$ giving  the classical (2+1)-dimensional $N$-wave equation 
\begin{eqnarray} \label{Nw}
B^{(2)} v_{t} B^{(1)}-B^{(1)} v_{t} B^{(2)} +
[v_{y_1},B^{(2)}]-[v_{y_2},B^{(1)}] + [[v,B^{(1)}],[v,B^{(2)}]] = 0,
\end{eqnarray}
where $B^{(i)}$ are constant diagonal matrices, $v$ is $Q\times Q$ matrix field, $t$ is time and $y_i$, $i=1,2$, are space variables.  
Its generalization, system of 4-dimensional 1-st order quasilinear PDEs solvable by  our algorithm with $\dim{\mbox{ker}}\hat \Psi =1$, is proposed in  Sec.\ref{Proposition}. The whole system of nonlinear  PDEs derived there  may be separated  into two subsystems. The first subsystem is the complete system of  evolution equations for the $Q\times Q$ matrix fields $w^{(p)}$, $q^{(p)}$, $p=1,2$, $v$ and $u$:
\begin{eqnarray}\label{sum_w}
&&
s^{(4;w;4)}_{\alpha \beta} \partial_{t}  w^{(p)}_{\alpha \beta}
+
\sum_{m=1}^3 s^{(4;w;m)}_{\alpha \beta}\partial_{y_{m}} 
 w^{(p)}_{\alpha \beta} - 
\sum_{{\gamma=1}\atop{\gamma\neq \beta}}^{Q}
 w^{(p)}_{\alpha \gamma} 
  v_{\gamma \beta} 
 T^{(4;wwv)}_{\alpha \gamma \beta}+ 
\\\nonumber
&&\sum_{{\gamma,\delta=1}\atop{\delta\neq\beta}}^{Q}
 q^{(p)}_{\alpha\gamma} 
  v_{\delta\beta} 
T^{(4;wqv)}_{\alpha \gamma\delta\beta } + 
\sum_{{\gamma=1}\atop{\gamma\neq \alpha}}^{Q}
\sum_{i_0=1}^2
 q^{(p)}_{\alpha\gamma} 
  w^{(i_0)}_{\gamma \beta}
 T^{(4;wqw;i_0)}_{\alpha \gamma \beta} 
 =0,
\\\label{sum_q}
&& 
s^{(4;q;4)}_\alpha \partial_t q^{(p)}_{\alpha \beta} +
\sum_{m=1}^2 s^{(4;q;m)}_\alpha  \partial_{y_{m}} 
 q^{(p)}_{\alpha \beta} -
\sum_{\gamma=1}^{Q}
 w^{(p)}_{\alpha \gamma} 
  u_{\gamma \beta} 
 T^{(4;qwu)}_{\alpha \gamma}  +
\\\nonumber
&&\sum_{{\gamma,\delta=1}}^{Q}
 q^{(p)}_{\alpha \gamma} 
  u_{\delta \beta}
  T^{(4;qqu)}_{\alpha \gamma \delta} 
  +
\sum_{{\gamma=1}\atop{
\gamma\neq \alpha}}^{Q}
\sum_{i_0=1}^2
 q^{(p)}_{\alpha\gamma} 
  q^{(i_0)}_{\gamma \beta}
  T^{(4;qqq;i_0)}_{\alpha \gamma}  
=0,
\end{eqnarray}
\begin{eqnarray}
\label{sum_v}
&&
s^{(4;v;4)}_{\alpha\beta}  \partial_{t}  v_{\alpha\beta} +
\sum_{m=1}^2  s^{(4;v;m)}_{\alpha\beta}  \partial_{y_{m}}  v_{\alpha\beta} -
\sum_{{\gamma=1}\atop{\gamma\neq \alpha\neq \beta}}^{Q}
 v_{\alpha \gamma}
  v_{\gamma\beta}
 T^{(4;vvv)}_{\alpha \gamma\beta}  +
\\\nonumber
&&
\sum_{{\gamma,\delta=1}\atop{ \delta\neq \alpha\neq \beta}}^{Q}
 u_{\alpha\gamma} 
  v_{\delta\beta}
 T^{(4;vuv)}_{\alpha\gamma \delta\beta }    +
\sum_{{\gamma=1}}^{Q}
 u_{\alpha\gamma} 
 \sum_{i_0=1}^2 w^{(i_0)}_{\gamma\beta}
 T^{(4;vuw;i_0)}_{\alpha\gamma\beta } 
  =0,\;\;\alpha\neq\beta,
  \\\label{sum_u}
&&  s^{(4;u;4)}_{\alpha} 
\partial_{t}  u_{\alpha\beta} +
s^{(4;u;1)}_{\alpha}
\partial_{y_1}  u_{\alpha\beta} -
\sum_{{\gamma=1}\atop{\gamma\neq \alpha}}^{Q}
 v_{\alpha \gamma} 
  u_{\gamma\beta}
T^{(4;uvu)}_{\alpha\gamma}
 +
\\\nonumber
&&
\sum_{{\gamma,\delta=1}\atop{\delta\neq \alpha}}^{Q}
 u_{\alpha\gamma} 
  u_{\delta\beta}
  T^{(4;uuu)}_{\alpha\gamma\delta}
  +
\sum_{{\gamma=1}}^{Q}
 \sum_{i_0=1}^2
 u_{\alpha\gamma} 
  q^{(i_0)}_{\gamma \beta}
  T^{(4;uuq;i_0)}_{\alpha\gamma}
 =0,
\end{eqnarray}
The equations of the second subsystem  must be viewed as a symmetry (i.e. compatible) constraints  to the system (\ref{sum_w}-\ref{sum_u}):
\begin{eqnarray}
\label{sum_qs}
&& 
s^{(3;q;3)}_\alpha \partial_{y_3} q^{(p)}_{\alpha \beta} +
\sum_{m=1}^2 s^{(3;q;m)}_\alpha  \partial_{y_{m}} 
 q^{(p)}_{\alpha \beta} -
\sum_{\gamma=1}^{Q}
 w^{(p)}_{\alpha \gamma} 
  u_{\gamma \beta} 
 T^{(3;qwu)}_{\alpha \gamma}  +
\\\nonumber
&&\sum_{{\gamma,\delta=1}}^{Q}
 q^{(p)}_{\alpha \gamma} 
  u_{\delta \beta}
  T^{(3;qqu)}_{\alpha \gamma \delta} 
  +
\sum_{{\gamma=1}\atop{
\gamma\neq \alpha}}^{Q}
\sum_{i_0=1}^2
 q^{(p)}_{\alpha\gamma} 
  q^{(i_0)}_{\gamma \beta}
  T^{(3;qqq;i_0)}_{\alpha \gamma}  
=0,
\end{eqnarray}
\begin{eqnarray}
\label{sum_vs}
&&
s^{(3;v;3)}_{\alpha\beta}  \partial_{y_3}  v_{\alpha\beta} +
\sum_{m=1}^2  s^{(3;v;m)}_{\alpha\beta}  \partial_{y_{m}}  v_{\alpha\beta} -
\sum_{{\gamma=1}\atop{\gamma\neq \alpha\neq \beta}}^{Q}
 v_{\alpha \gamma}
  v_{\gamma\beta}
 T^{(3;vvv)}_{\alpha \gamma\beta}  +
\\\nonumber
&&
\sum_{{\gamma,\delta=1}\atop{ \delta\neq \alpha\neq \beta}}^{Q}
 u_{\alpha\gamma} 
  v_{\delta\beta}
 T^{(3;vuv)}_{\alpha\gamma \delta\beta }    +
\sum_{{\gamma=1}}^{Q}
 u_{\alpha\gamma} 
 \sum_{i_0=1}^2 w^{(i_0)}_{\gamma\beta}
 T^{(3;vuw;i_0)}_{\alpha\gamma\beta } 
  =0,\;\;\alpha\neq\beta,\\\label{sum_us}
&&  s^{(j;u;j)}_{\alpha} 
\partial_{y_j}  u_{\alpha\beta} +
s^{(4;u;1)}_{\alpha}
\partial_{y_1}  u_{\alpha\beta} -
\sum_{{\gamma=1}\atop{\gamma\neq \alpha}}^{Q}
 v_{\alpha \gamma} 
  u_{\gamma\beta}
T^{(j;uvu)}_{\alpha\gamma}
 +
\\\nonumber
&&
\sum_{{\gamma,\delta=1}\atop{\delta\neq \alpha}}^{Q}
 u_{\alpha\gamma} 
  u_{\delta\beta}
  T^{(j;uuu)}_{\alpha\gamma\delta}
  +
\sum_{{\gamma=1}}^{Q}
 \sum_{i_0=1}^2
 u_{\alpha\gamma} 
  q^{(i_0)}_{\gamma \beta}
  T^{(j;uuq;i_0)}_{\alpha\gamma}
 =0,\;\;j=2,3.  
\end{eqnarray}
Although eqs.(\ref{sum_w}-\ref{sum_u}) represent the complete system for unknown fields,  the dressing method provides only such solutions  which satisfy the symmetry constraints (\ref{sum_qs}-\ref{sum_us}).
Here the constant parameters $T$ and $s$ are expressed in terms of the dressing data, see item (c) of Proposition,  eqs.(\ref{Prop:Ap1})  for details. Not all of them may be arbitrary. 
This system has the remarkable reduction:
if $u=0$
then the eq.(\ref{sum_v}) yields the $N$-wave equation (\ref{Nw}) while the eq.(\ref{sum_q}) yields the following linearizable system
\begin{eqnarray}
\label{Prop:3_Nw_w2_red0}
&&
\partial_{\tau_\alpha} 
q^{(p)}_{\alpha \beta}  +
\sum_{{\gamma=1}\atop{\gamma\neq \alpha}}^{Q}
\sum_{i_0=1}^2
q^{(p)}_{\alpha\gamma} 
 q^{(i_0)}_{\gamma \beta}
 T^{(j;qqq;i_0)}_{\alpha \gamma} =0,\;\;j=3,4,
\end{eqnarray}
where $\tau_\alpha$ are characteristic variables, see eq.(\ref{Prop:3_Nw_w2_red}) for more details.
Linearization of this system will be demonstrated  in  Sec.\ref{Section:expl_sol_1}.

Note, that another system of  PDEs having $S$- and $C$-integrable systems as particular reductions has been derived in \cite{Z}. 
Similar to the PDEs derived in \cite{Z}, the system (\ref{sum_w}-\ref{sum_u})  consist of 4 types of  nonlinear equations, which defer by dimensionalities of their linear parts. It is  2 in the eq.(\ref{sum_u}), 3 in the eqs.(\ref{sum_q},\ref{sum_v}) and 4 in  the eq.(\ref{sum_w}).  
However, the structure of the nonlinear PDEs is significantly different. In particular, some coefficients of the system (\ref{sum_w}-\ref{sum_u}) depend on the parameters $R^{(1\delta)}_{\gamma 1}$, $\gamma,\delta=1,\dots,Q$, reflecting the fact that $\dim{\mbox{ker}}\hat \Psi=1$ (see 
eq.(\ref{Prop:2_condition_Nw_red})). Such parameters do not appear in \cite{Z} since $\dim{\mbox{ker}}\hat \Psi=0$ therein.

 Let us clarify which kind of initial-boundary value  problem may be solved, in principle, numerically.
It may be easily seen that the fields $w^{(p)}$ may be given  arbitrary initial conditions (i.e. values at $t=0$)  in the whole  three dimensional space $(y_1,y_2,y_3)$.  The fields  $q^{(p)}$ and $v$ may be given arbitrary initial conditions only  on the plane (for instance, $y_3=0$) because of the constraints (\ref{sum_qs},\ref{sum_vs}). In order to define their initial conditions  in the whole space  $(y_1,y_2,y_3) $ we  must solve the system (\ref{sum_qs},\ref{sum_vs}).
 Field $u$ is even more restrictive because of the constraint (\ref{sum_us}). Initial condition for $u$ may be arbitrary on the line ( for instance, $y_3=y_2=0$). To define initial condition in the whole space $(y_1,y_2,y_3)$  we must solve the system  (\ref{sum_us}).  Finally,  $t$-evolution of all fields may be established solving the system (\ref{sum_w}-\ref{sum_u}). 
We conclude that correctly formulated initial-boundary value problem requires the following initial-boundary conditions:
two arbitrary matrix functions of 3 variables $w^{(p)}|_{t=0}$, $p=1,2$,  three  arbitrary matrix  functions of 2 variables $q^{(p)}|_{t=y_3=0}$, $p=1,2$ and  $v|_{t=y_3=0}$,  and single arbitrary matrix  function of one variable $u|_{t=y_3=y_2=0}$.
If such initial-boundary  data are provided by the dressing method we would consider the system (\ref{sum_w}-\ref{sum_u}) with compatible constraints (\ref{sum_qs}-\ref{sum_us}) as a completely integrable one.

However, we will show that the represented dressing algorithm may
 supply only  single arbitrary matrix  function of
  three variables, two arbitrary matrix functions of  two variables and single arbitrary matrix function of one variable. 
      For this reason, using our dressing algorithm,  we
      are not able to prescribe  arbitrary initial data to all
       fields. This fact   causes us to refer to such equations
        as partially integrable PDEs.
 It will be shown by construction that  they  admit infinitely many commuting flows.
 
 Remark, that another  system of nonlinear evolution PDEs  with symmetry constraints for some fields in the context of the dressing method has been derived in \cite{ZS3}.
 
  Detailed  derivation of the system (\ref{sum_w}-\ref{sum_us})   is given in  Appendix, Sec.\ref{Appendix1}.  
Sec.\ref{Solutions}  describes the solution space to the 4-dimensional system (\ref{sum_w}-\ref{sum_us}).  Conclusions are represented in  Sec.\ref{Conclusions}. 
 
%%%%%%%%%%%%%%%%%%%
\section{First order nonlinear PDEs: higher dimensional generalization of the $S$-integrable $N$-wave  equation}
%%%%%%%%%%%%%%%%%%%
\label{Section:Nw}

Dressing method for the (2+1)-dimensional $N$-wave equation can be generalized  to higher dimensions. Here we consider a variant of such generalization. For this purpose,  
function $\Psi^{(s)}(\lambda,\mu;x)$ in (\ref{U})  must be related with the  dressing 
functions $\Phi^{(sm)}(\lambda;x)$ and $C^{(m)}(\mu;x)$ by the following bi-linear system of 
equations introducing dependence on the variables $x_i$,
$i=1,2,\dots$:
\begin{eqnarray}\label{x}
\partial_{x_j}\Psi^{(s)}(\lambda,\mu;x) - {{A}}^{(j)} \partial_{x_1}\Psi^{(s)}(\lambda,\mu;x)   = \sum_{k=1}^{Q}\Phi^{(sk)}(\lambda;x) 
B^{(kj)} C^{(k)}(\mu;x),\\\nonumber
B^{(k2)}=I,\;\;j\ge 2.
\end{eqnarray}
Here $I$ is the identity matrix, $B^{(kj)}$ and ${{A}}^{(j)}$ are  constant diagonal matrices, dressing functions
$C^{(k)}(\lambda;x)$, $k=1,\dots,Q$,  will be specified below.
As far as the system  (\ref{x}) is overdetermined system of PDEs for the functions $\Psi^{(s)}$, this system  must be compatible. 
Compatibility condition reads
\begin{eqnarray}\label{comp0}
&&
\sum_{k=1}^{Q} \left[
\partial_{x_j}\Big(\Phi^{(sk)}(\lambda;x) B^{(ki)} C^{(k)}(\mu;x)\Big) -
\partial_{x_i}\Big(\Phi^{(sk)}(\lambda;x) B^{(kj)}
C^{(k)}(\mu;x)\Big)-\right.\\\nonumber
&&
\left.
{{A}}^{(j)}
\partial_{x_1}\Big(\Phi^{(sk)}(\lambda;x) B^{(ki)}
C^{(k)}(\mu;x)\Big) +
{{A}}^{(i)}\partial_{x_1}\Big(\Phi^{(sk)}(
\lambda;x) B^{(kj)} C^{(k)}(\mu;x)\Big)\right]=0.
\end{eqnarray}
Eq.(\ref{comp0}) consists of terms represented by  products of functions depending on  single spectral parameter. We would like to separate PDEs involving different spectral parameters, i.e. either parameter  $\lambda$ or $\mu$. This is possible due to the diagonal form of $\Phi^{(sm)}$ which provides commutativity of $A^{(i)}$ and  $\Phi^{(sk)}$:
\begin{eqnarray}\label{Phi0}
{{A}}^{(i)}\Phi^{(sk)}(\lambda;x) = 
\Phi^{(sk)}(\lambda;x)
A^{(i)},\;\;\forall \;i,s,k,
\end{eqnarray}
We put $i=2$ in (\ref{comp0}) without loss of generality and 
separate PDEs  involving different spectral parameters: 
\begin{eqnarray}\label{Phi}
\partial_{x_j}\Phi^{(sk)}(\lambda;x)  &=&  
\partial_{x_1}\Phi^{(sk)}(\lambda;x) 
P^{(kj)}+\partial_{x_2}\Phi^{(sk)}(\lambda;x) 
B^{(kj)},\;\;\\\label{C}
 \partial_{x_j}C^{(k)}(\mu;x)  &=&
P^{(kj)}\partial_{x_1}C^{(k)}(\mu;x)+B^{(kj)} \partial_{x_2}C^{(k)}(\mu;x)
\\\label{P_Nw}
&&
P^{(kj)} =   A^{(j)}-A^{(2)} B^{(kj)},\;\;j>2
\end{eqnarray}

Eqs.(\ref{Phi0}) and (\ref{Phi})  must be used  for the derivation of nonlinear PDEs as follows. First, substitute  $\Phi^{(sm)}$ from (\ref{U}) into (\ref{Phi0}):
\begin{eqnarray}\label{nl_01}
&&
\sum_{n=1}^{Q} {{A}}^{(j)}
\Psi^{(sn)}(\lambda,\mu;x)*U^{(nk)}(\mu;x) - \sum_{n=1}^{Q}  
\Psi^{(sn)}(\lambda,\mu;x)*U^{(nk)}(\mu;x)
A^{(j)}=0,
\end{eqnarray}
Second, substitute  $\Phi^{(sk)}$ from (\ref{U}) into
(\ref{Phi}) (we use (\ref{Phi0}) to result in (\ref{nl_02})):
\begin{eqnarray}
\label{nl_02}
&&
\sum_{n=1}^{Q}\Big[\partial_{x_j}\big(\Psi^{(sn)}(\lambda,\mu;x)*U^{(nk)}(\mu;x)\big) -{{A}}^{(j)}
\partial_{x_1} \big(\Psi^{(sn)}(\lambda,\mu;x)*U^{(nk)}(\mu;x) \big) -  \\\nonumber
&& 
\Big(\partial_{x_2}\big(\Psi^{(sn)}(\lambda,\mu;x)*U^{(nk)}(\mu;x)\big) -  {{A}}^{(2)}
\partial_{x_1}\big(\Psi^{(sn)}(\lambda,\mu;x)*U^{(nk)}(\mu;x)\big)\Big) B^{(kj)}\Big]=0.
\end{eqnarray}
Eqs.(\ref{nl_01},\ref{nl_02}) must be reduced  
to  homogeneous equations (see eq.(\ref{hom_Nw})). For this purpose we 
substitute $\Psi^{(s)}_{x_j}$ from (\ref{x}) into (\ref{nl_01},\ref{nl_02}) which results in the next homogeneous equations:

\begin{eqnarray}\label{hom_Nw}
\sum_{n=1}^Q \Psi^{(sn)}(\lambda,\mu;x) * L^{(ji;n k)}(\mu;x) =0,\;\;i=1,2,
\end{eqnarray}
where
\begin{eqnarray}\label{2_vec_Nw}\label{Q1E1}
L^{(j1;nk)}(\lambda;x)
&\equiv&  U^{(nk)}
 {\cal{B}}^{(j;n)},\;\;j\ge 2,
 \\\label{2_vec_Nw2}\label{Q1E2}
L^{(j2;nk)}(\lambda;x)&\equiv&\partial_{x_j} { U}^{(nk)}  + 
\partial_{x_1} { U}^{(nk)}{\cal{B}}^{(kj;n)}- 
\partial_{x_2} { U}^{(nk)}B^{(kj)} + 
\\\nonumber
&&
\sum_{i_1=1}^Q{ U}^{(ni_1)}
 \Big( B^{(i_1j)} v^{(ni_1k)} - v^{(ni_1k)}B^{(kj)}\Big) , \;\;j\ge 3,
\end{eqnarray}
fields $v^{(lik)}$ are introduced by the formulae
\begin{eqnarray}\label{v}
v^{(lik)}(x)=C^i(\mu;x)*U^{(lk)}(\mu;x), 
\end{eqnarray}
and the diagonal matrices ${\cal{B}}^{(j;n)}$ and ${\cal{B}}^{(kj; n)}$ are given by the formulae
\begin{eqnarray}\label{hatB_Nw}
  {\cal{B}}^{(j;n)} =
A^{(j)}_n I  -A^{(j)},
\;\;\;
 {\cal{B}}^{(kj; n)}=
B^{(kj)} A^{(2)}_n - A^{(j)}_n I.
\end{eqnarray}

In addition, applying 
 operators
 $(\partial_{x_j} - {{A}}^j \partial_{x_1})$ to the 
 eq.(\ref{hom_Nw}) with $i=1$ and $j=j_0$, one gets one more homogeneous equation:
\begin{eqnarray}\label{2_hom_Nw2}
\sum_{n=1}^Q \Psi^{(sn)}(\lambda,\mu;x) * L^{(j3;n k)}(\mu;x) =0,
\end{eqnarray}
where
\begin{eqnarray}\label{2_vec_Nw3}\label{Q2tE}
L^{(j3;nk)}(\lambda;x)\equiv\Big( \partial_{x_j}  U^{(nk)}  -
 \partial_{x_1}  U^{(nk)}
 A^{(j)}_n
 +
 \sum_{i_1=1}^Q U^{(ni_1)} B^{(i_1j)}
  v^{(ni_1k)}  \Big)
 {\cal{B}}^{(j_0;n)},\;\;j\ge 2.
 \end{eqnarray}
Comparing the eq.(\ref{P_Nw}) with the eq.(\ref{hatB_Nw}) we see that
\begin{eqnarray}
P^{(kj)}_\alpha=- {\cal{B}}^{(kj;\alpha)}_\alpha
.
\end{eqnarray}

Although this paper is devoted to the integral equation (\ref{U}) 
 with $\dim{\mbox{ker}}\;\hat \Psi^{(sn)}>0$, we consider the case 
 $\dim{\mbox{ker}}\;\hat\Psi^{(sn)}=0$ 
 in the next subsection for the sake of completeness. 
 The case $\dim{\mbox{ker}}\; \hat\Psi^{(sn)}=1$ will be considered 
 in  Sec.\ref{Section:Nw:dim1}.

%%%%%%%%%%%%%%
\subsection{ $\dim{\mbox{ker}}\;{\hat \Psi^{(sn)}} = 0$. $S$-integrable (2+1)-dimensional $N$-wave equation }
\label{dimker0}
If $\dim{\mbox{ker}}\;{\hat \Psi^{(sn)}} = 0$ then the
homogeneous equation 
\begin{eqnarray}\label{hom}
\sum_{n=1}^{Q}  \hat \Psi^{(sn)} {\bf H}^{(nk)} =0
\end{eqnarray}
has only the trivial solution ${\bf H}^{(nk)}=0$, i.e.
\begin{eqnarray}\label{h_2_vec_Nw}
L^{(j1;nk)}&:=& U^{(nk)}_{\alpha\beta}
 {\cal{B}}^{(j;n)}_\beta =0 
 \\\label{h_2_vec_Nw2}
L^{(j2;nk)}&:=&
\partial_{x_j} { U}^{(nk)}_{\alpha\beta}  + 
\partial_{x_1} { U}^{(nk)}_{\alpha\beta}{\cal{B}}^{(kj;n)}_\beta - 
\partial_{x_2} { U}^{(nk)}_{\alpha\beta}B^{(kj)}_\beta + 
\\\nonumber
&&
\sum_{i,\gamma=1}^Q{ U}^{(ni)}_{\alpha\gamma}
   v^{(nik)}_{\gamma\beta}\Big(B^{(ij)}_\gamma - B^{(kj)}_\beta\Big) =0,
\\\label{h_2_vec_Nw3}
L^{(j3;nk)}&:=&
\left( \partial_{x_j}  U^{(nk)}_{\alpha\beta}  -
 \partial_{x_1}  U^{(nk)}_{\alpha\beta} 
 A^{(j)}_n
 +
 %\\\nonumber
 %&&
 \sum_{i,\gamma=1}^Q U^{(ni)}_{\alpha\gamma}
  v^{(nik)}_{\gamma\beta}  B^{(ij)}_\gamma\right)
 {\cal{B}}^{(j_0;n)}_\beta =0,\;\;j\ge 2.
 \end{eqnarray}
Since $ {\cal{B}}^{j;n}_n =0$,  eq.(\ref{h_2_vec_Nw})
 tells us that  $ U^{(nk)}_{\alpha\beta}=0$ if $n\neq \beta$, and consequently $ v^{(nik)}_{\alpha\beta}=0$ if $n\neq \beta$.
Then, eq.(\ref{h_2_vec_Nw3}) is identical to zero, while
eqs.(\ref{h_2_vec_Nw2}) with different values of $\beta$ become decoupled:
\begin{eqnarray}\label{dimker0_sp}
V_{t_j} (\lambda;x)- V_{t_2}(\lambda;x) B^{(j)} -  V(\lambda;x) \;[v(x),B^{(j)}]=0,\;\;j=3,4,\dots,
\end{eqnarray}
where, for fixed $\beta$, 
\begin{eqnarray}
&&
V_{\alpha k}=U^{(\beta k)}_{\alpha\beta},\;\;v_{ik}= v^{(\beta ik)}_{\beta\beta},
B^{(j)}_k=B^{(kj)}_\beta,\;\;\alpha,i,k=1,\dots,Q,\\\nonumber
&&
\partial_{t_j}= \partial_{x_j} - A^{(j)}_\beta \partial_{x_1} ,\;\;\;j=2,4,\dots.
\end{eqnarray}
Eqs.(\ref{dimker0_sp}) represent the linear overdetermined system for the spectral function $V$. Compatibility condition of (\ref{dimker0_sp}) yields the classical $S$-integrable (2+1)-dimensional $N$-wave equation:
\begin{eqnarray}
[v_{t_k},B^{(j)}]-[v_{t_j},B^{(k)}] + B^{(j)} v_{t_2} B^{(k)}-B^{(k)} v_{t} B^{(j)} + [[v,B^{(k)}],[v,B^{(j)}]] = 0,\;\;k\neq j=3,4,\dots.
\end{eqnarray}
This example justifies the fact that our dressing algorithm with
$\dim{\mbox{ker}} \; \hat \Psi^{(sn)} =0$ gives rise to the classical $S$-integrable models.

%%%%%%%%%%%%%%%
\subsection{$\dim{\mbox{ker}} \;\hat\Psi^{(sn)} =1 $. Higher dimensional nonlinear PDEs}
\label{Section:Nw:dim1}
\label{Proposition}
Let  $\dim {\mbox{ker}} \;\hat\Psi^{(sn)}  =1 $. Then the solution
space of the homogeneous equation (\ref{hom}) is parametrized by the arbitrary  $Q\times Q$ matrix
 functions $f^{(ik)}(x)$, $i,k=1,\dots, Q$:
\begin{eqnarray}\label{h}
(U^h)^{(nk)}(\lambda;x) = \sum_{i=1}^{Q} 
H^{(ni)}(\lambda;x) f^{(ik)}(x).
\end{eqnarray}
This equation establishes a linear relation between any two solutions $U^{(nk)}$, 
$L^{(j1;nk)}$, $L^{(j3;nk)}$, $j\ge 2$ and $L^{(j2;nk)}$, 
$j\ge 3$ of the homogeneous
integral  equation (\ref{hom}). These linear relations represent a new overdetermined system of linear equations for the spectral functions $ U^{(nk)}$. Details of derivation of the system of linear PDEs for the spectral function $ U^{(nk)}$ as well as derivation of the assotiated system of nonlinear PDEs are given in Appendix, Sec.\ref{Appendix1}. Here we collect all basic results in the following Proposition. 

{\bf Proposition.}
Let $\dim {\mbox{ker}} \;\hat \Psi^{(sn)}  =1 $
in the integral equation (\ref{U}) where the dressing functions  $\Psi^{(s)}$, $\Phi^{(sk)}$ and $C^{(k)}$ are solutions of the  eqs.(\ref{x},\ref{Phi0},\ref{Phi},\ref{C}). Then

\noindent
a) Matrix functions of $x$ (in other words, fields)   may be introduced by the formulae
\begin{eqnarray}\label{vw}
&&
 v^{(nik)}(x)=C^{(i)}(\lambda;x)* U^{(nk)}(\lambda;x),\;\; v^{(nik;1)}(x)=C^{(i)}_{x_1}(\lambda;x)* U^{(nk)}(\lambda;x),\;\;\\\nonumber
&&
 w^{(nk)}(x)=G(\lambda;x)*  U^{(nk)}(\lambda;x),\;\;
 w^{(nk;p)}(x)= G_{x_p}(\lambda;x)*  U^{(nk)}(\lambda;x),\;\;\\\nonumber
&& w^{(nk;ps)}(x)= G_{x_px_s}(\lambda;x)*  U^{(nk)}(\lambda;x),\dots, 
\end{eqnarray}
where  $n,i,k=1,\dots,Q$, $p,s=1,2,\dots$,  function $G(\lambda;x)$ is the external dressing function, whose role will be clarified below, see Secs.\ref{Solutions} and \ref{Appendix1}. These fields are related with each other  by the system of nonlinear PDEs, see eqs.(\ref{2_vec_Nw_22}-\ref{2_vec_Nw_24},\ref{2_vec_Nw_w}-\ref{2_vec_Nw_w_p22}) in  Appendix. If $G$ is arbitrary, then the system of nonlinear PDEs may not be completed.

\noindent
b) Presence of  arbitrary matrix functions 
$f^{(ik)}(x)$, $i,k=1,\dots,Q$ in the solution space of the homogeneous eq.(\ref{hom}) (which is indicated in the  eq. (\ref{h})) allows one to impose a largely arbitrary relation among the matrix  fields (\ref{vw}):
\begin{eqnarray}\label{Prop:2_condition_Nw}
{\cal{F}}^{nk}({\mbox{all fields }}) =0,
\;\;n,k=1,\dots,Q,
\end{eqnarray}
where ${\cal{F}}^{nk}$ are $Q\times Q$ matrices.
The only requirement to this relation is its resolvability with respect to $f^{(ik)}(x)$. 

\noindent
c) Let relation (\ref{Prop:2_condition_Nw}) be taken in the next simple form
\begin{eqnarray}\label{Prop:2_condition_Nw_red}
  w^{(nk)} = R^{(nk)}
\end{eqnarray}
(where $R^{(nk)}$, $n,k=1,\dots,Q$, are constant non-degenerate
 $Q\times Q$ matrices), and  the external dressing function $G(\lambda;x)$
satisfies the following system of linear PDEs
\begin{eqnarray}\label{Prop:2_DimG}
G_{x_j}(\lambda;x)={\cal{P}}^{(j1)}  G_{x_1}(\lambda;x)+{\cal{P}}^{(j2)}  G_{x_2}(\lambda;x),\;\; j>2 
\end{eqnarray}
with  arbitrary diagonal matrices ${\cal{P}}^{(ji)}$, $i=1,2$. Replace $x_i$ by  the new independent variables 
\begin{eqnarray}
\label{Prop:y}\label{Prop:t}
\partial_{t_{j}}=\partial_{x_{j}}-A^{(j)}_1\partial_{x_1} ,\;\;j\ge 2\;\;\Rightarrow\;\;
t=t_5, \;\;\;y_j=t_{j+1},\;\;j=1,2,3.
\end{eqnarray}
Then the complete system of nonlinear PDEs can be derived for the
following fields 
\begin{eqnarray}\label{new_fields}
&& 
w^{(p)}_{\alpha \beta}\equiv
w^{(1\beta;p)}_{\alpha 1}, \;\;
\;\; 
 q^{(p)}_{\alpha \beta}\equiv \sum_{\gamma, \delta=1}^Q 
 w^{(1\delta;p)}_{\alpha \gamma} {\cal{B}}^{(2;1)}_\gamma (\hat 
 R^{-1})^{(\delta 1)}_{\gamma \beta}
 \;\;k,\alpha=1,\dots,Q, \;\;p=1,2.
\\\nonumber
&&
v_{\alpha\beta}\equiv v^{(1\alpha\beta)}_{11},\;\; 
u_{\alpha  \beta}\equiv \sum_{\gamma, \delta=1}^Q v^{(1\alpha\delta)}_{1\gamma} 
{\cal{B}}^{(2;1)}_\gamma (\hat R^{-1})^{(\delta1)}_{\gamma \beta},
 \;\;i, \beta=1,\dots,Q.
\end{eqnarray}
 This system of PDEs is naturally separated  into two 
 subsystems. 
The first subsystem is the complete system of   evolution  PDEs:
\begin{eqnarray}\label{Prop:3_Nw_w}\label{Prop:sum_w}
E^{(56;1\beta p)}_{\alpha 1}&:=&  
s^{(4;w;4)}_{\alpha \beta} \partial_{t}  w^{(p)}_{\alpha \beta}
+
\sum_{m=1}^3 s^{(4;w;m)}_{\alpha \beta}\partial_{y_{m}} 
 w^{(p)}_{\alpha \beta} - 
\sum_{{\gamma=1}\atop{\gamma\neq\beta}}^{Q}
 w^{(p)}_{\alpha \gamma} 
  v_{\gamma \beta} 
 T^{(4;wwv)}_{\alpha \gamma \beta}+ 
\\\nonumber
&&\sum_{{\gamma,\delta=1}\atop{\delta\neq\beta}}^{Q}
 q^{(p)}_{\alpha\gamma} 
  v_{\delta\beta} 
T^{(4;wqv)}_{\alpha \gamma\delta\beta } + 
\sum_{{\gamma=1}\atop{\gamma\neq \alpha}}^{Q}
\sum_{i_0=1}^2
 q^{(p)}_{\alpha\gamma} 
  w^{(i_0)}_{\gamma \beta}
 T^{(4;wqw;i_0)}_{\alpha \gamma \beta} 
 =0,
\\\label{Prop:3_Nw_w2}\label{Prop:sum_q}
\tilde E^{(57;1\beta p)}_{\alpha }&:=&  
s^{(4;q;4)}_\alpha \partial_t q^{(p)}_{\alpha \beta} +
\sum_{m=1}^2 s^{(4;q;m)}_\alpha  \partial_{y_{m}} 
 q^{(p)}_{\alpha \beta} -
\sum_{\gamma=1}^{Q}
 w^{(p)}_{\alpha \gamma} 
  u_{\gamma \beta} 
 T^{(4;qwu)}_{\alpha \gamma}  +
\\\nonumber
&&\sum_{{\gamma,\delta=1}}^{Q}
 q^{(p)}_{\alpha \gamma} 
  u_{\delta \beta}
  T^{(4;qqu)}_{\alpha \gamma \delta} 
  +
\sum_{{\gamma=1}\atop{
\gamma\neq \alpha}}^{Q}
\sum_{i_0=1}^2
 q^{(p)}_{\alpha\gamma} 
  q^{(i_0)}_{\gamma \beta}
  T^{(4;qqq;i_0)}_{\alpha \gamma}  
=0,
\end{eqnarray}
\begin{eqnarray}
\label{Prop:3_Nw_0}\label{Prop:sum_v}
E^{(52;1\alpha\beta)}_{11}&:=& 
s^{(4;v;4)}_{\alpha\beta}  \partial_{t}  v_{\alpha\beta} +
\sum_{m=1}^2  s^{(4;v;m)}_{\alpha\beta}  \partial_{y_{m}}  v_{\alpha\beta} -
\sum_{{\gamma=1}\atop{\gamma\neq \alpha\neq \beta}}^{Q}
 v_{\alpha \gamma}
  v_{\gamma\beta}
 T^{(4;vvv)}_{\alpha \gamma\beta}  +
\\\nonumber
&&
\sum_{{\gamma,\delta=1}\atop{ \delta\neq\alpha\neq \beta}}^{Q}
 u_{\alpha\gamma} 
  v_{\delta\beta}
 T^{(4;vuv)}_{\alpha\gamma \delta\beta }    +
\sum_{{\gamma=1}}^{Q}
 u_{\alpha\gamma} 
 \sum_{i_0=1}^2 w^{(i_0)}_{\gamma\beta}
 T^{(4;vuw;i_0)}_{\alpha\gamma\beta } 
  =0,\;\;\alpha\neq\beta
,\\\label{Prop:3_Nw_01}\label{Prop:sum_u}
\tilde E^{(53;1\alpha\beta)}_{1}&:=&  
 s^{(4;u;4)}_{\alpha} 
\partial_{t}  u_{\alpha\beta} +
s^{(4;u;1)}_{\alpha}
\partial_{y_1}  u_{\alpha\beta} -
\sum_{{\gamma=1}\atop{\gamma\neq \alpha}}^{Q}
 v_{\alpha \gamma} 
  u_{\gamma\beta}
T^{(4;uvu)}_{\alpha\gamma}
 +
\\\nonumber
&&
\sum_{{\gamma,\delta=1}\atop{\delta\neq \alpha}}^{Q}
 u_{\alpha\gamma} 
  u_{\delta\beta}
  T^{(4;uuu)}_{\alpha\gamma\delta}
  +
\sum_{{\gamma=1}}^{Q}
 \sum_{i_0=1}^2
 u_{\alpha\gamma} 
  q^{(i_0)}_{\gamma \beta}
  T^{(4;uuq;i_0)}_{\alpha\gamma}
 =0,
\end{eqnarray}
The second subsystem must be viewed as a system of symmetry (i.e. compatible) constraints to the system (\ref{Prop:sum_w}-\ref{Prop:sum_u}):
\begin{eqnarray}\label{Prop:sum_qs}
\label{Prop_s:3_Nw_w2}
\tilde E^{(47;1\beta p)}_{\alpha}&:=&  
s^{(3;q;3)}_\alpha \partial_{y_3} q^{(p)}_{\alpha \beta} +
\sum_{m=1}^2 s^{(3;q;m)}_\alpha  \partial_{y_{m}} 
 q^{(p)}_{\alpha \beta} -
\sum_{\gamma=1}^{Q}
 w^{(p)}_{\alpha \gamma} 
  u_{\gamma \beta} 
 T^{(3;qwu)}_{\alpha \gamma}  +
\\\nonumber
&&\sum_{{\gamma\delta=1}}^{Q}
 q^{(p)}_{\alpha \gamma} 
  u_{\delta \beta}
  T^{(3;qqu)}_{\alpha \gamma \delta} 
  +
\sum_{{\gamma=1}\atop{
\gamma\neq \alpha}}^{Q}
\sum_{i_0=1}^2
 q^{(p)}_{\alpha\gamma} 
  q^{(i_0)}_{\gamma \beta}
  T^{(3;qqq;i_0)}_{\alpha \gamma}  
=0,
\end{eqnarray}
\begin{eqnarray}
\label{Prop_s:3_Nw_0}\label{Prop:sum_vs}
E^{(42;1\alpha\beta)}_{11}&:=& 
s^{(3;v;3)}_{\alpha\beta}  \partial_{y_3}  v_{\alpha\beta} +
\sum_{m=1}^2  s^{(3;v;m)}_{\alpha\beta}  \partial_{y_{m}}  v_{\alpha\beta} -
\sum_{{\gamma=1}\atop{\gamma\neq \alpha\neq \beta}}^{Q}
 v_{\alpha \gamma}
  v_{\gamma\beta}
 T^{(3;vvv)}_{\alpha \gamma\beta}  +
\\\nonumber
&&
\sum_{{\gamma,\delta=1}\atop{ \delta\neq\alpha\neq \beta}}^{Q}
 u_{\alpha\gamma} 
  v_{\delta\beta}
 T^{(3;vuv)}_{\alpha\gamma \delta\beta }    +
\sum_{{\gamma=1}}^{Q}
 u_{\alpha\gamma} 
 \sum_{i_0=1}^2 w^{(i_0)}_{\gamma\beta}
 T^{(3;vuw;i_0)}_{\alpha\gamma\beta } 
  =0,\;\;\alpha\neq\beta
,\\\label{Prop_s:3_Nw_01}\label{Prop:sum_us}
\tilde E^{((j+1)3;1\alpha\beta)}_{1}&:=&  
s^{(j;u;j)}_{\alpha} 
\partial_{y_j}  u_{\alpha\beta} +
s^{(j;u;1)}_{\alpha}
\partial_{y_1}  u_{\alpha\beta} -
\sum_{{\gamma=1}\atop{\gamma\neq \alpha}}^{Q}
 v_{\alpha \gamma} 
  u_{\gamma\beta}
T^{(j;uvu)}_{\alpha\gamma}
 +
\\\nonumber
&&
\sum_{{\gamma,\delta=1}\atop{\delta\neq \alpha}}^{Q}
 u_{\alpha\gamma} 
  u_{\delta\beta}
  T^{(j;uuu)}_{\alpha\gamma\delta}
  +
\sum_{{\gamma=1}}^{Q}
 \sum_{i_0=1}^2
 u_{\alpha\gamma} 
  q^{(i_0)}_{\gamma \beta}
  T^{(j;uuq;i_0)}_{\alpha\gamma}
 =0
, \;\;j=2,3.
\end{eqnarray}
Coefficients of this system are following (see also  formulae (\ref{coef0})):
\begin{eqnarray}\label{Prop:Ap1}
 && s^{(j-1;w;1)}_{\alpha \beta}=
 s^{(j;w;2;\beta)}_{\alpha1}= -
\left|\begin{array}{ccc}
 B^{(\beta 3)}_1 &
 B^{(\beta 4)}_1 &
 B^{(\beta j)}_1 \cr
  B^{(\beta 3)}_1 -{\cal{P}}^{(31)}_{\alpha} &
  B^{(\beta 4)}_1-{\cal{P}}^{(41)}_{\alpha}  &
  B^{(\beta j)}_1 -{\cal{P}}^{(j1)}_{\alpha} \cr
B^{(\beta 3)}_{1} -{\cal{P}}^{(32)}_{\alpha}  &
 B^{(\beta 4)}_{1}  -{\cal{P}}^{(42)}_{\alpha} &
  B^{(\beta j)}_{1}  -{\cal{P}}^{(j2)}_{\alpha}
\end{array}\right|,
\end{eqnarray}
\begin{eqnarray}\nonumber
&&s^{(j-1;w;m-1)}_{\alpha \beta }=
s^{(j;w;m;\beta )}_{\alpha1}=
\left|\begin{array}{ccc}
 \delta^{(m3)} & \delta^{(m4)}&\delta^{(mj)}\cr
  B^{(\beta 3)}_1 -{\cal{P}}^{(31)}_{\alpha} &
 B^{(\beta 4)}_1 -{\cal{P}}^{(41)}_{\alpha}  &
B^{(\beta j)}_1 -{\cal{P}}^{(j1)}_{\alpha}   \cr
B^{(\beta 3)}_{1} -{\cal{P}}^{(32)}_{\alpha}  &
 B^{(\beta 4)}_{1} -{\cal{P}}^{(42)}_{\alpha}  &
  B^{(\beta j)}_{1} -{\cal{P}}^{(j2)}_{\alpha} 
\end{array}
\right|,\;\;
m=3,4,j,
\end{eqnarray}
\begin{eqnarray}\nonumber
&&T^{(j-1;wwv)}_{\alpha \gamma  \beta }=
T^{(j;wv;\gamma \beta )}_{\alpha1}
 =
\left|\begin{array}{ccc}
 B^{(\beta 3)}_1 -B^{(\gamma 3)}_1 
&B^{(\beta 4)}_1 -B^{(\gamma 4)}_1 
&B^{(\beta j)}_1 -B^{(\gamma j)}_1 
\cr
B^{(\beta 3)}_1 -{\cal{P}}^{(31)}_{\alpha}  &
 B^{(\beta 4)}_1  -{\cal{P}}^{(41)}_{\alpha} &
  B^{(\beta j)}_1 -{\cal{P}}^{(j1)}_{\alpha} \cr
 B^{(\beta 3)}_{1}-{\cal{P}}^{(32)}_{\alpha}  &
  B^{(\beta 4)}_{1} -{\cal{P}}^{(42)}_{\alpha} &
  B^{(\beta j)}_{1} -{\cal{P}}^{(j2)}_{\alpha} 
\end{array}
\right|,\\\nonumber
&&
T^{(j-1;wqv)}_{\alpha \gamma \delta \beta }= 
R^{(1 \delta)}_{\gamma1}T^{(j-1;wwv)}_{\alpha \delta \beta }
\end{eqnarray}
\begin{eqnarray}\nonumber
&&
 T^{(j-1;wqw;i_0)}_{\alpha\gamma \beta }
 =T^{(j;ww;\beta ;i_0)}_{\alpha\gamma 1} 
 =
\left|\begin{array}{ccc}
 B^{(\beta 3)}_1 -{\cal{P}}^{(3i_0)}_{\gamma }  
& B^{(\beta 4)}_1 -{\cal{P}}^{(4i_0)}_{\gamma } 
&B^{(\beta j)}_1 -{\cal{P}}^{(ji_0)}_{\gamma }  
\cr
B^{(\beta 3)}_1-{\cal{P}}^{(31)}_{\alpha}   &
 B^{(\beta 4)}_1 -{\cal{P}}^{(41)}_{\alpha}  &
 B^{(\beta j)}_1  -{\cal{P}}^{(j1)}_{\alpha} \cr
B^{(\beta 3)}_{1} -{\cal{P}}^{(32)}_{\alpha}  &
 B^{(\beta 4)}_{1} -{\cal{P}}^{(42)}_{\alpha}  &
  B^{(\beta j)}_{1}  -{\cal{P}}^{(j2)}_{\alpha}
\end{array}
\right|,
 \end{eqnarray}
\begin{eqnarray}\nonumber
&&s^{(j-1;q;m-1)}_{\alpha}=
\tilde s^{(j;w;m;1)}_{\alpha}= 
\left|\begin{array}{ccc}
 \delta^{(m2)} & \delta^{(m3)}& \delta^{(mj)}\cr
 1 &
{\cal{P}}^{(31)}_\alpha &
 {\cal{P}}^{(j1)}_\alpha \cr
1 &
 {\cal{P}}^{(32)}_\alpha &
 {\cal{P}}^{(j2)}_\alpha 
\end{array}
\right|,\;\;
m=2,3,j,\;\;
\\
\nonumber
&&
 T^{(j-1;qwu)}_{\alpha \gamma } =\tilde T^{(j;wv;\gamma 1)}_{\alpha}
 =-
\left|\begin{array}{ccc}
 1
&B^{(\gamma 3)}_1
&B^{(\gamma j)}_1 
\cr
1 &
{\cal{P}}^{(31)}_\alpha &
 {\cal{P}}^{(j1)}_\alpha \cr
1 &
 {\cal{P}}^{(32)}_\alpha &
 {\cal{P}}^{(j2)}_\alpha 
\end{array}
\right|,
 \;\;
 T^{(j-1;qqu)}_{\alpha \gamma \delta} =R^{(1\delta)}_{\gamma 1}
  T^{(j-1;qwu)}_{\alpha \delta} ,
\\\nonumber
&&T^{(j-1;qqq;i_0)}_{\alpha\gamma } = 
\tilde T^{(j;ww;1;i_0)}_{\alpha\gamma } 
=-
\left|\begin{array}{ccc}
1
&{\cal{P}}^{(3i_0)}_{\gamma } 
& {\cal{P}}^{(ji_0)}_{\gamma } 
\cr
1 &
 {\cal{P}}^{(31)}_\alpha &
 {\cal{P}}^{(j1)}_\alpha \cr
1 &
 {\cal{P}}^{(32)}_\alpha &
 {\cal{P}}^{(j2)}_\alpha 
\end{array}
\right|,\;\;
%\;\;\Rightarrow \;\;\tilde T^{(j;ww;n;2 )}_{\alpha\alpha} =0 
\end{eqnarray}

 \begin{eqnarray}\nonumber
  &&
s^{(j-1;v;1)}_{\alpha \beta }=s^{(j;v;2;\alpha \beta )}_{1}= 
\left|\begin{array}{ccc}
B^{(\beta 3)}_1 &
 B^{(\beta j)}_1 \cr
B^{(\alpha 3)}_1 & 
B^{(\alpha j)}_1
\end{array}\right|,\;\;
\\\nonumber
&&
s^{(j-1;v;m-1)}_{\alpha \beta }=s^{(j;v;m;\alpha \beta )}_{1}= 
\left|\begin{array}{ccc}
 \delta^{(3m)} & \delta^{(jm)}\cr
B^{(\beta 3)}_1 -B^{(\alpha 3)}_1& 
B^{(\beta j)}_1 -B^{(\alpha j)}_1 
\end{array}
\right|,\;\;
m=3,j,
\end{eqnarray}

\begin{eqnarray}\nonumber
&& T^{(j-1;vvv)}_{\alpha  \gamma  \beta }=
T^{(j;vv; \alpha  \gamma \beta )}_{1} 
=
\left|\begin{array}{ccc}
 B^{(\beta 3)}_1 -B^{(\gamma 3)}_1
&B^{(\beta j)}_1 -B^{(\gamma j)}_1  
\cr
B^{(\beta 3)}_1 -B^{(\alpha 3)}_1& 
B^{(\beta j)}_1-B^{(\alpha j)}_1
\end{array}
\right|, \;\; T^{(j-1;vuv)}_{\alpha  \gamma \delta \beta } =
R^{(1 \delta)}_{\gamma 1} T^{(j-1;vvv)}_{\alpha \delta\beta },
%\;\;\Rightarrow \;\;T^{(j;vv; \alpha  \alpha \beta )}_{1} =0,
%\end{eqnarray}
%\begin{eqnarray}\nonumber
\\\nonumber
&&
T^{(j-1;vuw;i_0)}_{\alpha \gamma \beta } =
T^{(j;vw;\alpha  \beta ;i_0)}_{1\gamma}
=
\left|\begin{array}{ccc}
  B^{(\beta 3)}_{1} -{\cal{P}}^{(3i_0)}_{\gamma} 
& B^{(\beta j)}_{1}-{\cal{P}}^{(ji_0)}_{\gamma}
\cr
B^{(\beta 3)}_1 -B^{(\alpha 3)}_1& 
B^{(\beta j)}_1 -B^{(\alpha j)}_1
\end{array}
\right|
\end{eqnarray}
 \begin{eqnarray}
\nonumber
&&
s^{(j-1;u;m-1)}_{\alpha }=\tilde s^{(j;v;m;\alpha )}_{1} = -
\left|
\begin{array}{cc}
\delta^{(2 m)} & \delta^{(j m)}\cr
1 & 
B^{(\alpha j)}_1 
\end{array}
 \right|,\;\;m=2,j,
 \\\nonumber
 && T^{(j-1;uvu)}_{\alpha \gamma }=
 \tilde  T^{(j;vv; \alpha  \gamma 
 )}_1 
  =\left|
 \begin{array}{cc}
 1& 
 B^{(\gamma j)}_1  \cr
1 & 
B^{(\alpha j)}_1
\end{array}
 \right|,\;\;
 T^{(j-1;uuu)}_{\alpha \gamma \delta}=R^{(1\delta)}_{\gamma 1} 
 T^{(j-1;uvu)}_{\alpha \delta}
 %\;\;\Rightarrow \;\; \tilde  T^{(j;vv; i i)}_1 =0
% \end{eqnarray}
 %\begin{eqnarray}
 \\\nonumber
 &&T^{(j-1;uuq;i_0)}_{\alpha\gamma}=
   \tilde T^{(j;vw; \alpha ;i_0
 )}_{1\gamma}=\left|
 \begin{array}{cc}
 1& 
 {\cal{P}}^{(ji_0)}_{\gamma}
  \cr
1 & 
B^{(\alpha j)}_1 
\end{array}
 \right|,
 %,\;\;\; \tilde T^{(j;vw; \alpha ;2
 %)}_{1\gamma_2}=\left|
 %\begin{array}{cc}
 %1& {\cal{P}}^{(j2)}_{\gamma_2}
 % \cr 1 & B^{(\alpha j)}_1
%\end{array}
% \rght|
 \end{eqnarray}
 where $R^{(1i)}_{\gamma 1}$, $B^{(ij)}_1$ and ${\cal{P}}^{(ij)}_1$  are arbitrary constant parameters.

Dimensionality of the system (\ref{Prop:3_Nw_0}-\ref{Prop_s:3_Nw_01}) is defined by the eqs.(\ref{Prop:3_Nw_w}) which involve derivatives with respect to 4 independent variables, while other equations involve derivatives with respect to two and three variables.
Dressing algorithm supplies one arbitrary function of three independent variables (which fixes the initial datum for one of the functions $w^{(p)}$, $p=1,2$), two arbitrary functions of two variables (fixing the initial data for $v$ and for one of the  functions $q^{(p)}$, $p=1,2$)  and single arbitrary function of one variable (fixing the initial datum for $u$). 
Namely the system (\ref{Prop:3_Nw_w}-\ref{Prop_s:3_Nw_01}) is written in  Introduction, see eqs.(\ref{sum_w}-\ref{sum_us}).

The  system (\ref{Prop:sum_w}-\ref{Prop_s:3_Nw_01}) has the following evident properties:

\noindent
d.1) There are infinitely many commuting flows to this system.

\noindent
d.2)
Equations of this system  have differential polynomial structure, also the number of equations  is rather big. So, one has 24 scalar PDEs in the simplest case $Q=2$.

\noindent
d.3) Independent variable $x_1$ appears only in the combinations (\ref{Prop:t}), so that it does not increase dimensionality of nonlinear PDEs. This happens due to the decoupling of equations for fields with different values of the first superscript, which, in turns, is a consequence of the eq.(\ref{Prop:2_condition_Nw_red}). However, if one considers  another
 relation ( \ref{Prop:2_condition_Nw}) mixing fields with different values of the first superscript, then derivatives with respect to $x_1$ will appear explicitly  in nonlinear PDEs which will  become 5-dimensional. But such system of nonlinear PDEs will have more complicated structure.

\noindent
d.4) Dimensionality of the nonlinear  PDEs is determined by the dimensionality of the linear PDE (\ref{Prop:2_DimG}) and may be increased without any problem.
 
\noindent 
d.5) {\it Reduction to  the 
linearizable system of nonlinear PDEs.}  The system
 (\ref{Prop:3_Nw_w}-\ref{Prop_s:3_Nw_01}) admits  reduction
  $ v=u= 0$, which corresponds to $C^{(i)}\equiv 0$
   in the dressing algorithm. In this case the  system of
    PDEs reduces to the single eq.(\ref{Prop:3_Nw_w2}) which now reads:
\begin{eqnarray}
\label{Prop:3_Nw_w2_red}
s^{(4;q;4)}_\alpha \partial_t q^{(p)}_{\alpha \beta} +
\sum_{m=1}^2 s^{(4;q;m)}_\alpha  \partial_{y_{m}} 
 q^{(p)}_{\alpha \beta} +
\sum_{{\gamma=1}\atop{
\gamma\neq \alpha}}^{Q}
\sum_{i_0=1}^2
 q^{(p)}_{\alpha\gamma} 
  q^{(i_0)}_{\gamma \beta}
  T^{(4;qqq;i_0)}_{\alpha \gamma}  
=0.
\end{eqnarray}
Eq.(\ref{Prop_s:3_Nw_w2}) is a symmetry of the eq.(\ref{Prop:3_Nw_w2_red}).
We may introduce "characteristics variables" 
 $\partial_{\tau_\alpha}=
 s^{(4;q;4)}_{\alpha}\partial_{t} 
+
\sum_{m=1}^2  s^{(4;q;m)}_{\alpha}\partial_{y_{m}}$, $\alpha=1,\dots,Q$
which indicate that  the dimensionality of the eq.(\ref{Prop:3_Nw_w2_red})  is, essentially,
$\min(3,Q)$. The eq.(\ref{Prop:3_Nw_w2_red})  is "linearizable" because the matrix fields
$q^{(p)}$ are algebraically expressible in
terms of solutions  of some 
linear PDE, see Sec.\ref{Section:expl_sol_1}. Eq. (\ref{Prop:3_Nw_w2_red}) 
 has been written in  Introduction, see eq.(\ref{Prop:3_Nw_w2_red0}).
This equation is also partially integrable in the same sense as the system (\ref{Prop:3_Nw_w}-\ref{Prop_s:3_Nw_01}) does: dressing algorithm provides arbitrary  initial condition  only for one of the matrix  fields   $q^{(p)}$, $p=1,2$.

\paragraph{Proof:} The proofs of the items (a)-(c) and (d.1, d.5) are given in 
Appendix,  Sec.\ref{Appendix1}. The items (d.2-d.4) are self-consistent. Relation between solutions of the eq. (\ref{Prop:3_Nw_w2_red})   and solutions of the appropriate linear  PDE  (item d.5) is shown in  Sec.\ref{Section:expl_sol_1}.

%%%%%%%%%%%
\section{Solutions}
\label{Solutions}

In this section all superscripts and subscripts take values from
$1$ to $Q$ unless different is specified.
Subscripts do not always denote elements of matrices. We use  Greek letters for  matrix indexes in order to distinguish them from others.

Solutions of the eqs. (\ref{Phi},\ref{C}) read:
\begin{eqnarray}\label{Sol_Phi}
\Phi^{(sm)}(\lambda;x)&=& \int \Phi^{(sm)}_0(\lambda,\varkappa)
e^{K^{(m)} (\varkappa_1,\varkappa_2;x)} d
\varkappa,\\\nonumber
C^{(m)}(\lambda;x)&=&\int  e^{K^m (q_1,q_2;x)} C^{(m)}_0(\lambda,q)d
q,\\\nonumber
&&
K^{(m)} (\varkappa_1,\varkappa_2;x) = \big(\varkappa_1 x_1 + \varkappa_2 x_2 \big) + 
\sum_{j=3}^5 \big(P^{(mj)} \varkappa_1  +B^{(mj)} \varkappa_2 \big)x_i ,
\end{eqnarray}
where $\varkappa=(\varkappa_1,\varkappa_2)$, $q=(q_1,q_2)$.
 The dressing function  $\Psi^{(s)}$ must be taken as  solution of the 
 eq.(\ref{x}) with $k=2$:
 \begin{eqnarray}
 \Psi^{(s)}(\lambda,\mu;x) &=&\Psi^{(s)}_p(\lambda,\mu;x) +
 \Psi^{(s)}_h(\lambda,\mu;x) +\Sigma^{(s)}(\lambda,\mu) \\\nonumber
\Psi^{(s)}_p(\lambda,\mu;x)
 &=&
  \int 
 \sum\limits_{m=1}^{Q}  \Phi^{(sm)}_0(\lambda,\varkappa) \Big(\varkappa_2+q_2-(\varkappa_1+q_1)A^{(2)}\Big)^{-1}e^{K^{(m)}
  (\varkappa_1+q_1,\varkappa_2+q_2;x)}\times\\\nonumber
&&
  C^{(m)}_0(\mu,q)dk d q
,
\\\nonumber
\Psi^{(s)}_h(\lambda,\mu;x)
 &=&
  \int e^{ \varkappa(\sum_{i=2}^5 A^{(i)}x_i + I x_1)}\Psi^{(s)}_{h0}(\lambda,\mu,\varkappa)d\varkappa
,
\end{eqnarray}
where $\Psi^{(s)}_p(\lambda,\mu;x)$ 
 is a particular solution of the eq.(\ref{x}) and $ \Psi^{(s)}_h(\lambda,\mu;x)$
 is a  solution of homogeneous equation associated with eq.(\ref{x}). 
In addition, we separate   matrix function $\Sigma^{(s)}(\lambda,\mu)$ independent on $x$ for our convenience, although it might be incorporated in 
$ \Psi^{(s)}_h(\lambda,\mu;x)$. Both $ \Psi^{(s)}_h(\lambda,\mu;x)$ and $\Sigma^{(s)}(\lambda,\mu)$ have both diagonal and off-diagonal parts.

The dressing function $G$ is a solution of the eq.(\ref{Prop:2_DimG}):
\begin{eqnarray}
&&
G(\lambda;x)=\int  e^{K^{(G)}(\varkappa_1,\varkappa_2;x)} G_0(\lambda,\varkappa) d\varkappa ,\\\nonumber
&&
K^{(G)}(\varkappa_1,\varkappa_2;x)={\varkappa_1 x_1 + \varkappa_2 x_2 +\sum_{j=3}^5 ({\cal{P}}^{(j1)}\varkappa_1+{\cal{P}}^{(j2)}\varkappa_2) x_j}.
\end{eqnarray}
It may not be diagonal, otherwise solution space will be poor.

It is quite standard to assume that the measure $d\Omega(\lambda)$ has support on an open domain 
${\cal{D}}$ of the $\lambda$-space, and on a disjoint discrete set of points 
$D=\{b_1,\dots,b_M\}$, $D\cap {\cal{D}}=\emptyset$. Correspondingly, we use the following notations 
for the dressing functions:
\begin{eqnarray}\label{def_Phi_C}
\Phi^{(sk)}(\lambda;x)&=&\left\{\begin{array}{ll}
\displaystyle \phi^{(sk)}(\lambda;x)=\int \phi^{(sk)}_0(\lambda,\varkappa)
e^{K^{(k)}(\varkappa_1,\varkappa_2;x)} d\varkappa, 
& \lambda \in {\cal{D}}, \cr
\displaystyle\phi^{(sk)}_n(x)=\int \phi^{(sk)}_{n0}(\varkappa) 
e^{K^{(k)}(\varkappa_1,\varkappa_2;x)}d\varkappa, 
&\lambda =b_n,\;\;n=1,\dots,M,
\end{array}\right.
\\\nonumber
\;\;\; 
C^{(k)}(\lambda;x)&=&\left\{\begin{array}{ll}
c^{(k)}(\lambda;x) = \int e^{K^{(k)}(q_1,q_2;x)} c^{(k)}_0(q,\lambda) dq, 
& \lambda \in {\cal{D}}, \cr
c^{(k)}_n(x)=\int e^{K^{(k)}(q_1,q_2;x)} c^{(k)}_{n0}(q) d q , 
&\lambda =b_n,\;\;n=1,\dots,M,
\end{array}\right.
 \\\nonumber
G(\lambda;x)&=&\left\{\begin{array}{ll}
g(\lambda;x) = \int e^{K^{(G)}(\varkappa_1,\varkappa_2;x)} g_0(\lambda,\varkappa) d\varkappa, 
& \lambda \in {\cal{D}}, \cr
g_n(x)=\int e^{K^{(G)}(\varkappa_1,\varkappa_2;x)} g_{n0}(\varkappa) d \varkappa , 
&\lambda =b_n,\;\;n=1,\dots,M,
\end{array}\right.
\;\;\; 
\end{eqnarray}   
\begin{eqnarray}\label{def_U}
U^{(lk)}(\lambda;x)&=&\left\{\begin{array}{ll}
u^{(lk)}(\lambda;x), & \lambda\in {\cal{D}}, \cr
u^{(lk)}_{n}(x), & \lambda =b_n,\;\;n=1,\dots,M
\end{array}\right. ,
\end{eqnarray}   
and we choose $\Sigma^{(s)}(\lambda,\mu)$ in the next form:
\begin{eqnarray}\label{def_A}
\Sigma^{(s)}_{\alpha\beta}(\lambda,\mu)&=&
\left\{\begin{array}{ll}
\sigma^{(s)}_{\alpha\beta}(\lambda,\mu)=\delta^{(s)}_\beta\delta(\lambda-\mu), & \lambda\in {\cal{D}},\;\;
 \mu\in {\cal{D}}\cr
(\sigma^{(s)}_{0m}(\lambda))_{\alpha\beta}=0, & \lambda \in {\cal{D}},
  \;\; \mu=b_m,\;\;m=1,\dots,M\cr
(\sigma^{(s)}_{n0}(\mu))_{\alpha\beta}=0, & \lambda=b_n,\;\;n=1,\dots,M,
  \;\; \mu\in {\cal{D}}\cr
(\sigma^{(s)}_{nm})_{\alpha\beta}= \delta^{(s)}_\beta\sigma_{nm}, &\lambda=b_n,\;\; \mu=b_m,\;\;n,m=1,\dots,M
\end{array}\right. ,
\\\nonumber
&&
\sigma_{nm}=\left\{\begin{array}{ll} \delta_{nm},& n=1,\dots,M-1,\;\;m=1,\dots M\cr
\sigma_{Mm},& n=M,\;\;m=1,\dots M
\end{array}\right.,
\end{eqnarray} 
where $\sigma_{Mm}$ are  scalar parameters.
Next,
\begin{eqnarray}\label{def_Psi}
 &&
 \Psi^{(s)}_p(\lambda,\mu;x) =\\\nonumber
&&
\left\{
\begin{array}{ll}
\psi^{(s)}_p(\lambda,\mu;x) =  \displaystyle \sum_{i=1}^{Q}
\int 
  \phi^{(si)}_0(\lambda,\varkappa)R^{(i)}(\varkappa,q;x)
  c^{(i)}_0(\mu,q)d\varkappa d q
,
 &
  \lambda,\mu \in {\cal{D}}
\cr
  \psi^{(s)}_{p;0m}(\lambda;x) =  
\displaystyle\sum_{i=1}^{Q} \int 
 \phi^{(si)}_0(\lambda,\varkappa)R^{(i)}(\varkappa,q;x) 
  c^{(i)}_{m0}(q)d\varkappa d q
,& \lambda \in {\cal{D}},\;\;
  \mu =b_m\cr
   \psi^{(s)}_{p;n0}(\mu;x) =  
\displaystyle 
\sum_{i=1}^{Q}\int 
  \phi^{(si)}_{n0}(\varkappa)R^{(i)}(\varkappa,q;x)
  c^{(i)}_0(\mu,q) d\varkappa dq
,& \lambda =b_n,\;\;
  \mu \in {\cal{D}}
  \cr
  \psi^{(s)}_{p;nm}(x) = 
\displaystyle \sum_{i=1}^{Q} \int 
 \phi^{(si)}_{n0}(\varkappa)R^{(i)}(\varkappa,q;x)
  c^{(i)}_{m0}(q)d\varkappa d q
,&\lambda =b_n,\;\;\mu=b_m
 \end{array}
 \right. ,
 \end{eqnarray}
 where $n,m=1,\dots,M$, 
 \begin{eqnarray}
 R^{(i)}(\varkappa,q;x)=
 \Big(\varkappa_2+q_2 - (\varkappa_1+q_1)A^{(2)}\Big)^{-1} e^{K^{(i)}
 (\varkappa_1+q_1,\varkappa_2+q_2;x)}.
 \end{eqnarray}
Remember that the functions $\Phi^{(sm)}$ are diagonal, i.e. $\phi^{(sm)}$ and 
$\phi^{(sm)}_n$ are diagonal as well.

The function $\Psi^{(s)}_h$  should be taken in the next form which provides the maximal possible richness of the solution space:
\begin{eqnarray}
\Psi^{(s)}_h(\lambda,\mu;x) =\left\{\begin{array}{ll}
\displaystyle \psi^{(s)}_h(\lambda,\mu;x) =
0,&\lambda,\mu\in{\cal{D}} \cr\displaystyle
\psi^{(s)}_{h;0m}(\lambda;x) =
\int e^{\varkappa(\sum_{i=2}^5 A^{(i)} x_i + I x_1 )}\psi^{(s)}_{h0;0m}
(\lambda,\varkappa)d\varkappa,&\lambda\in{\cal{D}}, \;\;\mu=b_m \cr\displaystyle
\psi^{(s)}_{h;n0}(\mu;x) =
0,&\lambda=b_n,\;\;\mu\in{\cal{D}}\cr\displaystyle
\psi^{(s)}_{h;nm}(x) =0
,&\lambda=b_n,\;\;\mu=b_m
\end{array}\right.,
\end{eqnarray}
where $n,m=1,\dots,M$.

Then eqs. (\ref{U3}) reduce to the following system of  equations
\begin{eqnarray}\label{Sec1:c_12}\label{U_epsilon31}
\phi^{(\gamma k)}_{\beta}(\lambda;x)&=& 
 \left(\int\limits_{{\cal{D}}} 
\psi^{(\gamma)}_p(\lambda,\nu;x) 
u^{(\beta k)}(\nu;x) d\nu+\right.\\\nonumber
&&
\left.\sum_{j=1}^M
\Big(\psi^{(\gamma)}_{p;0j}(\lambda;x)+\psi^{(\gamma)}_{h;0j}(\lambda;x) \Big)
u^{(\beta k)}_j(x) \right)_{\beta\beta}+
 u^{(\beta k)}_{\gamma\beta}(\lambda;x)
 ,\;\; \lambda \in {\cal {D}} \\\label{Sec1:c_13}\label{U_epsilon32}
0&=& 
\left(\int\limits_{{\cal{D}}} 
\psi^{(\gamma)}_p(\lambda,\nu;x) 
u^{(\alpha k)}(\nu;x) d\nu+\right.\\\nonumber
&&
\left.\sum_{j=1}^M
(\psi^{(\gamma)}_{p;0j}(\lambda;x)+\psi^{(\gamma)}_{h;0j}(\lambda;x) )
u^{(\alpha k)}_j(x) \right)_{\alpha\beta}+
 u^{(\alpha k)}_{\gamma\beta}(\lambda;x)
 ,\;\; \lambda \in {\cal {D}},\;\;\alpha\neq \beta \\\label{Sec22:Ub}\label{U_epsilon33}
(\phi^{(\gamma k)}_n(x))_\beta&=& 
\left(
\int\limits_{{\cal{D}}}
 \psi^{(\gamma)}_{p;n0}(\nu;x) u^{(\beta k)}(\nu;x) d\nu+
\right.\\\nonumber
&&
\left.
\sum_{j=1}^M \psi^{(\gamma)}_{p;nj}(x) u^{(\beta k)}_j(x)\right)_{\beta\beta} +
\sum_{m=1}^Q\sigma_{nm}( u^{(\beta k)}_{m}(x))_{\gamma\beta}
 ,\;\;
 n=1,\dots,M\\\label{Sec22:Ub2}\label{U_epsilon34}
0&=& 
\left(
\int\limits_{{\cal{D}}}
 \psi^{(\gamma)}_{p;n0}(\nu;x) u^{(\alpha k)}(\nu;x) d\nu+
\right.\\\nonumber
&&
\left.
\sum_{j=1}^M \psi^{(\gamma)}_{p;nj}(x) u^{(\alpha k)}_j(x)\right)_{\alpha\beta} +
\sum_{m=1}^Q\sigma_{nm}( u^{(\alpha k)}_{m}(x))_{\gamma\beta}
 ,\;\;
 n=1,\dots,M
\end{eqnarray}
for the unknown matrix functions 
$u^{(\alpha k)}(\lambda;x),~\lambda\in{\cal D}$ and $u^{(\alpha k)}_j(x)$, $\lambda=b_j$, $j=1,\dots,M$.

Following the strategy of \cite{ZS}, it is enough to provide  single linear relation among  equations 
(\ref{Sec22:Ub}) and single linear relation among  equations (\ref{U_epsilon34}) in order to satisfy the condition $\dim{\mbox{ker}}\;\hat\Psi = 1$. In turn, this requirement is equivalent to the  next two equations:
\begin{eqnarray}\label{condition:A}
\sum_{j=1}^M\sum_{s=1}^{Q} {{A}}^{(ms)}_j
\phi^{(si)}_j=0,\;\;\;\;\;
\sum_{j=1}^M {{A}}^{(mi)}_j
\sigma_{jn}=0,
\end{eqnarray}
where ${{A}}^{(ms)}_j$ are some constants.
Due to the eq.(\ref{condition:A}), eqs.(\ref{Sec22:Ub}) and (\ref{Sec22:Ub2})  with $n=1,\dots,M-1$
are independent  equations for the functions $u^{(\alpha k)}_n(x)$, $n=1,\dots,M-1$,
 where the matrix functions  $u^{(\alpha k)}_M(x)$  may be taken as
 arbitrary matrix functions 
 $f^{(\alpha k)}(x)$: $u^{(\alpha k)}_M(x)=f^{(\alpha k)}(x)$. 
In order to write  equations 
for $f^{(\alpha k)}$  we 
 involve the condition 
(\ref{Prop:2_condition_Nw_red}),  
 which reads
\begin{eqnarray}\label{condition_res}
&&
G(\lambda;x)* U^{(\alpha k)}(\lambda;x) \equiv  
g(\lambda;x)*  u^{(\alpha k)}(\lambda;x) +\sum_{n=1}^{M} 
g_n(x)  u^{(\alpha k)}_n(x) = R^{(\alpha k)},\\\nonumber
&&
\alpha,k=1,\dots,Q. 
\end{eqnarray}
Thus, the system (\ref{Sec22:Ub},\ref{Sec22:Ub2}), $n=1,\dots,M-1$  and (\ref{condition_res})  represent the complete system for the functions  $u^{(\alpha k)}_n(x)$, $n=1,\dots,M$.

Once $ u^{(\alpha k)}$ and $ u^{(\alpha k)}_n$, $n=1,\dots,M$ have been found, one constructs the matrix 
fields $ v^{(nik)}$ and $w^{nk;p}$ 
using the definitions (\ref{vw}):
\begin{eqnarray}
\label{fields}
 v^{nik}(x)&=&
\int\limits_{{\cal{D}}} c^{(i)}(\lambda;x) u^{(nk)}(\lambda;x)
d\lambda+\sum\limits_{j=1}^M c^{(i)}_j(x) 
 u^{(nk)}_j(x),\\\nonumber
 w^{(nk;p)}(x)&=&\int\limits_{{\cal{D}}} 
 g_{x_p}(\lambda;x) u^{(nk)}(\lambda;x)
d\lambda+\sum\limits_{j=1}^M {g_j}_{x_p}(x) 
 u^{(nk)}_j(x),\;\;p=1,2,
\end{eqnarray}
$n,i,k=1,\dots,Q$.

In  Secs.\ref{Section:expl_sol_1}, \ref{Section:expl_sol_2}, we consider two examples when  solutions to the nonlinear PDEs can be constructed explicitly taking into account that  the independent variables of the nonlinear equations written in  Proposition are $t$, $y_i$, $i=1,2,3$, see eqs.(\ref{Prop:t}).

%%%%%%%%%%%%%%%%%
\subsection{Solutions to the system (\ref{Prop:3_Nw_w2_red})}
\label{Section:expl_sol_1}

Consider the  particular case 
 $C^{(i)}=0$ leading to the eq.(\ref{Prop:3_Nw_w2_red}). 
 One has  $ \psi^{(\gamma)}_p=0$ and $ \psi^{(\gamma)}_{p;nj}=0$.
Thus the system (\ref{U_epsilon31}-\ref{U_epsilon34}) reduces to the next one:
\begin{eqnarray}\label{r_U_epsilon31}
\phi^{(\gamma k)}_{\beta}(\lambda;x)&=& 
  \left(\sum_{j=1}^M
\psi^{(\gamma)}_{h;0j}(\lambda;x) 
u^{(\beta k)}_j(x) \right)_{\beta\beta}+
 u^{(\beta k)}_{\gamma\beta}(x)
 ,\;\; \lambda \in {\cal {D}} \\\label{r_U_epsilon32}
0&=& 
  \left(\sum_{j=1}^M
\psi^{(\gamma)}_{h;0j}(\lambda;x) 
u^{(\alpha k)}_j(x) \right)_{\alpha\beta}+
 u^{(\alpha k)}_{\gamma\beta}(x)
 ,\;\; \lambda \in {\cal {D}},\;\;\alpha\neq \beta \\\label{r_U_epsilon33}
(\phi^{(\gamma k)}_n(x))_\beta&=& 
( u^{(\beta k)}_{n}(x))_{\gamma\beta}
 ,\;\;
 n=1,\dots,M-1\\\label{r_U_epsilon34}
0&=& 
( u^{(\alpha k)}_{n}(x))_{\gamma\beta}
 ,\;\;
 n=1,\dots,M-1
\end{eqnarray}
supplemented by the eq.(\ref{condition_res})
where we take $g_n=\delta_{nM}$ without loss of generality. 
Then eq.(\ref{condition_res}) gives
\begin{eqnarray}\label{u_M}
 u^{(\alpha k)}_M(x) = R^{(\alpha k)} - \bar u^{(\alpha k)}(x), \;\; \bar u^{(\alpha k)}(x) =g(\lambda;x)* u^{(\alpha k)}(\lambda;x).
\end{eqnarray}
We need only $u^{(1 k)}_{\gamma\beta}$ with $\beta\neq 1$ in order to construct the fields $w^{(1k)}_{\gamma\beta}$, $\beta\neq 1$. Thus, 
eqs.(\ref{r_U_epsilon32},\ref{r_U_epsilon34},\ref{u_M}) yield:
 \begin{eqnarray}\label{r_U}
 u^{(1 k)}_{\gamma\beta}(\lambda;x)&=& 
- \left(
\psi^{(\gamma)}_{h;0M}(\lambda;x) \Big(R^{(1 k)} - \bar u^{(1 k)}(x)\Big)
\right)_{1\beta},\;\;\beta\neq 1.
 \end{eqnarray}
The function  $\bar u^{(1 m)}(x) $ can be found as a solution of the following linear algebraic system which appears after applying $\displaystyle\sum_{\gamma=1}^Q \int_{{\cal{D}}} d\lambda\;g_{\tilde\gamma \gamma }(\lambda;x)\cdot $ to (\ref{r_U})
and replacing $\tilde \gamma$ by $\gamma$ in the result:
\begin{eqnarray}\label{r_U_epsilon31_red}
\\\label{r_U_epsilon32_red}
\bar u^{(1 k)}_{\gamma\beta}&=&-\Big(
\xi^{(0)}(\lambda)(R^{(1 k)} - \bar u^{(1  k)})\Big)_{\gamma\beta},\;\;\beta\neq 1,
\end{eqnarray}
where 
\begin{eqnarray}
\xi^{(0)}_{\alpha\beta}=\sum_{\gamma=1}^Q\int\limits_{{\cal{D}}} d\lambda\;
g_{\alpha\gamma}(\lambda;x)
  \Big(\psi^{(\gamma)}_{h;0M}(\lambda;x)\Big)_{1\beta}.
\end{eqnarray}
The functions $w^{(1k;p)}_{\gamma\beta}$ can be found by definition as follows:
\begin{eqnarray}\label{sol_w}
w^{(1k;p)}_{\gamma\beta}(x)&\equiv& \Big(g_{x_p}(\lambda;x)* u^{(1k)}(\lambda;x)\Big)_{\gamma\beta} = \\\nonumber
&&-
\Big(
 \xi^{(p)}(x) 
 (R^{(1k)} -\bar u^{(1k)}(x))\Big)_{\gamma\beta} ,\;\; \beta\neq 1, ,\;\;p=1,2,
\end{eqnarray}
where
\begin{eqnarray}\xi^{(p)}_{\alpha\beta}=\sum_{\gamma=1}^Q\int\limits_{{\cal{D}}} d\lambda\;
\Big(g_{\alpha\gamma}(\lambda;x)\Big)_{x_p}
  \Big(\psi^{(\gamma)}_{h;0M}(\lambda;x)\Big)_{1\beta}.
\end{eqnarray}
Now  solution $q^{(p)}_{\alpha\beta}$ of (\ref{Prop:3_Nw_w2_red}) can be found by the definition (\ref{new_fields}).
Formulae of this section suggest us to choose 
$M=1$ without loss of generality.

We see that the fields $ w^{(1k;p)}$ given by (\ref{vw})
are expressed in terms of $\xi^{(p)}$, $p=0,1,2$. By construction, the
  functions $\xi^{(p)}$ are solutions of
   the  linear PDE  due to the fact that $g(\lambda;x)$
  is 
    solution of the linear PDE (by definition)  and functions $ \Big(\psi^{(\gamma)}_{h;0j}(\lambda;x)\Big)_{1\beta}$ do not depend on the variables $t, y_i$, $i=1,2,3$ (\ref{Prop:t}) by definition.
Thus, the system (\ref{Prop:3_Nw_w2_red}) is $C$-integrable. 
      
 By construction,  functions 
$\xi^{(p)}$ admit  arbitrary dependence on 2 variables $y_i$, $i=1,2$. Since functions $\xi^{(p)}$ with   different values of $p$ introduce the same arbitrary matrix   function of 2 variables, we have $Q^2$ arbitrary scalar functions of $2$ variables  and $2 Q^2$ independent scalar fields, which are elements of  the matrices  $ q^{(p)}$, $p=1,2$. 
Thus, solving, for instance, an initial-boundary  value problem, only $Q^2$ scalar fields may be given arbitrary  initial conditions.
For this reason, we consider eq.(\ref{Prop:3_Nw_w2_red}) as a  partially integrable system.

%%%%%%%%%%%%%%%%%%%%%%%%%%%%%%%%%
\subsection{Degenerate kernel, $C^{(i)}\neq 0$.}
\label{Section:expl_sol_2}
The system of linear equations (\ref{U_epsilon31}-\ref{U_epsilon34},\ref{condition_res})
has a rich manifold of explicit solutions. To construct them, we choose, as 
usual, a degenerate kernel:
\begin{eqnarray}
c^{(i)}_0(q,\mu) = \sum_{j=1}^{\tilde M} 
\tilde c^{(i)}_{1j}(q) \tilde c^{(i)}_{2j}(\mu), 
\end{eqnarray} 
where $\tilde c^{(i)}_{2j}(\mu)$ are  diagonal matrix functions.
Then
\begin{eqnarray}
 \psi^{(s)}_p(\lambda,\mu;x)&=& \sum_{i=1}^{Q} 
\sum_{j=1}^{\tilde M}
  \psi^{(si)}_{p;j}(\lambda;x) \tilde c^{(i)}_{2j}(\mu),\\\nonumber
  \psi^{(s)}_{p;n0}(\mu;x)&=& \sum_{i=1}^{Q} 
\sum_{j=1}^{\tilde M}
  \psi^{(si)}_{p;n0;j}(x) \tilde c^{(i)}_{2j}(\mu),
\end{eqnarray}
where
\begin{eqnarray}\nonumber
&& \psi^{(si)}_{p;j}(\lambda;x)=
\int\phi_0^{(si)}(\lambda,\varkappa)R^{(i)}(\varkappa,q;x)\tilde c^{(i)}_{1j}(q) d\varkappa dq,\\\nonumber
&&
 \psi^{(ki)}_{p;n0;j}(x)=
 \int\phi^{(si)}_{n0}(\varkappa)R^{(i)}(\varkappa,q;x)\tilde c^{(i)}_{1j}(q) d\varkappa dq
\end{eqnarray}
In this case, equations (\ref{Sec1:c_12}-\ref{U_epsilon34}) reduce to
the following  system of linear equations for the matrix functions $ u^{(\alpha k)}_n(x),~\tilde u^{(\alpha ik)}_j(x)$:
\begin{eqnarray}\label{Sec1:c_12_red}
( \phi^{(l\gamma k)}_n(x))_{\beta}&=&
\Big[\sum_{j=1}^{\tilde M}\sum_{i=1}^{Q}
\tilde \nu^{(l\gamma i)}_{nj}(x) \tilde
u^{(\beta ik)}_j(x)+
\sum_{j=1}^M 
\nu^{(l\gamma)}_{nj}(x)  u^{(\beta k)}_j(x)\Big]_{\beta\beta}+
\\\nonumber
&&
(\tilde u^{(\beta l k)}_n(x))_{\gamma\beta},\;\;
n=1,\dots,\tilde M,\\
\label{Sec1:c_12_red2}
0&=&
\Big[\sum_{j=1}^{\tilde M}\sum_{i=1}^{Q}
\tilde \nu^{(l\gamma i)}_{nj}(x) \tilde
u^{(\alpha ik)}_j(x)+
\sum_{j=1}^M 
\nu^{(l\gamma)}_{nj}(x)  u^{(\alpha k)}_j(x)\Big]_{\alpha\beta}+
\\\nonumber
&&
(\tilde u^{(\alpha l k)}_n(x))_{\gamma\beta},\;\;
n=1,\dots,\tilde M,\;\;\alpha \neq \beta
\\\label{Sec22:Ub_red}
(\phi^{(\gamma k)}_n(x))_{\alpha\beta}&=&
\Big[\sum_{j=1}^{\tilde M}\sum_{i=1}^{Q}
\tilde\rho^{(l\gamma i)}_{nj}(x)\tilde u^{(\beta ik)}_j(x)+
\sum_{j=1}^M
 \rho^{(\gamma)}_{nj}(x) u^{(\beta k)}_j(x)\Big]_{\beta\beta}+\\\nonumber
&&
  (u^{(\beta k)}_n)_{\gamma\beta},\;\;n=1,\dots, M-1,
\\\label{Sec22:Ub_red2}
0&=&
\Big[\sum_{j=1}^{\tilde M}\sum_{i=1}^{Q}
\tilde\rho^{(\gamma i)}_{nj}(x)\tilde u^{(\alpha ik)}_j(x)+
\sum_{j=1}^M
 \rho^{(\gamma)}_{nj}(x) u^{(\alpha k)}_j(x)\Big]_{\alpha\beta}+\\\nonumber
&&
  (u^{(\alpha k)}_n)_{\gamma\beta},\;\;n=1,\dots, M-1,\;\;\alpha\neq \beta,
\end{eqnarray}
$k,\alpha,\beta,\gamma=1,\dots,Q$,
$n=1,\dots,M,~j=1,\dots,\tilde M $, 
where 
\begin{eqnarray}
\tilde u^{(\alpha lk)}_j(x) = \int\limits_{{\cal{D}}} \tilde c^{(l)}_{2j}(\lambda)  u^{(\alpha k)}
(\lambda;x)d\lambda,\;\;
\end{eqnarray}
and where the given coefficients 
$\nu^{(l\gamma )}_{nj},\tilde \nu^{(l\gamma i)}_{nj},\rho^{(\gamma)}_{nj},\tilde\rho^{(\gamma i)}_{nj},
 \phi^{(l\gamma k)}_n$ 
are defined in terms 
of the dressing functions:
\begin{eqnarray}
\label{def_rho}
\nu^{(l\gamma)}_{nj}(x)&=&
\int\limits_{{\cal{D}}}d\lambda\;
\tilde c^{(l)}_{2n}(\lambda) 
 \Big(\psi^{(\gamma)}_{p;0j}(\lambda;x)+ \psi^{(\gamma)}_{h;0j}(\lambda;x)\Big),
\\\nonumber
\tilde \nu^{(l\gamma i)}_{nj}(x)&=&
\int\limits_{{\cal{D}}}d\lambda\;
\tilde c^{(l)}_{2n}(\lambda) 
 \psi^{(\gamma i)}_{p;j}(\lambda;x)  ,\\\nonumber
\rho^{(\gamma)}_{nj}(x)&=&
\psi^{(\gamma)}_{p;nj}(x),\\\nonumber
\tilde\rho^{(\gamma i)}_{nj}(x)&=& \psi^{\gamma i}_{p;n0;j}(x), \\\nonumber
 \phi^{(l\gamma k)}_{n}(x) &=& 
\int\limits_{{\cal{D}}} d\lambda\;\tilde c^{(l)}_{2n}(\lambda)
  \phi^{(\gamma k)}(\lambda;x).
\end{eqnarray}
Eqs.(\ref{Sec1:c_12_red}) and (\ref{Sec1:c_12_red2})
are obtained applying 
$\int\limits_{{\cal{D}}}d\lambda\; \tilde c^{(l)}_{2n}(\lambda;x)  \cdot$
to the eqs.(\ref{U_epsilon31}) and (\ref{U_epsilon32}) respectively  from the left.

Having constructed, from (\ref{Sec1:c_12_red}-\ref{Sec22:Ub_red2}),
 the $ u^{(\alpha k)}_j(x)$, $j=1,\dots,M-1$ and the $\tilde u^{(\alpha ik)}_j(x)$ in terms of $ u^{(\alpha k)}_M(x)$, 
one obtains the functions $ u^{(\alpha k)}(\lambda;x)$ 
via the formulae (\ref{U_epsilon31}) and (\ref{U_epsilon32}):
\begin{eqnarray}\label{u_res}
u^{(\beta k)}_{\gamma\beta}(\lambda;x)&=&\phi^{(\gamma k)}_{\beta}(\lambda;x)-
   \left( \sum_{j=1}^{\tilde M}\sum_{i=1}^Q
\psi^{(\gamma i)}_{p;j}(\lambda;x) 
\tilde u^{(\beta i k)}_j(x) +\right.\\\nonumber
&&
\left.\sum_{j=1}^M
(\psi^{(\gamma)}_{p;0j}(\lambda;x)+\psi^{(\gamma)}_{h;0j}(\lambda;x) )
u^{(\beta k)}_j(x) \right)_{\beta\beta}
 ,\;\; \lambda \in {\cal {D}} \\\nonumber
u^{(\alpha k)}_{\gamma\beta}(\lambda;x)&=&- 
\left(
 \sum_{j=1}^{\tilde M}\sum_{i=1}^Q \psi^{(\gamma i)}_{p;j}(\lambda;x) 
\tilde u^{(\alpha i k)}_j(x) +\right.\\\nonumber
&&
\left.\sum_{j=1}^M
(\psi^{(\gamma)}_{p;0j}(\lambda;x)+\psi^{(\gamma)}_{h;0j}(\lambda;x) )
u^{(\alpha k)}_j(x) \right)_{\alpha\beta}
 ,\;\; \lambda \in {\cal {D}},\;\;\alpha\neq \beta.
\end{eqnarray}
Substituting $ u^{(\alpha k)}$ and $ u^{(\alpha k)}_j$,
$j=1,\dots,M-1$ into (\ref{condition_res}), one obtains
the expressions for $ u^{(\alpha k)}_M$. If $g_n=\delta_{nM}$, then $u^{(\alpha k)}_M$ are defined by the eqs.(\ref{u_M}).
Equations for the functions $\bar u^{(\alpha k)}$ appearing in (\ref{u_M}) can be derived applying $\displaystyle\sum_{\gamma=1}^Q\int_{{\cal{D}}}d\lambda \; g_{\tilde \gamma \gamma} \cdot$ to the eqs. (\ref{u_res}) and replacing $\tilde \gamma $ 
by $\gamma$ in the result:
\begin{eqnarray}\label{bar_u_1}
\bar u^{(\beta k)}_{\gamma\beta}(x)&=&\eta^{(\beta k)}_{\gamma\beta}(x)-
   \left( \sum_{j=1}^{\tilde M}\sum_{i=1}^Q
\xi^{(\beta i)}_{j}(x) 
\tilde u^{(\beta i k)}_j(x) +
%\right.\\\nonumber
%&&
%\left.
\sum_{j=1}^M
\xi^{(\beta)}_{j}(x)
u^{(\beta k)}_j(x) \right)_{\gamma\beta}
 , \\\nonumber
\bar u^{(\alpha k)}_{\gamma\beta}(x)&=&- 
\left(
 \sum_{j=1}^{\tilde M}\sum_{i=1}^Q \xi^{(\alpha i)}_{j}(x) 
\tilde u^{(\alpha i k)}_j(x) +
%\right.\\\nonumber
%&&
%\left.
\sum_{j=1}^M
\xi^{(\alpha)}_{j}(x)
u^{(\alpha k)}_j(x) \right)_{\gamma\beta}
 ,\;\;\alpha\neq \beta.
\end{eqnarray}
where
\begin{eqnarray}
&&
\eta^{(\beta k)}_{\gamma \beta}(x)=\sum_{\gamma_1=1}^Q \int\limits_{{\cal{D}}} d\lambda\; 
g_{\gamma\gamma_1}(\lambda;x) \phi^{(\gamma_1 k)}_\beta(\lambda;x),\\\nonumber
&&
\Big( \xi^{(\alpha i)}_j(x)\Big)_{\gamma\beta} =\sum_{\gamma_1=1}^Q\int\limits_{{\cal{D}}} d\lambda\; g_{\gamma\gamma_1}(\lambda;x)\Big( \psi^{(\gamma_1 i)}_{p;j}(\lambda;x)\Big)_{\alpha\beta},\;\;\;\\\nonumber
&&
 \Big(\xi^{(\alpha)}_j(x)\Big)_{\gamma\beta} = \sum_{\gamma_1=1}^Q\int\limits_{{\cal{D}}} d\lambda \;
 g_{\gamma\gamma_1}(\lambda;x) \Big(\psi^{(\gamma_1)}_{p;0j}(\lambda;x)+\psi^{(\gamma_1)}_{h;0j}(\lambda;x)\Big)_{\alpha\beta}.
\end{eqnarray}
Now the system (\ref{Sec1:c_12_red}-\ref{Sec22:Ub_red2}, 
\ref{u_M},\ref{bar_u_1}) should be considered as the
 complete linear algebraic system for the matrix fields 
$\tilde u^{(\gamma lk)}_j$, $ u^{(\gamma k)}_i$, $ \bar u^{(\gamma k)}$, $j=1,\dots, \tilde M$, $i=1,\dots,  M$,
$k,l,\gamma=1,\dots,Q$
where  $ u^{(\gamma k)}_M=f^{(\gamma k)}$.  This system is solvable, in general. 

At last, one constructs the  fields $ v^{(nik)},   w^{(nk;p)}$ using
(\ref{fields}), and the fields $w^{(p)}$, $q^{(p)}$, $v$, $u$, solutions of the nonlinear PDEs (\ref{Prop:sum_w}-\ref{Prop:sum_us}), using the formulae (\ref{new_fields}).

Simplest case corresponds to $M=\tilde M=1$. Then the system (\ref{u_res}) yields (using (\ref{u_M}) for $u^{(\alpha k)}_1$):
\begin{eqnarray}\label{u_res_2}
u^{(\beta k)}_{\gamma\beta}(x)&=&\phi^{(\gamma k )}_{\beta}(\lambda;x)-
 \left(
\sum_{i=1}^Q\psi^{(\gamma i)}_{p;1}(\lambda;x) 
\tilde u^{(\beta i k)}_1(x) +\right.\\\nonumber
&&
\left.
(\psi^{(\gamma)}_{p;01}(\lambda;x)+\psi^{(\gamma)}_{h;01}(\lambda;x) )
(R^{(\beta k)} -\bar u^{(\beta k)}(x) )\right)_{\beta\beta}
 ,\;\; \lambda \in {\cal {D}} \\\nonumber
u^{(\alpha k)}_{\gamma\beta}(x)&=&- 
 \left( 
\sum_{i=1}^Q \psi^{(\gamma i)}_{p;1}(\lambda;x) 
\tilde u^{(\alpha i k)}_1(x) +\right.\\\nonumber
&&
\left.
(\psi^{(\gamma)}_{p;01}(\lambda;x)+\psi^{(\gamma)}_{h;01}(\lambda;x) )
(R^{(\alpha k)} -\bar u^{(\alpha k)} (x)) \right)_{\alpha\beta}
 ,\;\; \lambda \in {\cal {D}},\;\;\alpha\neq \beta.
\end{eqnarray}
The eqs. (\ref{Sec1:c_12_red},\ref{Sec1:c_12_red2} ) read (substitute (\ref{u_M}) for $u_1^{(\alpha k)}$)
\begin{eqnarray}\label{Sec1:c_12_red_2}\label{u_res_30}
( \phi^{(l\gamma k)}_n(x))_{\beta}&=&
\Big[\sum_{i=1}^{Q}
\tilde \nu^{(l\gamma i)}_{n1}(x) \tilde
u^{(\beta ik)}_1(x)+
\nu^{(l\gamma)}_{n1}(x) (R^{(\beta k)} -\bar u^{(\beta k)} (x))\Big]_{\beta\beta}+
%\\\nonumber
%&&
(\tilde u^{(\beta l k)}_1(x))_{\gamma\beta},\\
\nonumber
0&=&
\Big[\sum_{i=1}^{Q}
\tilde \nu^{(l\gamma i)}_{n1}(x) \tilde
u^{(\alpha ik)}_1(x)+ 
\nu^{(ls)}_{n1}(x)(R^{(\alpha k)} -\bar u^{(\alpha k)} (x)) \Big]_{\alpha\beta}+
%\\\nonumber
%&&
(\tilde u^{(\alpha lk)}_1(x))_{\gamma\beta},\;\;\alpha \neq \beta
\end{eqnarray}
while the eqs. (\ref{Sec22:Ub_red},\ref{Sec22:Ub_red2} ) disappear.
 At last, the eqs.(\ref{bar_u_1}) read:
\begin{eqnarray}\label{u_res_3}
\bar u^{(\beta k)}_{\gamma\beta}(x)&=&\eta^{(\beta k)}_{\gamma \beta}(x)-
 \left(
\sum_{i=1}^Q\xi^{(\beta i)}_{1}(x) 
\tilde u^{(\beta i m)}_1(x) +
\xi^{(\beta)}_1(x) 
(R^{(\beta k)} -\bar u^{(\beta k)} (x)) \right)_{\gamma\beta}
 ,\\\nonumber
\bar u^{(\alpha k)}_{\gamma\beta}(x)&=&- 
 \left(
\sum_{i=1}^Q \xi^{(\alpha i)}_{1}(x) 
\tilde u^{(\alpha i k)}_1(x) +
\xi^{(\alpha)}_{1}(x)
(R^{(\alpha k)} -\bar u^{(\alpha k)} (x)) \right)_{\gamma\beta}
 ,\;\;\alpha\neq \beta.
\end{eqnarray}
The system of linear equations (\ref{u_res_30},\ref{u_res_3}) may be solved for $\tilde u^{(\alpha l k)}$ and $\bar u^{(\alpha k)}$. Then, the matrix 
functions $ u^{(\alpha k)}$ will be  explicitly 
given by the eqs.(\ref{u_res_2}). Next, the fields $ v^{(nik)}$, $ w^{(nk;p)}$ can be constructed using the
eqs.(\ref{fields}). Finally, the fields of the nonlinear system can be found by their definitions (\ref{new_fields}).
 For instance, the explicit solutions in the form of rational functions of exponents have been constructed.  We do not represent them here since they are too cumbersome.

%%%%%%%%%%%%%%%%%%
\subsubsection{Richness of the solution space to the system (\ref{Prop:3_Nw_w}-\ref{Prop_s:3_Nw_01})}

Let us discuss the richness of solution space, taking into account that the fields of the
 nonlinear equations underlined in 
Proposition (Sec.\ref{Proposition}) are expressed in terms of  the scalar fields $v^{(1ik)}_{1\beta}$ and $w^{(1k;p)}_{\alpha \beta}$, $i,k,\beta=1,\dots, Q$, $p=1,2$. 
Using eqs.(\ref{U_epsilon31}, \ref{U_epsilon32}) as the definitions of the function $u(\lambda;x)$ and eqs.(\ref{fields})  for the fields $w^{(1k;p)}_{\alpha \beta}$ and $v^{(1ik)}_{1\beta}$ we observe  in the small field limit:
\begin{eqnarray}\label{linlim}
w^{(1k;p)}_{\alpha \beta}(x) &\sim& \sum_{\gamma=1}^Q {g_{\alpha\gamma}}_{x_p}(\lambda;x) *\Big( \sum_{j=1}^M\psi^{(\gamma)}_{h;0j}(\lambda;x)u^{(1k)}_j(x)\Big)_{1 \beta},\;\;\beta\neq 1,\\\nonumber
w^{(1k;p)}_{\alpha 1}(x) &\sim&  \sum_{\gamma=1}^Q {g_{\alpha\gamma}}_{x_p}(\lambda;x)*\phi^{(\gamma k)}_1(\lambda;x),\\\nonumber
v^{(1ik)}_{11}(x) &\sim& \sum_{\gamma=1}^Q c^{(i)}_{1\gamma}(\lambda;x)*\phi^{(\gamma k)}_1(\lambda;x),
\\\nonumber
v^{(1ik)}_{1\beta}(x) &\sim& \sum_{\gamma=1}^Q {c^{(i)}_{1\gamma}}_{x_p}(\lambda;x) * \Big(\sum_{j=1}^M \psi^{(\gamma)}_{h;0j}(\lambda;x)
u^{(1k)}_j(x)\Big)_{1 \beta},\;\;\beta\neq 1.
\end{eqnarray}
By construction, the function  $g$ has arbitrary dependence on $2$ variables $y_1$ and $y_2$. Functions  $\phi^{(\gamma k)}_1$  and  $c^{(i)}_{1\gamma}$ have arbitrary dependence on single variable $y_1$. Finally, $\psi_{h;0j}$ do not depend on $t,y_i$.

As we have seen in Introduction, the correctly formulated 
initial-boundary value problem involves  two arbitrary  matrix  functions of three variables $y_i$, $i=1,2,3$, for the
fields  $w^{(p)}$, $p=1,2$, three arbitrary  matrix  
functions of two variables $y_i$, $i=1,2$, for the fields  
$q^{(p)}$, $p=1,2$, and $v$ and one arbitrary  matrix 
function of single  variable $y_1$ for the field  $u$ (see 
eqs.(\ref{new_fields}) for  definitions of these fields). However,
eqs.(\ref{linlim}) show that we have single arbitrary matrix 
function of three variables (formula (\ref{linlim}a) after applying  $\sum_{k,\beta=1}^Q \cdot {\cal{B}}^{(2;1)}_\beta (\hat 
R^{-1})^{k1}_{\beta \tilde \beta}$), two arbitrary matrix 
function of two variables (formulae (\ref{linlim}b,c))  and one 
arbitrary matrix function of single variables (formula (\ref{linlim}d)
after applying $\sum_{k,\beta=1}^Q \cdot
 {\cal{B}}^{(2;1)}_\beta (\hat R^{-1})^{k1}_{\beta \tilde 
 \beta}$). 
Thus we may not supply all necessary initial data in order to 
completely formulate an initial-boundary value problem. For this 
reason, the system of PDEs underlined in  Proposition 
is considered as a partially integrable system.

%%%%%%%%%%%%%%%%%%%%%%%

\section{Conclusions}
\label{Conclusions}
We have represented a new type of nonlinear PDEs integrable by a new  version   of the  dressing method. 
This algorithm  is based on  two principal novelties   introduced in \cite{ZS}:
\begin{enumerate}
\item
The nontrivial kernel of the integral operator used in the dressing method: $\dim{\mbox{ker}} \;\hat \Psi^{(sn)} =1$. This allows one to increase the dimensionality of nonlinear PDEs. In addition, it generates arbitrary functions of $x$ in  solution space, which allow one to introduce different relations among the fields and,
consequently, to increase  variety of nonlinear PDEs treatable by the dressing method. 
\item
The external dressing function $G(\lambda;x)$ which allows one (a) to increase  dimensionality of solution space and, similarly to the previous item, (b) to increase  variety of  nonlinear PDEs solvable by the dressing method.
\end{enumerate}
These two novelties seemed out to be  extremely promising in development of  theory of  integrable PDEs. 
This paper is devoted to the investigation of 
two particular  problems outlined  in \cite{ZS}:
\begin{enumerate}
\item
Find  reductions of nonlinear PDEs derived in \cite{ZS} which simplify they structure.
\item
Enrich  solution space  of  the constructed nonlinear PDEs (remember that $n$-dimensional nonlinear PDEs derived in \cite{ZS} have at most $(n-2)$ dimensional solution space, which is not enough for full integrability).
\end{enumerate}

We succeeded in simplification of nonlinear PDEs. The systems (\ref{Prop:sum_w}-\ref{Prop_s:3_Nw_01}) and (\ref{Prop:3_Nw_w2_red}) have differential polynomial form, although the number of equations is rather big. We also have enriched solution space. Thus,  solution space of the constructed  $4$-dimensional PDEs is parametrized by one  arbitrary  matrix function of $3$ independent variables, by two arbitrary   matrix function of $2$ independent variables and by one  arbitrary   matrix function of single independent variable. However, we need one more arbitrary matrix function of 3 independent variables and one more  arbitrary  matrix function of two independent variables in order to achieve   full integrability.
In other words, we achieve only partial integrability of the derived nonlinear PDEs.

A natural generalization of this algorithm is an increase of  dimensionality of the kernel of the integral operator: 
 \begin{eqnarray}
 \dim {\mbox{ker}}\;\hat \Psi^{(sn)}  >1.
 \end{eqnarray}   
The study of associated (partially) integrable equations is postponed to future investigation.

Author thanks Prof. P.M.Santini for  useful discussions and referee for  important comments. The work was supported by INTAS Young Scientists Fellowship Nr. 04-83-2983, RFBR grants 04-01-00508, 06-01-90840, 06-01-92053 and grant Ns. 7550.2006.2.

%%%%%%%%%%%%%%%%%%%%
\section{Appendix:  proof of Proposition, items (a-c) and (d.1, d.5)} 
\label{Appendix1}

As we  have seen in Sec.\ref{Section:Nw:dim1},  the eq.(\ref{h})
leads to the linear relations between any two solutions  
 $U^{(nk)}$, $L^{(j1;nk)}$, $L^{(j3;nk)}$, $j\ge 2$ and $L^{(j2;nk)}$, $j\ge 3$ of the homogeneous integral equation (\ref{hom}). In orther words, one has
the following system of independent linear equations for the spectral functions $U^{(nk)}$:
\begin{eqnarray}\label{2_sp_1}\label{2_vec_Nw_d}
&&
L^{(j2;nk)}_{\alpha\beta}=\sum_{i_1,\gamma=1}^Q L^{(21;ni_1)}_{\alpha\gamma} F^{(j;i_1k)}_{\gamma\beta} \Rightarrow
\\\nonumber
&&
\partial_{x_j} { U}^{(nk)}_{\alpha\beta}
 + 
\partial_{x_1} { U}^{(nk)}_{\alpha\beta}{\cal{B}}^{(kj;n)}_\beta -
\partial_{x_2} { U}^{(nk)}_{\alpha\beta}
B^{(kj)}_\beta + 
\\\nonumber
&&
\sum_{i_1,\gamma=1}^Q{ U}^{(ni_1)}_{\alpha\gamma}
  v^{(ni_1k)}_{\gamma\beta}\Big(B^{(i_1j)}_\gamma - B^{(kj)}_\beta\Big) =
\sum_{i_1,\gamma=1}^Q
 U^{(ni_1)}_{\alpha\gamma}
 {\cal{B}}^{(2;n)}_\gamma  F^{(j;i_1k)}_{\gamma\beta},\;\;j\ge 3,
\end{eqnarray}
 \begin{eqnarray}
&&\label{2_sp_12}\label{2_vec_Nw_d2}
L^{(j3;lk)}_{\alpha\beta}=\sum_{i_1,\gamma=1}^Q L^{(21;li_1)}_{\alpha\gamma}  F^{(j;i_1k)}_{\gamma\beta} \Rightarrow\\
 \nonumber
&&
\Big( \partial_{x_j}  U^{(nk)}_{\alpha\beta}  -
 \partial_{x_1}  U^{(nk)}_{\alpha\beta} 
 A^{(j)}_n
 +
 \sum_{i_1,\gamma=1}^Q U^{(ni_1)}_{\alpha\gamma}
  v^{(ni_1k)}_{\gamma\beta}  B^{(i_1j)}_\gamma\Big) {\cal{B}}^{(2;n)}_\beta=\\\nonumber
 &&
 \sum_{i_1,\gamma=1}^Q
  U^{(n i_1)}_{\alpha\gamma} 
  {\cal{B}}^{(2;n)}_\gamma \tilde
F^{(j;i_1k)}_{\gamma\beta},\;\;j\ge 2.
 \\
 \label{2_sp_0}\label{2_sp_0_star}
&&
L^{(j1;lk)}_{\alpha\beta}=\sum_{i_1,\gamma=1}^Q L^{(21;li_1)}_{\alpha\gamma} \hat F^{(j;i_1k)}_{\gamma\beta} \Rightarrow\\\nonumber
&&  
  U^{(n k)}_{\alpha\beta} 
 {\cal{B}}^{(j;n)}_\beta=
 \sum_{i_1,\gamma=1}^Q
  U^{(n i_1)}_{\alpha\gamma} 
  {\cal{B}}^{(2;n)}_\gamma  \hat F^{(j;i_1k)}_{\gamma\beta}
 \end{eqnarray}
 where  $F^{j;i_1k}_{\gamma\beta}$, $\tilde F^{j;i_1k}_{\gamma\beta}$ and $\hat F^{j;i_1k}_{\gamma\beta}$
  are  scalar functions of $x$ to be
 defined. 
This set of equations can be taken as a basis. It is simple to check that any other linear equation for the spectral functions $ U^{(nk)}$ is a  linear combination of (\ref{2_sp_1}-\ref{2_sp_0}).
 Remark, that if $n=\beta$, then LHS of the eq. (\ref{2_vec_Nw_d2}) disappears yielding  algebraic relations among elements of matrix  spectral functions:
 \begin{eqnarray}\label{2_vec_Nw_d3}
 \sum_{i_1,\gamma=1}^Q
  U^{(\beta i_1)}_{\alpha\gamma} 
  {\cal{B}}^{(2;\beta)}_\gamma \tilde
F^{(j;i_1k)}_{\gamma\beta} = 0,
\end{eqnarray}
which must be involved into consideration as compatible constraints.
We consider  equations 
(\ref{2_vec_Nw_d},\ref{2_vec_Nw_d2}) and (\ref{2_vec_Nw_d3}) as an overdetermined system of linear equations for the spectral functions $ U^{(nk)}$ disregarding equation (\ref{2_sp_0_star}). It will be shown in the
  Sec.(\ref{Remarks}) that eq.(\ref{2_sp_0_star}) follows from the eq.(\ref{2_vec_Nw_d3}).
 
Now we are going  (a) to derive nonlinear equations for the  fields (\ref{vw}) 
and 
(b) to express functions $F(x)$, $\tilde F(x)$ and $ \hat F(x)$ in terms of the fields (\ref{vw}).

Nonlinear equations for the fields $ v^{(nik)}, v^{(nik;1)}$ result from the 
eqs.(\ref{2_vec_Nw_d},\ref{2_vec_Nw_d2},\ref{2_vec_Nw_d3}) 
 after applying $ \displaystyle\sum_{\alpha=1}^Q C^{(i)}_{\tilde \alpha\alpha}*$ to them and using (\ref{C}) for  $C^{(i)}_{x_k}$, $k>2$ (we replace $\tilde \alpha$ by $\alpha$ in the result):
 \begin{eqnarray}\label{2_vec_Nw_22}
&&
\partial_{x_j} { v}^{(n ik)}_{\alpha\beta} + 
\partial_{x_1} { v}^{(n ik)}_{\alpha\beta}{\cal{B}}^{(kj;n)}_\beta -
\partial_{x_2} { v}^{(n ik)}_{\alpha\beta}
B^{(kj)}_\beta+\\\nonumber
 &&
  { v}^{(n ik;1)}_{\alpha\beta} 
 {\cal S}^{(v;j1;n ik)}_{\alpha\beta}
   +
  { v}^{(\beta ik;2)}_{\alpha\beta}
  {\cal S}^{(v;j2;ik)}_{\alpha\beta}
  + 
\sum_{i_1,\gamma=1}^Q{ v}^{(n ii_1)}_{\alpha\gamma} 
  v^{(n i_1k)}_{\gamma\beta}\Big(B^{(i_1j)}_\gamma - B^{(kj)}_\beta\Big) 
  =\\\nonumber 
 &&
\sum_{i_1,\gamma=1}^Q
 v^{(n ii_1)}_{\alpha\gamma} 
 {\cal{B}}^{(2;n)}_\gamma  F^{(j;i_1k)}_{\gamma\beta},\;\; j\ge 3,\;\;n\neq \beta,
\end{eqnarray}
\begin{eqnarray}\label{2_vec_Nw_23}
&&
\Big(\partial_{x_j}  v^{(nik)}_{\alpha\beta} -
 \partial_{x_1}  v^{(nik)}_{\alpha\beta} 
 A^{(j)}_n
 +
 \\\nonumber
 &&
  { v}^{(nik;1)}_{\alpha\beta} 
\tilde  {\cal S}^{(v;j1;ni)}_{\alpha}
   +
  { v}^{(nik;2)}_{\alpha\beta}
 \tilde  {\cal S}^{(v;j2;i)}_{\alpha}
  + 
 \sum_{i_1,\gamma=1}^Q  v^{(nii_1)}_{\alpha\gamma}
  v^{(ni_1k)}_{\gamma\beta}  B^{(i_1j)}_\gamma\Big){\cal{B}}^{(2;n)}_\beta =\\\nonumber
 &&
 \sum_{i_1,\gamma=1}^Q
  v^{(n i i_1)}_{\alpha\gamma} 
 {\cal{B}}^{(2;n)}_\gamma
 \tilde
F^{(j;i_1k)}_{\gamma\beta},\;\;j\ge 2,\;\;n\neq \beta,
 \end{eqnarray}
 \begin{eqnarray}\label{2_vec_Nw_24}
 \sum_{i_1,\gamma=1}^Q
 v^{(\beta i i_1)}_{\alpha\gamma} 
 {\cal{B}}^{(2;\beta)}_\gamma
 \tilde
F^{(j;i_1k)}_{\gamma\beta} =0,
 \end{eqnarray}
where
\begin{eqnarray}\label{Ap3}
&&
{\cal S}^{(v;j1;nik)}_{\alpha\beta} = {\cal{B}}^{ij;\alpha}_\alpha  - 
{\cal{B}}^{(kj;n)}_\beta =A^{(j)}_n - A^{(j)}_\alpha +
B^{(ij)}_\alpha A^{(2)}_\alpha - B^{(kj)}_\beta A^{(2)}_n
   ,\\\nonumber
&&
{\cal S}^{(v;j2;ik)}_{\alpha\beta} =
      B^{(kj)}_\beta-B^{(ij)}_\alpha,\\\nonumber
&&
\tilde {\cal S}^{(v;21;ni)}_\alpha=
 A^{(2)}_n,\;\;\tilde {\cal S}^{(v;22;i)}_{\alpha}=
-1,
\\\nonumber
&&
\tilde {\cal S}^{(v;j1;ni)}_{\alpha}=
 A^{(j)}_n + {\cal{B}}^{(ij;\alpha)}_\alpha=A^{(j)}_n -
 A^{(j)}_\alpha + B^{(ij)}_\alpha A^{(2)}_\alpha,\;\;
 \tilde {\cal S}^{(v;j2;i)}_{\alpha}=-
B^{(ij)}_\alpha,\;\;j>2.
\end{eqnarray}

 In order to fix functions ${\cal{F}}^{(j;i_1k)}(x)$
we introduce the external dressing matrix function $G(\lambda;x)$
and the associated fields
\begin{eqnarray}\label{2_hatw}\label{hatw}
 w^{(nk)}=G*  U^{(nk)},\;\;
 w^{(nk;p)}= G_{x_p}*  U^{(nk)},\;\;
 w^{(nk;ps)}= G_{x_p x_s}*  U^{(nk)},\dots,  w^{(nk;ps)}= w^{(nk;sp)}
\end{eqnarray}
(see eq.(\ref{vw}) in  Proposition ).
Applying $\sum_{\alpha=1}^Q G_{\tilde \alpha \alpha}*$ to
the eqs.(\ref{2_vec_Nw_d},\ref{2_vec_Nw_d2},\ref{2_vec_Nw_d3}) 
  one gets the first set of
equations for these fields (we replace $\tilde \alpha$ by $\alpha$ in the result):
\begin{eqnarray}\label{2_vec_Nw_w}
&&
\partial_{x_j} { w}^{(n k)}_{\alpha\beta} + 
\partial_{x_1} { w}^{(n k)}_{\alpha\beta}{\cal{B}}^{(kj;n)}_\beta - 
\partial_{x_2} { w}^{(n k)}_{\alpha\beta}B^{(kj)}_\beta -
   { w}^{(n k;j)}_{\alpha\beta}  
   \\\nonumber
&& - { w}^{(n k;1)}_{\alpha\beta}
{\cal{B}}^{(kj;n)}_\beta +
 { w}^{(n k;2)}_{\alpha\beta}B^{(kj)}_\beta
  + 
\sum_{i_1,\gamma=1}^Q{ w}^{(n i_1)}_{\alpha\gamma} 
  v^{(n i_1k)}_{\gamma\beta}\Big(B^{(i_1j)}_\gamma - B^{(kj)}_\beta\Big) =\\\nonumber
&&
\sum_{i_1,\gamma=1}^Q
 w^{(ni_1)}_{\alpha\gamma}
 {\cal{B}}^{(2;n)}_\gamma  F^{(j;i_1k)}_{\gamma\beta}.
\end{eqnarray}
\begin{eqnarray}\label{2_vec_Nw_w2}
&&
\Big(\partial_{x_j}  w^{(nk)}_{\alpha\beta} -
 \partial_{x_1}  w^{(nk)}_{\alpha\beta} 
 A^{(j)}_n
 -\\\nonumber
 &&   w^{(nk;j)}_{\alpha\beta}  + w^{(nk;1)}_{\alpha\beta} A^{(j)}_n  +
 \sum_{i_1,\gamma=1}^Q  w^{(ni_1)}_{\alpha\gamma}
  v^{(ni_1k)}_{\gamma\beta}  B^{(i_1j)}_\gamma\Big){\cal{B}}^{(2;n)}_\beta =\\\nonumber
 &&
 \sum_{i_1,\gamma=1}^Q
  w^{(n i_1)}_{\alpha\gamma} 
 {\cal{B}}^{(2;n)}_\gamma \tilde
F^{(j;i_1k)}_{\gamma\beta},\;\;j\ge 2,\;\;n\neq \beta.
 \end{eqnarray}
 \begin{eqnarray}\label{2_vec_Nw_w22}
 \sum_{i_1=1}^{Q}
 \sum_{\gamma=1}^Q
  w^{(\beta i_1)}_{\alpha\gamma} 
  {\cal{B}}^{(2;\beta)}_\gamma \tilde
F^{(j;i_1k)}_{\gamma\beta} = 0.
\end{eqnarray}
 
 Applying  
$\sum_{\alpha=1}^Q (G_{\tilde \alpha \alpha})_{x_p}*$
to the eqs.(\ref{2_vec_Nw_d},\ref{2_vec_Nw_d2},\ref{2_vec_Nw_d3})  and replacing $\tilde \alpha$ by $\alpha$ in the result, one gets
 the second set of equations for the fields (\ref{2_hatw}):
\begin{eqnarray}\label{2_vec_Nw_w_p}
&&
\partial_{x_j} { w}^{(n k;p)}_{\alpha\beta} + 
\partial_{x_1} { w}^{(n k;p)}_{\alpha\beta}{\cal{B}}^{(kj;n)}_\beta - 
\partial_{x_2} { w}^{(n k;p)}_{\alpha\beta}B^{(kj)}_\beta -
   { w}^{(n k;pj)}_{\alpha\beta}  
   \\\nonumber
&& - { w}^{(n k;p1)}_{\alpha\beta}
{\cal{B}}^{(kj;n)}_\beta +
 { w}^{(n k;p2)}_{\alpha\beta}B^{(kj)}_\beta
  + 
\sum_{i_1,\gamma=1}^Q{ w}^{(n i_1;p)}_{\alpha\gamma} 
  v^{(n i_1k)}_{\gamma\beta}\Big(B^{(i_1j)}_\gamma - B^{(kj)}_\beta\Big) =\\\nonumber
&& 
\sum_{i_1,\gamma=1}^Q
 w^{(ni_1;p)}_{\alpha\gamma}
 {\cal{B}}^{(2;n)}_\gamma  F^{(j;i_1k)}_{\gamma\beta}.
\end{eqnarray}
\begin{eqnarray}\label{2_vec_Nw_w_p2}
&&
\Big(\partial_{x_j}  w^{(nk;p)}_{\alpha\beta} -
 \partial_{x_1}  w^{(nk;p)}_{\alpha\beta} 
 A^{(j)}_n
 -\\\nonumber
 &&   w^{(nk;pj)}_{\alpha\beta}  + w^{(nk;p1)}_{\alpha\beta} A^{(j)}_n +
 \sum_{i_1,\gamma=1}^Q  w^{(ni_1;p)}_{\alpha\gamma}
  v^{(ni_1k)}_{\gamma\beta}  B^{(i_1j)}_\gamma\Big){\cal{B}}^{(2;n)}_\beta =\\\nonumber
 &&
 \sum_{i_1,\gamma=1}^Q
  w^{(n i_1;p)}_{\alpha\gamma} 
 {\cal{B}}^{(2;n)}_\gamma \tilde
F^{(j;i_1k)}_{\gamma\beta},\;\;j\ge 2,\;\;n\neq \beta.
 \end{eqnarray}
\begin{eqnarray}\label{2_vec_Nw_w_p22}
 \sum_{i_1,\gamma=1}^Q
  w^{(\beta i_1;p)}_{\alpha\gamma} 
  {\cal{B}}^{(2;\beta)}_\gamma \tilde
F^{(j;i_1k)}_{\gamma\beta} = 0.
\end{eqnarray}
We refer to the eqs.(\ref{2_vec_Nw_w}-\ref{2_vec_Nw_w_p22}) as 
 equations for the fields $ w^{(nk;p)}$ and $ w^{(nk;ps)}$, also the nonlinear parts of these equations involve  fields $ v^{(nik)}$ as well.

The set of nonlinear equations may be continued applying $\displaystyle\sum\limits_{\alpha=1}^Q  (G_{\tilde \alpha \alpha})_{x_px_s}*$ to  the eqs.(\ref{2_vec_Nw_d},\ref{2_vec_Nw_d2},\ref{2_vec_Nw_d3}), but the system of nonlinear PDEs constructed in this way may not be completed since (a) 
 its solution space has arbitrary functions of all variables $x$, (see  functions $f^{(ik)}$ in the eq.(\ref{h})) and (b) the dressing function $G$ is arbitrary, i.e. all  derivatives of $G$ with respect to $x_p$, $p=1,2,\dots$ are independent.
 This ends the proof of the item (a) of Proposition, i.e. we have derived a set of nonlinear PDEs (\ref{2_vec_Nw_22}-\ref{2_vec_Nw_24}), (\ref{2_vec_Nw_w}-\ref{2_vec_Nw_w_p22}) for the fields (\ref{vw})
 which does not represent a complete system of PDEs.
 
 Now we pay attention to the  completeness of the derived nonlinear system.

 In order to fix  arbitrary functions in the  solution space  we must introduce an extra largely
arbitrary relation among the fields
\begin{eqnarray}\label{2_condition_Nw}
{\cal{F}}^{rs}({\mbox{all fields }}) =0,
\;\;r,s=1,\dots,Q,
\end{eqnarray}
where ${\cal{F}}^{rs}$ are the $Q\times Q$ matrices. (See the item (b) of Proposition).
For instance, in order to obtain a system of nonlinear PDEs having differential polynomial form, we choose 
the eq. (\ref{2_condition_Nw}) in the form (\ref{Prop:2_condition_Nw_red})
(see the item (c) in   Proposition ).

We use the eqs.(\ref{2_vec_Nw_w},\ref{2_vec_Nw_w2}) to express 
$F^{(j;i_1k)}_{\gamma\beta}$, $\tilde F^{(j;i_1k)}_{\gamma\beta}$  in terms of the fields (\ref{vw}). In view of the eq.(\ref{Prop:2_condition_Nw_red}),
the eqs.(\ref{2_vec_Nw_w},\ref{2_vec_Nw_w2}) read:
\begin{eqnarray}
\label{2_vec_Nw_red_d}
&&
 -
   { w}^{(n k;j)}_{\alpha\beta}  
    - { w}^{(n k;1)}_{\alpha\beta}
{\cal{B}}^{(kj;n)}_\beta +
 { w}^{(n k;2)}_{\alpha\beta}B^{(kj)}_\beta
  + \sum_{i_1,\gamma=1}^Q{ R}^{(ni_1)}_{\alpha\gamma} 
  v^{(ni_1k)}_{\gamma\beta}\Big(B^{(i_1j)}_\gamma - B^{(kj)}_\beta\Big)
   =\\\nonumber
&&
\sum_{i_1,\gamma=1}^Q
R^{(ni_1)}_{\alpha\gamma}
 {\cal{B}}^{(2;n)}_\gamma  F^{(j;i_1k)}_{\gamma\beta},\;\;j\ge 3,\\\nonumber
&&
\Big(-  w^{(nk;j)}_{\alpha\beta}  + w^{(nk;1)}_{\alpha\beta} A^{(j)}_n  +
\sum_{i_1,\gamma=1}^Q{ R}^{(ni_1)}_{\alpha\gamma} 
  v^{(ni_1k)}_{\gamma\beta}   B^{(i_1j)}_\gamma\Big){\cal{B}}^{(2;n)}_\beta=\\\nonumber
&&  
\sum_{i_1,\gamma=1}^Q
R^{(ni_1)}_{\alpha\gamma}
 {\cal{B}}^{(2;n)}_\gamma \tilde  F^{(j;i_1k)}_{\gamma\beta},\;\;j\ge 2,\;\;n\neq\beta.
\end{eqnarray}
The matrices $R^{(lk)}$ must provide unique solvability of
(\ref{2_vec_Nw_red_d}) with respect to 
$F^{(j;i_1k)}_{\gamma\beta}$ and $\tilde
F^{(j;i_1k)}_{\gamma\beta}$, which become linear functions of the fields $ v$ and $ w$:
\begin{eqnarray}
F^{(j;lk)}_{\alpha\beta} &=& 
 \sum_{i_1,i_2,\gamma=1}^Q
S^{(j;l i_1 i_2  k)}_{\alpha\gamma\beta}  v^{(i_1 i_2
k)}_{\gamma\beta}+\\\nonumber
&&
\sum_{i_1,\gamma=1}^Q( \hat R^{-1})^{(li_1)}_{\alpha\gamma}\Big(-
   { w}^{(i_1 k;j)}_{\gamma\beta}  
    - { w}^{(i_1 k;1)}_{\gamma\beta}
{\cal{B}}^{(kj;i_1)}_\beta +
 { w}^{(i_1 k;2)}_{\gamma\beta}B^{(kj)}_\beta
  \Big), \;\;j\ge3,\\\nonumber
\tilde F^{(j;lk)}_{\alpha\beta} &=& 
 \sum_{i_1,i_2,\gamma=1}^Q
\tilde S^{(j;l i_1 i_2 ) }_{\alpha\gamma\beta}  v^{(i_1 i_2
k)}_{\gamma\beta}+\\\nonumber
&&
\sum_{i_1,\gamma=1}^Q
( \hat R^{-1})^{(li_1)}_{\alpha\gamma}\Big(-  w^{(i_1k;j)}_{\gamma\beta}  + w^{(i_1k;1)}_{\gamma\beta} A^{(j)}_{i_1} \Big) {\cal{B}}^{(2;i_1)}_\beta , \;\;j\ge 2,
\end{eqnarray}
where 
\begin{eqnarray}\label{Ap4}
&&
S^{(j;l i_1 i_2  k)}_{\alpha\gamma\beta} = \sum_{\gamma_1=1}^Q
( \hat R^{-1})^{(li_1)}_{\alpha\gamma_1} 
{ R}^{(i_1 i_2)}_{\gamma_1 \gamma} 
  (B^{(i_2 j)}_\gamma - B^{(kj)}_\beta),\\\nonumber
&&
  \tilde S^{(j;l i_1 i_2 ) }_{\alpha\gamma\beta} = \sum_{\gamma_1=1}^Q
( \hat R^{-1})^{(li_1)}_{\alpha\gamma_1} 
{ R}^{(i_1 i_2)}_{\gamma_1 \gamma} 
  {\cal{B}}^{(2;i_1)}_\beta B^{(i_2j)}_\gamma
  \end{eqnarray}
  and the operator $ \hat R^{-1}$ is the inverse 
of the operator $R^{(ni)}{\cal{B}}^{(2;n)}$, i.e.
 \begin{eqnarray}\label{2_RinvR}\label{RinvR}
&&\sum_{n,\gamma_1=1}^{Q}
( \hat R^{-1})^{(ln)}_{\alpha\gamma_1} 
R^{(ni)}_{\gamma_1\gamma}
 {\cal{B}}^{(2;n)}_\gamma = \delta^{(l i)} 
\delta_{\alpha\gamma}
,\;\;\;{\mbox{or}}\\\nonumber
 \label{2_RinvR_right}
 &&
 \sum_{i_1,\gamma_1=1}^{Q}
R^{(ni_1)}_{\gamma\gamma_1}
 {\cal{B}}^{(2;n)}_{\gamma_1} 
( \hat R^{-1})^{(i_1l)}_{\gamma_1 \alpha}  = \delta^{(n  l )} 
\delta_{\gamma\alpha}.
\end{eqnarray}

The eq.(\ref{2_vec_Nw_d3})
may be transformed  as follows.
Substituting $\tilde F^{(j;ik)}$ into (\ref{2_vec_Nw_d3}) one gets
\begin{eqnarray}\label{2_UBR0}
\sum_{i_1,i_2,\gamma_1,\gamma_2=1}^{ Q} 
 U^{(\beta i_1)}_{\alpha\gamma_1} 
{\cal{B}}^{(2;\beta)}_{\gamma_1}  
( \hat R^{-1})^{(i_1 i_2)}_{\gamma_1\gamma_2} 
{\cal{B}}^{(2;i_2)}_{\beta} 
{\cal{E}}^{(i_2 j)}_{\gamma_2 k}=0
\end{eqnarray}
where
\begin{eqnarray}
{\cal{E}}^{(i_2 j)}_{\gamma_2 k} =\sum_{i_3,\gamma_3=1}^{Q} 
R^{(i_2 i_3)}_{\gamma_2\gamma_3} 
 B^{(i_3 j)}_{\gamma_3}  v^{(i_2i_3k)}_{\gamma_3\beta}
-  w^{(i_2k;j)}_{\gamma_2\beta}  + w^{(i_2k;1)}_{\gamma_2\beta} A^{(j)}_{i_2}  ,\;\;\\\nonumber
i_2=1,\dots,Q,
\;\;j=2,\dots,Q+1,\;\;k,\gamma_2= 1,\dots,Q.
\end{eqnarray}
Assuming large arbitrariness of   $ v^{(i_2i_3k)}_{\gamma_3\beta}$ and, consequently, of ${\cal{E}}^{(i_2 j)}_{\gamma_2 k}$, we conclude that  coefficients ahead of ${\cal{E}}^{(i_2 j)}_{\gamma_2 k}$ are zeros:
\begin{eqnarray}\label{2_UBR}
&&\sum_{{i_1,\gamma_1=1}\atop{\gamma_1\neq\beta}}^{ Q}
 U^{(\beta i_1)}_{\alpha\gamma_1}  {\cal{B}}^{(2;\beta)}_{\gamma_1}  
( \hat R^{-1})^{(i_1 i_2)}_{\gamma_1\gamma_2} 
{\cal{B}}^{(2;i_2)}_{\beta}  =0,\\\nonumber
&&i_2,\alpha,\beta,\gamma_2=1,\dots,Q,\;\;i_2\neq \beta.
\end{eqnarray}
We take $i_2\neq \beta$ in (\ref{2_UBR}) since LHS of this equation is identical to zero if $i_2=\beta$ due to the equality ${\cal{B}}^{(2;\beta)}_\beta=0$.
Thus we have $Q^3 (Q-1)$ equations and the same number of the elements of the spectral functions $ U^{(\beta i_1)}_{\alpha\gamma_1}$, $\gamma_1\neq\beta$ in (\ref{2_UBR}). 
Remark that  eqs. (\ref{2_UBR}) are linearly dependent. 
 Namely, they admit $Q$ linear  algebraic relations
 for each pair $(\alpha,\beta)$. To show this, one applies  
 $\sum_{i_2=1}^{Q}\sum_{\gamma_2=1}^Q \cdot R^{(i_2 k)}_{\gamma_2\beta}$, $k=1,\dots,Q$, to the eq.(\ref{2_UBR}).
This yields identity due to the eq.(\ref{2_RinvR}).
  One concludes that 
  the system  (\ref{2_UBR}) with fixed $\alpha$ and $\beta$   consists of 
  $Q(Q-1) -Q$ equations and consequently 
  $Q(Q-1) -[Q(Q-1) -Q] = Q$ elements of the spectral functions 
  $ U^{(\beta i_1)}_{\alpha\gamma_1}$, $\gamma_1\neq\beta$, may be independent for each pair  $(\alpha,\beta)$. If $\beta\neq Q$, then $ U^{(\beta i_1)}_{\alpha Q}$, $i_1=1,\dots,Q$ may be taken as independent functions in the set  $ U^{(\beta i_1)}_{\alpha\gamma_1}$, $i_1,\gamma_1=1,\dots,Q$, $\gamma_1\neq\beta$.
 
Applying $\displaystyle\sum_{\alpha=1}^QC^{(i)}_{\tilde \alpha\alpha}*$ and $\displaystyle\sum_{\alpha=1}^Q({G_{\tilde \alpha\alpha}})_{x_p}*$ to the eq.(\ref{2_UBR}) and replacing $\tilde \alpha$ by $\alpha$ in the result one gets the following 
 algebraic relations among the fields $ v^{(\beta m i_1)}_{\alpha\gamma_1}$ and $ w^{(\beta i_1;p)}_{\alpha\gamma_1}$:
 \begin{eqnarray}\label{2_UBRv}
&&\sum_{{i_1,\gamma_1=1}\atop{\gamma_1\neq \beta}}^{ Q}
 v^{(\beta i i_1)}_{\alpha\gamma_1}  {\cal{B}}^{(2;\beta)}_{\gamma_1}  
( \hat R^{-1})^{(i_1 i_2)}_{\gamma_1\gamma_2} 
{\cal{B}}^{(2;i_2)}_{\beta}  =0,\\
\label{2_UBRw}
&&\sum_{{i_1,\gamma_1=1}\atop{\gamma_1\neq \beta}}^{ Q}
 w^{(\beta i_1;p)}_{\alpha\gamma_1}  {\cal{B}}^{(2;\beta)}_{\gamma_1}  
( \hat R^{-1})^{(i_1 i_2)}_{\gamma_1\gamma_2} 
{\cal{B}}^{(2;i_2)}_{\beta}  =0,\\\nonumber
&&i,i_2,\alpha,\beta,\gamma_2=1,\dots,Q,\;\;i_2\neq \beta.
\end{eqnarray}
 
The last step which must be done to complete the system of nonlinear PDEs is introducing additional PDEs for the dressing functions $G$ establishing relations among derivatives $G_{x_p}$ and, consequently, among the fields $ w^{(ij;p)}$. The simplest case leading to 4-dimensional nonlinear PDEs  is following:
\begin{eqnarray}\label{2_DimG}
\partial_{x_k}G={\cal{P}}^{(k1)} \partial_{x_1} G+{\cal{P}}^{(k2)} \partial_{x_2} G,\;\; k>2. 
\end{eqnarray}
Then
\begin{eqnarray}
 w^{(ij;k)}={\cal{P}}^{k1}  w^{(ij;1)}+{\cal{P}}^{k2}  w^{(ij;2)},\;\;k>2.
\end{eqnarray}

Not all equations of the system (\ref{2_vec_Nw_22}-\ref{2_vec_Nw_24}), (\ref{2_vec_Nw_w}-\ref{2_vec_Nw_w_p22})  are independent. Namely, although 
 the linear equations (\ref{2_vec_Nw_d})  can be written for any spectral function $ U^{(nj)}$, it is not necessary to use all these equations because the eqs.(\ref{2_vec_Nw_d}) 
 with  $n\neq \beta$ are
combinations of the eqs.(\ref{2_vec_Nw_d2}) by construction.
Thus, we need only those equations  (\ref{2_vec_Nw_d})  which correspond to 
 $n=\beta$, while the equations for $ U^{(nj)}_{\alpha\beta}$, $n\neq \beta$, are represented by the system (\ref{2_vec_Nw_d2}), which is simpler. Since the
 eq.(\ref{2_vec_Nw_d}) generates nonlinear equations
 (\ref{2_vec_Nw_22}), (\ref{2_vec_Nw_w}) and (\ref{2_vec_Nw_w_p}), one needs only
 those
 of them which correspond to $n=\beta$, while the rest  equations of the nonlinear system will be generated by 
 the eq.(\ref{2_vec_Nw_d2}), see eqs.(\ref{2_vec_Nw_23},\ref{2_vec_Nw_w2},
 \ref{2_vec_Nw_w_p2})
 
Now we write down the complete system of nonlinear PDEs  for the fields (\ref{vw}) using 
 eqs.(\ref{2_UBRv},\ref{2_UBRw}) and (\ref{RinvR}) (see
eqs.(\ref{2_vec_Nw_red}-\ref{2_vec_Nw_red_w2})). 
Thus, eq.(\ref{2_vec_Nw_22}), $n=\beta$, yields:
\begin{eqnarray}\label{2_vec_Nw_red}
E^{(j;\beta ik)}_{\alpha\beta} &:=& \partial_{x_j}
 { v}^{(\beta ik)}_{\alpha\beta}  + 
\partial_{x_1} { v}^{(\beta ik)}_{\alpha\beta}{\cal{B}}^{(kj;\beta)}_\beta - 
\partial_{x_2}
 { v}^{(\beta ik)}_{\alpha\beta}B^{(kj)}_\beta+ 
\\\nonumber
&&  
    { v}^{(\beta ik;1)}_{\alpha\beta} {\cal{S}}^{(v;j1;\beta ik)}_{\alpha\beta}
   +
      { v}^{(\beta ik;2)}_{\alpha\beta} {\cal{S}}^{(v;j2;ik)}_{\alpha\beta}
\\\nonumber
&& 
\sum_{{i_1,\gamma_1=1}\atop{i_1\neq i}}^{Q}
  v^{(\beta i i_1)}_{\alpha \gamma_1} \Big(\sum_{{i_3=1}\atop{i_3\neq k}}^{Q}
  v^{(\beta i_3k)}_{\beta \beta} S^{(v;j; i_1  i_3
 k)}_{\gamma_1 \beta }+\sum_{\gamma_2=1}^Q\sum_{i_0=1}^2  w^{(\beta k;i_0)}_{\gamma_2 \beta} S^{(w;i_0;j; i_1
 k)}_{\gamma_1\gamma_2\beta}\Big)=0,\;\;j\ge 3, 
 \\\label{2_def_S}
 && S^{(v;j; i_1  i_3 k)}_{\gamma_1\beta}=  \Big({\cal{E}}^{(i_1i_3\beta)}_{\gamma_1}-
 \delta^{(i_3i_1)}\delta_{\gamma_1\beta}\Big)\; {\cal{S}}^{(v;j2;i_3k)}_{\beta\beta},\\
 &&
 {\cal{E}}^{(i_1i_3n)}_{\gamma_1}=
{\cal{B}}^{(2;n)}_{\gamma_1}\sum_{\gamma_2=1}^Q ( \hat R^{-1})^{(i_1n)}_{\gamma_1\gamma_2} R^{(n i_3)}_{\gamma_2n}
 ,\;\;j\ge 3,
 \\
 && 
 S^{(w;1;j;  i_1 k)}_{\gamma_1\gamma_2\beta}=
 {\cal{B}}^{(2;\beta)}_{\gamma_1}
 ( \hat R^{-1})^{(i_1\beta)}_{\gamma_1\gamma_2}{\cal{S}}^{(w;1;j; k)}_{\gamma_2\beta},\\
 &&
 S^{(w;2;j;  i_1 k)}_{\gamma_1\gamma_2\beta}=
 {\cal{B}}^{(2;\beta)}_{\gamma_1}
 ( \hat R^{-1})^{(i_1\beta)}_{\gamma_1\gamma_2}{\cal{S}}^{(w;2;j; k)}_{\gamma_2\beta},
\end{eqnarray}
where
\begin{eqnarray}
\label{Ap52}
&&
{\cal{S}}^{(w;1;j; k)}_{\alpha\beta}= -{\cal{P}}^{(j1)}_\alpha -{\cal{B}}^{(kj;\beta)}_\beta=-{\cal{P}}^{(j1)}_\alpha -B^{(kj)}_\beta A^{(2)}_\beta + A^{(j)}_\beta ,\\\nonumber
&&
{\cal{S}}^{(w;2;j; k)}_{\alpha\beta}= -{\cal{P}}^{(j2)}_\alpha  +  B^{(kj)}_\beta,\;\;j\ge 3.
\end{eqnarray}
Writing the expression (\ref{2_def_S})  we used eq.(\ref{2_UBRv}) and eq.(\ref{RinvR_cons}b) from the following list of identities:
\begin{eqnarray}
\label{RinvR_cons}
&&
\sum_{i_1,\gamma,\gamma_2=1}^Q U^{(ni_1)}_{\alpha\gamma} {\cal{B}}^{(2;n)}_\gamma (\hat R^{-1})^{(i_1n)}_{\gamma\gamma_2} R^{(ni_3)}_{\gamma_2\gamma_1}=\sum_{i_1,\gamma=1}^Q U^{(ni_1)}_{\alpha\gamma}\delta^{(i_1i_3)}\delta_{\gamma\gamma_1},\;\;\;\gamma_1\neq n,\;\;\;\Rightarrow\\\nonumber
&&
\sum_{i_1,\gamma,\gamma_2=1}^Q v^{(nii_1)}_{\alpha\gamma} {\cal{B}}^{(2;n)}_\gamma (\hat R^{-1})^{(i_1n)}_{\gamma\gamma_2} R^{(ni_3)}_{\gamma_2\gamma_1}=\sum_{i_1,\gamma=1}^Q v^{(nii_1)}_{\alpha\gamma}\delta^{(i_1i_3)}\delta_{\gamma\gamma_1},
\\\nonumber
&&
\sum_{i_1,\gamma,\gamma_2=1}^Q w^{(ni_1;p)}_{\alpha\gamma} {\cal{B}}^{(2;n)}_\gamma (\hat R^{-1})^{(i_1n)}_{\gamma\gamma_2} R^{(ni_3)}_{\gamma_2\gamma_1}=\sum_{i_1,\gamma=1}^Q w^{(ni_1;p)}_{\alpha\gamma}\delta^{(i_1i_3)}\delta_{\gamma\gamma_1},\\\nonumber
&&n,i,\alpha,\gamma_1=1,\dots,Q,\;\;p=1,2,\;\;\gamma_1\neq n
.
\end{eqnarray}
To prove equality (\ref{RinvR_cons}a) one must apply $\displaystyle\sum_{\gamma_1,i_3=1}^Q \cdot{\cal{B}}^{(2;n)}_{\gamma_1} 
(\hat R^{-1})^{i_3 m}_{\gamma_1 \beta}$ to (\ref{RinvR_cons}) and use (\ref{2_UBR}) to simplify the RHS of the resulting equation.
Eqs.(\ref{RinvR_cons}b,c) follows from the eq. (\ref{RinvR_cons}a) after applying  $\displaystyle\sum_{ \alpha=1}^Q C^{(i)}_{\tilde \alpha \alpha} *$ and $\displaystyle\sum_{ \alpha=1}^Q (G_{\tilde \alpha \alpha})_{x_p} *$, $p=1,2$ to the eq.(\ref{RinvR_cons}a) and replacing $\tilde \alpha$ by $\alpha$ in the result.

The eq. (\ref{2_vec_Nw_23}) yields
\begin{eqnarray}\label{2_vec_Nw_red2}
\tilde E^{(j;nik)}_{\alpha\beta} &:=&
\Big(\partial_{x_j}  v^{(nik)}_{\alpha\beta} -
 \partial_{x_1}  v^{(nik)}_{\alpha\beta} 
 A^{(j)}_n
 +
 \\\nonumber
 &&
  { v}^{(nik;1)}_{\alpha\beta} 
\tilde  {\cal S}^{(v;j1;ni)}_{\alpha}
   +
  { v}^{(nik;2)}_{\alpha\beta}
 \tilde  {\cal S}^{(v;j2;i)}_{\alpha}\Big)
 {\cal{B}}^{(2;n)}_\beta
  + 
 \\\nonumber
&&
\sum_{i_1,\gamma_1=1}^{Q}
  v^{(nii_1)}_{\alpha\gamma_1}\Big(\sum_{i_2=1}^{Q} 
  v^{(ni_2k)}_{n\beta} \tilde S^{(v;j;n i_1  i_2
 )}_{\gamma_1\beta}+\sum_{\gamma_2=1}^Q \sum_{i_0=1}^2  w^{(nk;i_0)}_{\gamma_2\beta} \tilde S^{(w;i_0;j;n  i_1  
 )}_{\gamma_1\gamma_2\beta}\Big)
,\;\;j\ge 2,\;\;n\neq \beta,\\\label{Ap5}
&&
\tilde S^{(v;j;n i_1  i_2 )}_{\gamma_1\beta}= {\cal{B}}^{(2;n)}_\beta \Big({\cal{E}}^{(i_1i_2n)}_{\gamma_1}-\delta^{(i_1i_2)} \delta_{\gamma_1 n}\Big)\tilde{\cal{S}}^{(v;j2;i_2)}_{n}
 ,\\
&&
\tilde S^{(w;1;j;n i_1 )}_{\gamma_1\gamma_2\beta}=
 {\cal{B}}^{(2;n)}_{\gamma_1}
 ( \hat R^{-1})^{(i_1n)}_{\gamma_1\gamma_2}  {\cal{B}}^{(2;n)}_\beta \tilde {\cal{S}}^{(w;1;j;n)}_{\gamma_2},\\
&&
\tilde S^{(w;2,j;n i_1 )}_{\gamma_1\gamma_2\beta}=
 {\cal{B}}^{(2;n)}_{\gamma_1}
 ( \hat R^{-1})^{(i_1n)}_{\gamma_1\gamma_2} 
   {\cal{B}}^{(2;n)}_\beta\tilde {\cal{S}}^{(w;2;j;n)}_{\gamma_2},
 \end{eqnarray}
 where
 \begin{eqnarray}
 \label{Ap6}
&&
\tilde {\cal{S}}^{(w;1;2;n)}_{\alpha}=A^{(2)}_n,\;\; 
\tilde {\cal{S}}^{(w;2;2;n)}_{\alpha}=-1,\\\nonumber
&&
\tilde {\cal{S}}^{(w;1;j;n)}_{\alpha}= -{\cal{P}}^{(j1)}_\alpha + A^{(j)}_n,\;\;
\tilde {\cal{S}}^{(w;2;j;n)}_{\alpha}= -{\cal{P}}^{(j2)}_\alpha ,\;\;j\ge 3.
\end{eqnarray} 
Writing the expression (\ref{Ap5}) we used the eqs.(\ref{2_UBRv}) and (\ref{RinvR_cons}b).

The eqs.(\ref{2_vec_Nw_w_p}) with $n=\beta$ gives 
\begin{eqnarray}\label{2_vec_Nw_red_w}
E^{(j;\beta k;p)}_{\alpha\beta} &:=&\partial_{x_j} { w}^{(\beta k;p)}_{\alpha\beta} + 
\partial_{x_1} { w}^{(\beta k;p)}_{\alpha\beta}{\cal{B}}^{(kj;\beta)}_\beta - 
\partial_{x_2} { w}^{(\beta k;p)}_{\alpha\beta}B^{(kj)}_\beta
 +\\\nonumber
 &&
 \sum_{i_0=1}^2  { w}^{(\beta k;pi_0)}_{\alpha\beta}{\cal{S}}^{(w;i_0;j;k)}_{\alpha\beta}
{\cal{B}}^{(kj;\beta)}_\beta 
  + 
  \\\nonumber
&&
\sum_{i_1,\gamma_1=1}^{Q}
  w^{(\beta i_1;p)}_{\alpha\gamma_1} \Big(\sum_{i_2=1}^{Q}
  v^{(\beta i_2k)}_{\beta\beta} S^{(v;j; i_1  i_2
 k)}_{\gamma_1\beta}+\sum_{\gamma_2=1}^Q
 \sum_{i_0=1}^2
  w^{(\beta k;i_0)}_{\gamma_2\beta} S^{(w;i_0;j; i_1 
 k)}_{\gamma_1 \gamma_2\beta}\Big)
=0.
\end{eqnarray}
Finally, the eq.(\ref{2_vec_Nw_w_p2}) reads:
\begin{eqnarray}\label{2_vec_Nw_red_w2}
E^{(j;nk;p)}_{\alpha\beta} &:=&
\Big(\partial_{x_j}  w^{(nk;p)}_{\alpha\beta} -
 \partial_{x_1}  w^{(nk;p)}_{\alpha\beta} 
 A^{(j)}_n
  +\sum_{i_0=1}^2 w^{(nk;pi_0)}_{\alpha\beta} \tilde {\cal{S}}^{(w;i_0;j;n)}_{\alpha} \Big) {\cal{B}}^{(2;n)}_\beta +
 \\\nonumber
&&
\sum_{i_1,\gamma_1=1}^{Q}
  w^{(ni_1;p)}_{\alpha\gamma_1}\Big(\sum_{i_2=1}^{Q} 
  v^{(ni_2k)}_{n\beta} \tilde S^{(v;j;n i_1 i_2
 )}_{\gamma_1\beta}+\sum_{\gamma_2=1}^Q \sum_{i_0=1}^2
  w^{(nk;i_0)}_{\gamma_2\beta} \tilde S^{(w;i_0;j;n i_1 
 )}_{\gamma_1 \gamma_2\beta}\Big)
 ,\;\;j\ge 2,\;\;n\neq \beta.
\end{eqnarray}
Deriving the coefficients in the nonlinear parts of the eqs.(\ref{2_vec_Nw_red_w}, \ref{2_vec_Nw_red_w2}) we used the eqs. (\ref{2_UBRw}) and (\ref{RinvR_cons}c).

 Now we consider such combinations of the eqs. (\ref{2_vec_Nw_red}), (\ref{2_vec_Nw_red2}) which do not involve
 the fields  $ v^{(nik;j)}$, $j=1,2$, and such combinations of the eqs.
 (\ref{2_vec_Nw_red_w}) and  (\ref{2_vec_Nw_red_w2}) which do not involve the  fields $ w^{(nk;pi_0)}$,
 $p,i_0=1,2$.  There are several types of such 
combinations depending on the relations among the coefficients 
ahead of the fields $ v^{(nik;j)}$ in the 
eqs.(\ref{2_vec_Nw_red}), (\ref{2_vec_Nw_red2}) and ahead of the 
fields $ w^{(nk;pi_0)}$ in the eqs.(\ref{2_vec_Nw_red_w}), 
(\ref{2_vec_Nw_red_w2}). Remember that the parameters 
${\cal{S}}^{(v;j1;i_1i_2i_3)}_{\alpha\beta}$, 
${\cal{S}}^{(v;j2;i_1i_2)}_{\alpha\beta}$, 
$\tilde{\cal{S}}^{(v;j1;i_1i_2)}_{\alpha\beta}$, 
$\tilde{\cal{S}}^{(v;j2;i_1)}_{\alpha\beta}$, ${\cal{S}}^{(w;i;j;k)}_{\alpha\beta}$, ${\cal{S}}^{(w;i;j;k)}_{\alpha\beta}$ and ${\cal{B}}^{(ij;k)}_\alpha$
are given by the 
eqs.(\ref{Ap3},\ref{Ap52},\ref{Ap6},\ref{hatB_Nw}).

1. Let $n=\beta=\alpha$, $i=k$. Then ${\cal S}^{(v;j1;\alpha kk)}_{\alpha\alpha} =
{\cal S}^{(v;j2; kk)}_{\alpha\alpha} =0$. The  eq. (\ref{2_vec_Nw_red}) reads:
\begin{eqnarray}\label{2_Nw_00}
E^{(j1;\alpha kk)}_{\alpha\alpha} &:=&\partial_{x_j}
 { v}^{(\alpha kk)}_{\alpha\alpha}  + 
\partial_{x_1} { v}^{(\alpha kk)}_{\alpha\alpha}{\cal{B}}^{(kj;\alpha)}_\alpha - 
\partial_{x_2} { v}^{(\alpha kk)}_{\alpha\alpha}B^{(kj)}_\alpha+ 
 \\\nonumber
&&
\sum_{i_1,\gamma_1=1}^{Q}
 v^{(\alpha ki_1)}_{\alpha\gamma_1}
 \Big(\sum_{{i_2=1}}^{Q}
  v^{(\alpha i_2k)}_{\alpha\alpha} S^{(v;j;i_1 i_2
 k)}_{\gamma_1\alpha}+\sum_{\gamma_2=1}^Q \sum_{i_0=1}^2
  w^{(\alpha k;i_0)}_{\gamma_2\alpha} S^{(w;i_0;j;i_1  
 k)}_{\gamma_1\gamma_2\alpha}\Big)
=0.
\end{eqnarray}

2. Let $\beta=\alpha$ in the eq.(\ref{2_vec_Nw_red}) and $n=\alpha$ in the eq. (\ref{2_vec_Nw_red2}).  Then 
 ${\cal S}^{(v;j1;\alpha ik)}_{\alpha\alpha} = - A^{(2)}_\alpha {\cal S}^{(v;j2; ik)}_{\alpha\alpha}$,
 $\tilde {\cal S}^{(v;j1;\alpha i)}_{\alpha\beta} = - A^{(2)}_\alpha \tilde {\cal S}^{(v;j2; i)}_{\alpha\beta}$.
So that the following combinations of the eqs. (\ref{2_vec_Nw_red}) and (\ref{2_vec_Nw_red2}) have no fields $v^{(nik;j)}_{\alpha\alpha}$:
\begin{eqnarray}
E^{(j2;\alpha ik)}_{\alpha\alpha}&:=&\left|
\begin{array}{cc}
E^{(3;\alpha ik)}_{\alpha\alpha} &  E^{(j;\alpha ik)}_{\alpha\alpha} \cr
 {\cal{S}}^{(v;3 2;ik)}_{\alpha\alpha} &  
 {\cal{S}}^{(v;j 2;ik)}_{\alpha\alpha} 
\end{array}
 \right|,\;\;j \ge 4,\\\nonumber
E^{(j3;\alpha ik)}_{\alpha\beta}&:=& \left|
\begin{array}{cc}
\tilde E^{(2;\alpha ik)}_{\alpha\beta} & \tilde E^{(j;\alpha ik)}_{\alpha\beta} \cr
\tilde {\cal{S}}^{(v;2 2; i)}_{\alpha} &
 \tilde {\cal{S}}^{(v;j 2; i)}_{\alpha} 
\end{array}
 \right|,\;\;j \ge 3,
\end{eqnarray}
which have the following explicit forms:
\begin{eqnarray}\label{2_Nw_0}
 E^{(j2;\alpha ik)}_{\alpha\alpha} &:=& s^{(j;v;j;ik)}_{\alpha} \partial_{x_{j}}  v^{(\alpha ik)}_{\alpha\alpha}+\sum_{m=1}^3 s^{(j;v;m;ik)}_{\alpha} \partial_{x_m}  v^{(\alpha ik)}_{\alpha\alpha}  - 
\\\nonumber
&&
\sum_{{i_1=1}\atop{ i_1\neq i\neq k}}^Q
 v^{(\alpha ii_1)}_{\alpha\alpha} 
  v^{(\alpha i_1k)}_{\alpha \alpha} 
 T^{(j;vv;i i_1k)}_{\alpha}  +
%\\\nonumber
%&&
\sum_{{i_1,i_2,\gamma=1}\atop{i_2\neq i\neq k,\gamma\neq \alpha}}^Q
 v^{(\alpha ii_1)}_{\alpha\gamma} 
  v^{(\alpha i_2k)}_{\alpha \alpha} {\cal{E}}^{(i_1i_2 \alpha)}_{\gamma}
 T^{(j;vv;i i_2k)}_{\alpha}  +
\\\nonumber
&&
\sum_{{i_1,\gamma_1,\gamma_2=1}\atop{\gamma_1\neq \alpha}}^Q
 v^{(\alpha ii_1)}_{\alpha\gamma_1} 
 \sum_{i_0=1}^2 w^{(\alpha k;i_0)}_{\gamma_2\alpha}
  {\cal{B}}^{(2;\alpha)}_{\gamma_1}(\hat R^{-1})^{(i_1\alpha)}_{\gamma_1\gamma_2}
 T^{(j;vw;i k;i_0)}_{\alpha\gamma_2} =0,\;\;i\neq k,\;\;j \ge 4,\\\label{2_Nw_01}
  E^{(j3;\alpha ik)}_{\alpha\beta} &:=& \tilde s^{(j;v;j;i)}_{\alpha} \partial_{x_j}  v^{(\alpha ik)}_{\alpha\beta} +
\sum_{m=1}^2 \tilde s^{(j;v;m;i)}_{\alpha} 
\partial_{x_{m}}  v^{(\alpha ik)}_{\alpha\beta} -
\\\nonumber
&&
\sum_{{i_1=1}\atop{i_1\neq i}}^Q
 v^{(\alpha ii_1)}_{\alpha\alpha} 
  v^{(\alpha i_1k)}_{\alpha\beta}
\tilde T^{(j;vv;i i_1)}_{\alpha} 
  +
%\\\nonumber
%&&
\sum_{{i_1,i_2,\gamma=1}\atop{i_2\neq i,\gamma\neq\alpha}}^Q
 v^{(\alpha ii_1)}_{\alpha\gamma} 
  v^{(\alpha i_2k)}_{\alpha\beta}
  {\cal{E}}^{(i_1i_2 \alpha)}_{\gamma} 
\tilde T^{(j;vv;i i_2)}_{\alpha} 
  +
\\\nonumber
&&
\sum_{{i_1,\gamma_1,\gamma_2=1}\atop{\gamma_1\neq \alpha}}^Q \sum_{i_0=1}^2
 v^{(\alpha ii_1)}_{\alpha\gamma_1} 
  w^{(\alpha k;i_0)}_{\gamma_2\beta}
  {\cal{B}}^{(2;\alpha)}_{\gamma_1}(\hat R^{-1})^{(i_1\alpha)}_{\gamma_1\gamma_2}
\tilde T^{(j;vw;i;i_0
 )}_{\alpha\gamma_2} =0,\\\nonumber
 &&j \ge 3,\;\;\alpha\neq\beta,
\end{eqnarray}
where 
\begin{eqnarray}\label{Ap1}
&&
s^{(j;v;1;ik)}_{\alpha}= 
\left|\begin{array}{ccc}
  {\cal{B}}^{(k3;\alpha )}_\alpha &
  {\cal{B}}^{(kj;\alpha )}_\alpha  \cr
 B^{(k3)}_\alpha -B^{(i3)}_\alpha & 
 B^{(kj)}_\alpha -B^{(ij)}_\alpha 
\end{array}\right|
 ,\\\nonumber
 &&
s^{(j;v;2;ik)}_{\alpha}= -
\left|\begin{array}{ccc}
B^{(k3)}_\alpha &
 B^{(kj)}_\alpha  \cr
 B^{(k3)}_\alpha -B^{(i3)}_\alpha & 
 B^{(kj)}_\alpha -B^{(ij)}_\alpha 
\end{array}\right|,
%\end{eqnarray}
%\begin{eqnarray}
\\\nonumber
&&
s^{(j;v;m;ik)}_{\alpha}= 
\left|\begin{array}{ccc}
 \delta^{(3m)} & \delta^{(jm)}\cr
B^{(k3)}_\alpha -B^{(i3)}_\alpha& 
B^{(kj)}_\alpha -B^{(ij)}_\alpha 
\end{array}
\right|,\;\;
m=3,j,
\end{eqnarray}

\begin{eqnarray}\nonumber
&&
T^{(j;vv; i i_1k)}_{\alpha} 
=
\left|\begin{array}{ccc}
 B^{(k3)}_\alpha -B^{(i_13)}_\alpha 
&B^{(kj)}_\alpha -B^{(i_1j)}_\alpha  
\cr
B^{(k3)}_\alpha -B^{(i3)}_\alpha& 
B^{(kj)}_\alpha -B^{(ij)}_\alpha
\end{array}
\right|,\;\;\Rightarrow \;\;T^{(j;vv; i ik)}_{\alpha} =
T^{(j;vv; k i_1 k)}_{\alpha} =T^{(j;vv; i kk)}_{\alpha} =0,
\end{eqnarray}
\begin{eqnarray}\nonumber
&&
T^{(j;vw;i k;1)}_{\alpha\gamma }
=-
\left|\begin{array}{ccc}
  {\cal{B}}^{(k3;\alpha)}_{\alpha} +{\cal{P}}^{(31)}_{\gamma } \;\;
&\;\; {\cal{B}}^{(kj;\alpha)}_{\alpha}+{\cal{P}}^{(j1)}_{\gamma }
\cr
B^{(k3)}_\alpha -B^{(i3)}_\alpha& 
B^{(kj)}_\alpha -B^{(ij)}_\alpha
\end{array}
\right|,\;\;\Rightarrow \;\;T^{(j;vw;i i;1)}_{\alpha\gamma }=0, \\\nonumber
&&
T^{(j;vw;i k;2)}_{\alpha\gamma }
=
\left|\begin{array}{ccc}
  B^{(k3)}_{\alpha}  -{\cal{P}}^{(32)}_{\gamma }
& B^{(kj)}_{\alpha} -{\cal{P}}^{(j2)}_{\gamma } 
\cr
B^{(k3)}_\alpha -B^{(i3)}_\alpha& 
B^{(kj)}_\alpha -B^{(ij)}_\alpha
\end{array}
\right|,\;\;\Rightarrow \;\;T^{(j;vw;i i;2)}_{\alpha\gamma }=0,
\end{eqnarray}
 \begin{eqnarray}
\\\nonumber
&&
\tilde s^{(j;v;1;i)}_{\alpha} =
\left|
\begin{array}{cc}
\ A^{(2)}_\alpha &  A^{(j)}_\alpha\cr
1 & 
B^{(ij)}_\alpha 
\end{array}
 \right|,
%\end{eqnarray}
%\begin{eqnarray}\nonumber
\;\;\;
\tilde s^{(j;v;m;i)}_{\alpha} = -
\left|
\begin{array}{cc}
\delta^{(2 m)} & \delta^{(j m)}\cr
1 & 
B^{(ij)}_\alpha 
\end{array}
 \right|,\;\;m=2,j,
 %\end{eqnarray}
 %\begin{eqnarray}
 \\\nonumber
 &&
 \tilde  T^{(j;vv; i i_1
 )}_{\alpha} 
  =\left|
 \begin{array}{cc}
 1& 
 B^{(i_1j)}_\alpha  \cr
1 & 
B^{(ij)}_\alpha 
\end{array}
 \right|,\;\;\Rightarrow \;\; \tilde  T^{(j;vv; i i
 )}_{\alpha} =0
% \end{eqnarray}
 %\begin{eqnarray}
 \\\nonumber
 &&
   \tilde T^{(j;vw; i;1
 )}_{\alpha\gamma }=-\left|
 \begin{array}{cc}
 A^{(2)}_\alpha& 
 A^{(j)}_\alpha-{\cal{P}}^{(j1)}_{\gamma }
  \cr
1 & 
B^{(ij)}_\alpha 
\end{array}
 \right|,\;\;\; \tilde T^{(j;vw; i;2
 )}_{\alpha\gamma }=\left|
 \begin{array}{cc}
 1& {\cal{P}}^{(j2)}_{\gamma }
  \cr
1 & 
B^{(ij)}_\alpha 
\end{array}
 \right|
 \end{eqnarray}

3. As for other values of $n,i,k,\alpha,\beta$,
the  following combinations of eqs.(\ref{2_vec_Nw_red}) and (\ref{2_vec_Nw_red2}) have no fields 
$v^{(nik;j)}_{\alpha\beta}$:
\begin{eqnarray}\nonumber
&&
E^{(j4;\beta ik)}_{\alpha\beta}:=\left|\begin {array}{ccc}
E^{(3;\beta ik)}_{\alpha\beta} &E^{(4;\beta ik)}_{\alpha\beta} & E^{(j;\beta ik)}_{\alpha\beta} \cr
{\cal{S}}^{(v;31;\beta ik)}_{\alpha\beta} &
 {\cal{S}}^{(v;41;\beta ik)}_{\alpha\beta} &
  {\cal{S}}^{(v;j1;\beta ik)}_{\alpha\beta} \cr
{\cal{S}}^{(v;32; ik)}_{\alpha\beta} &
 {\cal{S}}^{(v;42;ik)}_{\alpha\beta} &
  {\cal{S}}^{(v;j2; ik)}_{\alpha\beta} 
\end{array}
\right|,\;\;j\ge 5,\\\nonumber
&&
E^{(j5;n ik)}_{\alpha\beta}:=\left|\begin{array}{ccc}
\tilde E^{(2;nik)}_{\alpha\beta} &\tilde E^{(3;nik)}_{\alpha\beta} & \tilde E^{(j;nik)}_{\alpha\beta} \cr
\tilde {\cal{S}}^{(v;21;ni)}_{\alpha} &
 \tilde {\cal{S}}^{(v;31;ni)}_{\alpha} & 
 \tilde {\cal{S}}^{(v;j1;ni)}_{\alpha}\cr
\tilde {\cal{S}}^{(v;22;i)}_{\alpha} & 
\tilde {\cal{S}}^{(v;32;i)}_{\alpha} &
\tilde {\cal{S}}^{(v;j2;i)}_{\alpha} 
\end{array}
\right|, \;\; j\ge 4,\;\;n\neq \beta.
\end{eqnarray}
These expressions have the following explicit forms:
\begin{eqnarray}\label{2_Nw}
 E^{(j4;\beta ik)}_{\alpha\beta} &:=& s^{(j;v;j;ik)}_{\alpha\beta}\partial_{x_j}  v^{(\beta ik)}_{\alpha\beta} +
\sum_{m=1}^4 s^{(j;v;m;ik)}_{\alpha\beta}\partial_{x_{m}} 
 v^{(\beta ik)}_{\alpha\beta} -
\\\nonumber
&&
\sum_{{i_1,\gamma_1=1}\atop{i_1 \neq  k}}^Q
 v^{(\beta ii_1)}_{\alpha\beta} 
  v^{(\beta i_1k)}_{\beta\beta}
 T^{(j;vv;ii_1k)}_{\alpha\beta} +
 % \\\nonumber
%&&
\sum_{{i_1,i_2,\gamma=1}\atop{i_2\neq   k,\gamma\neq \beta}}^Q
 v^{(\beta ii_1)}_{\alpha\gamma} 
  v^{(\beta i_2k)}_{\beta\beta}
 {\cal{E}}^{(i_1i_2\beta)}_{\gamma}
 T^{(j;vv;ii_2k)}_{\alpha\beta} +
\\\nonumber
&&
\sum_{{i_1,\gamma_1,\gamma_2=1}\atop{\gamma_1\neq \beta}}^Q\sum_{i_0=1}^2
 v^{(\beta ii_1)}_{\alpha\gamma_1} 
  w^{(\beta k;i_0)}_{\gamma_2\beta}
 {\cal{B}}^{(2;\beta)}_{\gamma_1}
 (\hat R^{-1})^{(i_1\beta)}_{\gamma_1\gamma_2}
 T^{(j;vw;ik;i_0)}_{\alpha\gamma_2\beta} 
 =0,\;\;j\ge 5
\\\label{2_Nw2}
E^{(j5;n ik)}_{\alpha\beta} &:=& \tilde s^{(j;v;j;ni)}_{\alpha}\partial_{x_j}  v^{(nik)}_{\alpha\beta} +
\sum_{m=1}^3 \tilde s^{(j;v;m;ni)}_{\alpha}\partial_{x_{m}} 
 v^{(nik)}_{\alpha\beta} -
\\\nonumber
&&\sum_{{i_1=1}}^Q
 v^{(nii_1)}_{\alpha n} 
  v^{(ni_1k)}_{n\beta}
 \tilde T^{(j;vv;ii_1n)}_{\alpha}
 +
 %\\\nonumber
%&&
\sum_{{i_1,i_2,\gamma=1}\atop{\gamma\neq n}}^Q
 v^{(nii_1)}_{\alpha\gamma} 
  v^{(ni_2k)}_{n\beta}
  {\cal{E}}^{(i_1i_2n)}_{\gamma}
 \tilde T^{(j;vv;ii_2n)}_{\alpha}
 +
\\\nonumber
&&\sum_{{i_1,\gamma_1,\gamma_2=1}\atop{\gamma_1\neq n}}^Q \sum_{i_0=1}^2
 v^{(nii_1)}_{\alpha\gamma_1} 
  w^{(nk;i_0)}_{\gamma_2\beta}
 {\cal{B}}^{(2;n)}_{\gamma_1}
 (\hat R^{-1})^{(i_1 n)}_{\gamma_1\gamma_2}
 \tilde T^{(j;vw;in;i_0 
 )}_{\alpha\gamma_2}=0,\;\;
 j\ge 4,\;\;n\neq\beta,\;\;n\neq \alpha,
\end{eqnarray}
where
\begin{eqnarray}\nonumber
&&
s^{(j;v;1;ik)}_{\alpha\beta}= 
\left|\begin{array}{ccc}
  {\cal{B}}^{(k3;\beta )}_\beta &
  {\cal{B}}^{(k4;\beta )}_\beta &
  {\cal{B}}^{(kj;\beta )}_\beta \cr
 {\cal{S}}^{(v;31;\beta ik)}_{\alpha\beta} &
  {\cal{S}}^{(v;41;\beta ik)}_{\alpha\beta} &
{\cal{S}}^{(v;j1;\beta ik)}_{\alpha\beta} \cr
{\cal{S}}^{(v;32; ik)}_{\alpha\beta} & 
{\cal{S}}^{(v;42; ik)}_{\alpha\beta} &
{\cal{S}}^{(v;j2; ik)}_{\alpha\beta} 
\end{array}\right|
 ,\\\nonumber
 &&
 s^{(j;v;2;ik)}_{\alpha\beta}= -
\left|\begin{array}{ccc}
 B^{(k3)}_\beta &
 B^{(k4)}_\beta &
 B^{(kj)}_\beta \cr
 {\cal{S}}^{(v;31;\beta ik)}_{\alpha\beta} &
  {\cal{S}}^{(v;41;\beta ik)}_{\alpha\beta} &
{\cal{S}}^{(v;j1;\beta ik)}_{\alpha\beta} \cr
{\cal{S}}^{(v;32; ik)}_{\alpha\beta} & 
{\cal{S}}^{(v;42; ik)}_{\alpha\beta} &
{\cal{S}}^{(v;j2; ik)}_{\alpha\beta} 
\end{array}\right|,
\end{eqnarray}
\begin{eqnarray}\nonumber
&&
s^{(j;v;m;ik)}_{\alpha\beta}=
\left|\begin{array}{ccc}
 \delta^{(1m)} & \delta^{(2m)}&\delta^{(3m)}\cr
 {\cal{S}}^{(v;31;\beta ik)}_{\alpha\beta} &
  {\cal{S}}^{(v;41;\beta ik)}_{\alpha\beta} &
{\cal{S}}^{(v;j1;\beta ik)}_{\alpha\beta} \cr
{\cal{S}}^{(v;32; ik)}_{\alpha\beta} &
{\cal{S}}^{(v;42; ik)}_{\alpha\beta} &
{\cal{S}}^{(v;j2; ik)}_{\alpha\beta} 
\end{array}
\right|,\;\;
m=2,3,j,\;\;s^{(j;v;m;ik)}_{\alpha\beta} = s^{(j;v;m;ki)}_{\beta\alpha},
\end{eqnarray}

\begin{eqnarray}\nonumber
&&
T^{(j;vv;i i_1k)}_{\alpha\beta} 
=
\left|\begin{array}{ccc}
{\cal{S}}^{(v;32;i_1k)}_{\beta\beta}&
{\cal{S}}^{(v;42;i_1k)}_{\beta\beta}&
{\cal{S}}^{(v;j2;i_1k)}_{\beta\beta}
\cr
{\cal{S}}^{(v;31;\beta ik)}_{\alpha\beta} &
 {\cal{S}}^{(v;41;\beta ik)}_{\alpha\beta} &
  {\cal{S}}^{(v;j1;\beta ik)}_{\alpha\beta} \cr
{\cal{S}}^{(v;32; ik)}_{\alpha\beta} &
 {\cal{S}}^{(v;42; ik)}_{\alpha\beta} &
  {\cal{S}}^{(v;j2; ik)}_{\alpha\beta} 
\end{array}
\right|,\;\;\Rightarrow \;\; T^{(j;vv;i kk)}_{\alpha\beta} =0,
\end{eqnarray}

\begin{eqnarray}\nonumber
&&
T^{(j;vw;ik;i_0)}_{\alpha\gamma \beta}
 =
\left|\begin{array}{ccc}
 {\cal{S}}^{(w;i_0;3;  k)}_{\gamma  \beta} 
&{\cal{S}}^{(w;i_0;4;  k)}_{\gamma  \beta} 
&{\cal{S}}^{(w;i_0;j;  k)}_{\gamma  \beta} 
\cr
{\cal{S}}^{(v;31;\beta ik)}_{\alpha\beta} &
 {\cal{S}}^{(v;41;\beta ik)}_{\alpha\beta} &
  {\cal{S}}^{(v;j1;\beta ik)}_{\alpha\beta} \cr
{\cal{S}}^{(v;32; ik)}_{\alpha\beta} &
 {\cal{S}}^{(v;42; ik)}_{\alpha\beta} &
  {\cal{S}}^{(v;j2; ik)}_{\alpha\beta} 
\end{array}
\right|,
\end{eqnarray}

\begin{eqnarray}\nonumber
&&
\tilde s^{(j;v;1;ni)}_{\alpha}=-  
\left|\begin{array}{ccc}
 A^{(j_1)}_n& A^{(j_2)}_n &A^{(j_3)}_n \cr
\tilde  {\cal{S}}^{(v;21;ni)}_{\alpha} & 
\tilde {\cal{S}}^{(v;31;ni)}_{\alpha} & 
\tilde {\cal{S}}^{(v;j1;ni)}_{\alpha} \cr
\tilde {\cal{S}}^{(v;22;i)}_{\alpha} & 
\tilde {\cal{S}}^{(v;32;i)}_{\alpha} & 
\tilde {\cal{S}}^{(v;j2;i)}_{\alpha} 
\end{array}\right|
 ,
\end{eqnarray}
\begin{eqnarray}\nonumber
&&
\tilde s^{(j;v;m;ni)}_{\alpha}= 
\left|\begin{array}{ccc}
 \delta^{(1m)} & \delta^{(2m)}&\delta^{(3m)}\cr
  \tilde{\cal{S}}^{(v;21;ni)}_{\alpha} & 
   \tilde{\cal{S}}^{(v;31;ni)}_{\alpha} & 
    \tilde{\cal{S}}^{(v;j1;ni)}_{\alpha} \cr
 \tilde{\cal{S}}^{(v;22;i)}_{\alpha} & 
  \tilde{\cal{S}}^{(v;32;i)}_{\alpha} & 
   \tilde{\cal{S}}^{(v;j2;i)}_{\alpha} 
\end{array}
\right|,\;\;
m=3,4,j
\end{eqnarray}

\begin{eqnarray}\nonumber
&&
\tilde T^{(j;vv;ii_1n
 )}_{\alpha}
 =
\left|\begin{array}{ccc}
 \tilde {\cal{S}}^{(v;22; i_1  )}_{n} 
&\tilde {\cal{S}}^{(v;32; i_1  )}_{n} 
&\tilde {\cal{S}}^{(v;j2; i_1  )}_{n} 
\cr
\tilde{\cal{S}}^{(v;21;ni)}_{\alpha} &
 \tilde{\cal{S}}^{(v;31;ni)}_{\alpha} &
  \tilde{\cal{S}}^{(v;j1;ni)}_{\alpha} \cr
\tilde{\cal{S}}^{(v;22;i)}_{\alpha} &
 \tilde{\cal{S}}^{(v;32;i)}_{\alpha} &
  \tilde{\cal{S}}^{(v;j2;i)}_{\alpha} 
\end{array}
\right|,
\end{eqnarray}

\begin{eqnarray}\nonumber
&&
\tilde T^{(j;vw;in;i_0 
 )}_{\alpha\gamma }
=
\left|\begin{array}{ccc}
 \tilde {\cal{S}}^{(w;i_0;2;n   )}_{\gamma } 
&\tilde {\cal{S}}^{(w;i_0;3;n    )}_{\gamma } 
&\tilde {\cal{S}}^{(w;i_0;j;n  
  )}_{\gamma } 
\cr
\tilde{\cal{S}}^{(v;21;ni)}_{\alpha} &
 \tilde{\cal{S}}^{(v;31;ni)}_{\alpha} &
  \tilde{\cal{S}}^{(v;j1;ni)}_{\alpha} \cr
\tilde{\cal{S}}^{(v;22;i)}_{\alpha} &
 \tilde{\cal{S}}^{(v;32;i)}_{\alpha} &
  \tilde{\cal{S}}^{(v;j2;i)}_{\alpha} 
\end{array}
\right|.
\end{eqnarray}
As we shall see in the  end of this section, the eqs.(\ref{2_Nw},\ref{2_Nw2}) do not appear in the final complete system  of nonlinear PDEs. 
For this reason, we leave coefficients
 $T$, $\tilde T$, $s$ and $\tilde s$ in general form.
 
4. As for the eqs. (\ref{2_vec_Nw_red_w}) and (\ref{2_vec_Nw_red_w2}), the following combinations should be considered:

\begin{eqnarray}\nonumber
&&
E^{(j6;\beta kp)}_{\alpha\beta}:=\left|\begin {array}{ccc}
E^{(3;\beta k;p)}_{\alpha\beta} &E^{(4;\beta k;p)}_{\alpha\beta}&E^{(j;\beta k;p)}_{\alpha\beta}  \cr
{\cal{S}}^{(w;1;3; k)}_{\alpha\beta} &
 {\cal{S}}^{(w;1;4; k)}_{\alpha\beta} &
 {\cal{S}}^{(w;1;j; k)}_{\alpha\beta} \cr
{\cal{S}}^{(w;2;3;k)}_{\alpha\beta} &
 {\cal{S}}^{(w;2;4; k)}_{\alpha\beta} &
 {\cal{S}}^{(w;2;j; k)}_{\alpha\beta} 
 \end{array}
\right|,\;\;j\ge 5,\;\;p=1,2\\\nonumber
&&
E^{(j7;n kp)}_{\alpha\beta}:=\left|\begin{array}{ccc}
\tilde E^{(2;nk;p)}_{\alpha\beta} &\tilde E^{(3;nk;p)}_{\alpha\beta} &\tilde E^{(j;nk;p)}_{\alpha\beta} \cr
\tilde {\cal{S}}^{(w;1;2;n)}_{\alpha} &
 \tilde {\cal{S}}^{(w;1;3;n)}_{\alpha} &
 \tilde {\cal{S}}^{(w;1;j;n)}_{\alpha} \cr
\tilde {\cal{S}}^{(w;2;2;n)}_{\alpha} &
 \tilde {\cal{S}}^{(w;2;3;n)}_{\alpha} &
 \tilde {\cal{S}}^{(w;2;j;n)}_{\alpha} 
 \end{array}
\right|, \;\; j\ge 4,\;\;p=1,2,\;\;n\neq \beta.
\end{eqnarray}
These expressions have the following explicit forms:
\begin{eqnarray}\label{2_Nw_w}
E^{(j6;\beta k p)}_{\alpha\beta} &:=&s^{(j;w;1;k)}_{\alpha\beta}\partial_{x_j}  w^{(\beta k;p)}_{\alpha\beta} +
\sum_{m=1}^4 s^{(j;w;m;k)}_{\alpha\beta}\partial_{x_{m}} 
 w^{(\beta k;p)}_{\alpha\beta} -
\\\nonumber
&&
\sum_{{i_1=1}\atop{i_1\neq k}}^Q
 w^{(\beta i_1;p)}_{\alpha\beta} 
  v^{(\beta i_1k)}_{\beta\beta}
 T^{(j;wv;
 i_1k)}_{\alpha\beta} + 
\\\nonumber
&&
\sum_{{i_1,i_2,\gamma=1}\atop{i_2\neq k, \gamma\neq\beta}}^Q
 w^{(\beta i_1;p)}_{\alpha\gamma} 
  v^{(\beta i_2k)}_{\beta\beta}
{\cal{E}}^{(i_1i_2\beta)}_{\gamma}
 T^{(j;wv;
 i_2k)}_{\alpha\beta} + 
\\\nonumber
&&
\sum_{{i_1,\gamma_1,\gamma_2=1}\atop{\gamma_1\neq\beta,\gamma_2\neq\alpha}}^Q \sum_{i_0=1}^2
 w^{(\beta i_1;p)}_{\alpha\gamma_1} 
  w^{(\beta k;i_0)}_{\gamma_2\beta}
 {\cal{B}}^{(2;\beta)}_{\gamma_1}
 (\hat R^{-1})^{(i_1\beta)}_{\gamma_1\gamma_2}
 T^{(j;ww;k;i_0)}_{\alpha\gamma_2\beta} 
 =0,\;\;j\ge 5,
\\\label{2_Nw_w2} 
E^{(j7;n k p)}_{\alpha\beta} &:=&\tilde s^{(j;w;j;n)}_{\alpha}\partial_{x_j}  w^{(nk;p)}_{\alpha\beta} +
\sum_{m=1}^3 \tilde s^{(j;w;m;n)}_{\alpha}\partial_{x_{m}} 
 w^{(nk;p)}_{\alpha\beta} -
\\\nonumber
&&
\sum_{i_1=1}^Q
 w^{(ni_1;p)}_{\alpha n} 
  v^{(ni_1k)}_{n\beta}
 \tilde T^{(j;wv;i_1n)}_{\alpha}
  +
\\\nonumber
&&
\sum_{{i_1,i_2,\gamma=1}\atop{\gamma\neq n}}^Q
 w^{(ni_1;p)}_{\alpha\gamma} 
  v^{(ni_2k)}_{n\beta}
  {\cal{E}}^{(i_1i_2n)}_{\gamma}
 \tilde T^{(j;wv;i_2n)}_{\alpha}
  +
\\\nonumber
&&
\sum_{{i_1,\gamma_1,\gamma_2=1}\atop{\gamma_1\neq n,\gamma_2\neq\alpha}}^Q\sum_{i_0=1}^2
 w^{(ni_1;p)}_{\alpha\gamma_1} 
  w^{(nk;i_0)}_{\gamma_2\beta}
 {\cal{B}}^{(2;n)}_{\gamma_1}
 (\hat R^{-1})^{(i_1 n)}_{\gamma_1\gamma_2}
 \tilde T^{(j;ww;n;i_0 
 )}_{\alpha\gamma_2} 
 =0,\\\nonumber
 &&
 j\ge 4,\;\;n\neq\beta,
\end{eqnarray}
where 
\begin{eqnarray}\label{Ap2}
&&
s^{(j;w;1;k)}_{\alpha\beta}= 
-\left|\begin{array}{ccc}
  {\cal{B}}^{(k3;\beta )}_\beta &
  {\cal{B}}^{(k4;\beta )}_\beta  &
  {\cal{B}}^{(kj;\beta )}_\beta \cr
  {\cal{B}}^{(k3;\beta)}_\beta +{\cal{P}}^{(31)}_{\alpha} &
  {\cal{B}}^{(k4;\beta)}_\beta +{\cal{P}}^{(41)}_{\alpha} &
  {\cal{B}}^{(kj;\beta)}_\beta +{\cal{P}}^{(j1)}_{\alpha} \cr
 B^{(k3)}_{\beta}-{\cal{P}}^{(32)}_{\alpha}  &
  B^{(k4)}_{\beta} -{\cal{P}}^{(42)}_{\alpha} &
  B^{(kj)}_{\beta} -{\cal{P}}^{(j2)}_{\alpha} 
 \end{array}\right|
 ,\\\nonumber
 &&
 s^{(j;w;2;k)}_{\alpha\beta}= 
\left|\begin{array}{ccc}
 B^{(k3)}_\beta &
 B^{(k4)}_\beta &
 B^{(kj)}_\beta \cr
  {\cal{B}}^{(k3;\beta)}_\beta +{\cal{P}}^{(31)}_{\alpha} &
  {\cal{B}}^{(k4;\beta)}_\beta+{\cal{P}}^{(41)}_{\alpha}  &
  {\cal{B}}^{(kj;\beta)}_\beta +{\cal{P}}^{(j1)}_{\alpha} \cr
B^{(k3)}_{\beta} -{\cal{P}}^{(32)}_{\alpha}  &
 B^{(k4)}_{\beta}  -{\cal{P}}^{(42)}_{\alpha} &
  B^{(kj)}_{\beta}  -{\cal{P}}^{(j2)}_{\alpha}
\end{array}\right|,
\end{eqnarray}
\begin{eqnarray}\nonumber
&&
s^{(j;w;m;k)}_{\alpha\beta}=-
\left|\begin{array}{ccc}
 \delta^{(m3)} & \delta^{(m4)}&\delta^{(mj)}\cr
  {\cal{B}}^{(k3;\beta)}_\beta +{\cal{P}}^{(31)}_{\alpha} &
 {\cal{B}}^{(k4;\beta)}_\beta +{\cal{P}}^{(41)}_{\alpha}  &
{\cal{B}}^{(kj;\beta)}_\beta +{\cal{P}}^{(j1)}_{\alpha}   \cr
B^{(k3)}_{\beta} -{\cal{P}}^{(32)}_{\alpha}  &
 B^{(k4)}_{\beta} -{\cal{P}}^{(42)}_{\alpha}  &
  B^{(kj)}_{\beta} -{\cal{P}}^{(j2)}_{\alpha} 
\end{array}
\right|,\;\;
m=3,4,j,
\end{eqnarray}

\begin{eqnarray}\nonumber
&&
T^{(j;wv;
  i_1k)}_{\alpha\beta}
 =-
\left|\begin{array}{ccc}
 B^{(k3)}_\beta -B^{(i_13)}_\beta 
&B^{(k4)}_\beta -B^{(i_14)}_\beta 
&B^{(kj)}_\beta -B^{(i_1j)}_\beta 
\cr
{\cal{B}}^{(k3;\beta)}_\beta +{\cal{P}}^{(31)}_{\alpha}  &
 {\cal{B}}^{(k4;\beta)}_\beta  +{\cal{P}}^{(41)}_{\alpha} &
  {\cal{B}}^{(kj;\beta)}_\beta +{\cal{P}}^{(j1)}_{\alpha} \cr
 B^{(k3)}_{\beta}-{\cal{P}}^{(32)}_{\alpha}  &
  B^{(k4)}_{\beta} -{\cal{P}}^{(42)}_{\alpha} &
  B^{(kj)}_{\beta} -{\cal{P}}^{(j2)}_{\alpha} 
\end{array}
\right|,\;\;\Rightarrow \;\; T^{(j;wv;
  kk)}_{\alpha\beta}=0,
\end{eqnarray}

\begin{eqnarray}\nonumber
&&
 T^{(j;ww;k;1)}_{\alpha\gamma \beta} 
 =
\left|\begin{array}{ccc}
 {\cal{B}}^{(k3;\beta)}_\beta +{\cal{P}}^{(31)}_{\gamma }  
& {\cal{B}}^{(k4;\beta)}_\beta +{\cal{P}}^{(41)}_{\gamma } 
&{\cal{B}}^{(kj;\beta)}_\beta +{\cal{P}}^{(j1)}_{\gamma }  
\cr
{\cal{B}}^{(k3;\beta)}_\beta+{\cal{P}}^{(31)}_{\alpha}   &
 {\cal{B}}^{(k4;\beta)}_\beta +{\cal{P}}^{(41)}_{\alpha}  &
 {\cal{B}}^{(kj;\beta)}_\beta  +{\cal{P}}^{(j1)}_{\alpha} \cr
B^{(k3)}_{\beta} -{\cal{P}}^{(32)}_{\alpha}  &
 B^{(k4)}_{\beta} -{\cal{P}}^{(42)}_{\alpha}  &
  B^{(kj)}_{\beta}  -{\cal{P}}^{(j2)}_{\alpha}
\end{array}
\right|,\;\;\Rightarrow \;\; T^{(j;ww;k;1)}_{\alpha\alpha\beta} =0,\\\nonumber
&&
 T^{(j;ww;k;2)}_{\alpha\gamma \beta} 
 =-
\left|\begin{array}{ccc}
  B^{(k3)}_{\beta}  -{\cal{P}}^{(32)}_{\gamma }
& B^{(k4)}_{\beta} -{\cal{P}}^{(42)}_{\gamma } 
&B^{(kj)}_{\beta}  -{\cal{P}}^{(j2)}_{\gamma } 
\cr
{\cal{B}}^{(k3;\beta)}_\beta+{\cal{P}}^{(31)}_{\alpha}   &
 {\cal{B}}^{(k4;\beta)}_\beta  +{\cal{P}}^{(41)}_{\alpha} &
 {\cal{B}}^{(kj;\beta)}_\beta +{\cal{P}}^{(j1)}_{\alpha}  \cr
 B^{(k3)}_{\beta} -{\cal{P}}^{(32)}_{\alpha} &
  B^{(k4)}_{\beta} -{\cal{P}}^{(42)}_{\alpha} &
 B^{(kj)}_{\beta} -{\cal{P}}^{(j2)}_{\alpha}  
\end{array}
\right|,\;\;\Rightarrow \;\; T^{(j;ww;k;2)}_{\alpha\alpha\beta} =0
%\end{eqnarray}
%\begin{eqnarray}
\\\nonumber
&&
\tilde s^{(j;w;1;n)}_{\alpha}=  
\left|\begin{array}{ccc}
 A^{(2)}_n& A^{(3)}_n & A^{(j)}_n \cr
A^{(2)}_n &
 A^{(3)}_n-{\cal{P}}^{(31)}_\alpha &
 A^{(j)}_n-{\cal{P}}^{(j1)}_\alpha \cr
1 &
 {\cal{P}}^{(32)}_\alpha &
 {\cal{P}}^{(j2)}_\alpha 
\end{array}\right|
 ,
%\end{eqnarray}
%\begin{eqnarray}
\\\nonumber
&&
\tilde s^{(j;w;m;n)}_{\alpha}= -
\left|\begin{array}{ccc}
 \delta^{(m2)} & \delta^{(m3)}& \delta^{(mj)}\cr
 A^{(2)}_n &
 A^{(3)}_n-{\cal{P}}^{(31)}_\alpha &
 A^{(j)}_n-{\cal{P}}^{(j1)}_\alpha \cr
1 &
 {\cal{P}}^{(32)}_\alpha &
 {\cal{P}}^{(j2)}_\alpha 
\end{array}
\right|,\;\;
m=2,3,j,
%\end{eqnarray}
%\begin{eqnarray}
\\\nonumber
&&
 \tilde T^{(j;wv;i_1n
 )}_{\alpha}
 =
\left|\begin{array}{ccc}
 1
&B^{(i_13)}_n 
&B^{(i_1j)}_n 
\cr
A^{(2)}_n &
 A^{(3)}_n-{\cal{P}}^{(31)}_\alpha &
 A^{(j)}_n-{\cal{P}}^{(j1)}_\alpha \cr
1 &
 {\cal{P}}^{(32)}_\alpha &
 {\cal{P}}^{(j2)}_\alpha 
\end{array}
\right|.
\end{eqnarray}

\begin{eqnarray}\nonumber
&&\tilde T^{(j;ww;n;1 
 )}_{\alpha\gamma } 
=-
\left|\begin{array}{ccc}
A^{(2)}_n
& A^{(3)}_n-{\cal{P}}^{(31)}_{\gamma } 
& A^{(j)}_n-{\cal{P}}^{(j1)}_{\gamma } 
\cr
A^{(2)}_n &
 A^{(3)}_n-{\cal{P}}^{(31)}_\alpha &
 A^{(j)}_n-{\cal{P}}^{(j1)}_\alpha \cr
1 &
 {\cal{P}}^{(32)}_\alpha &
 {\cal{P}}^{(j2)}_\alpha 
\end{array}
\right|,\;\;\Rightarrow \;\;\tilde T^{(j;ww;n;1 
 )}_{\alpha\alpha} =0,\\\nonumber
  &&
  \tilde T^{(j;ww;n;2 
 )}_{\alpha\gamma } 
=
\left|\begin{array}{ccc}
1
& {\cal{P}}^{(32)}_{\gamma } 
& {\cal{P}}^{(j2)}_{\gamma } 
\cr
A^{(2)}_n &
 A^{(3)}_n-{\cal{P}}^{(31)}_\alpha &
 A^{(j)}_n-{\cal{P}}^{(j1)}_\alpha \cr
1 &
 {\cal{P}}^{(32)}_\alpha &
 {\cal{P}}^{(j2)}_\alpha 
\end{array}
\right|,\;\;\Rightarrow \;\;\tilde T^{(j;ww;n;2 
 )}_{\alpha\alpha} =0 
\end{eqnarray}

It is remarkable that the system of nonlinear PDEs 
(\ref{2_Nw_00},\ref{2_Nw_0},\ref{2_Nw_01},\ref{2_Nw},
\ref{2_Nw2},\ref{2_Nw_w},\ref{2_Nw_w2}) has the complete subsystem of  nonlinear PDEs. Let us show this. 
First of all, we remark, that the fields $ v^{(\alpha kk)}_{\alpha\alpha}$ appear only in the eq.(\ref{2_Nw_00}). 
Moreover, the fields $ v^{(n ik)}_{\alpha\beta} $, $n\neq \alpha$
do not appear in the eqs.(\ref{2_Nw_0}, \ref{2_Nw_01}, \ref{2_Nw_w}, \ref{2_Nw_w2}). Thus namely these equations compose  the complete system of  nonlinear PDEs for the fields $ v^{(\alpha ik)}_{\alpha\beta}$ and $ w^{(nk)}_{\alpha\beta}$. This system is compatible with the system of  algebraic relations (\ref{2_UBRv},\ref{2_UBRw}). Eqs.(\ref{2_Nw},
\ref{2_Nw2}) must be considered as  linear PDEs for the functions $v^{(nik)}_{\alpha\beta}$, $n\neq\alpha$, with coefficients depending on the fields $ v^{(\alpha ik)}_{\alpha\beta}$ and $ w^{(nk)}_{\alpha\beta}$. In addition, the fields $v^{(\alpha ik)}_{\alpha\beta}$  and $w^{(nk;p)}_{\alpha\beta}$
with different values of the first superscript appear in the decoupled subsystems.
For this reason, we fix the value of the first superscript in the
 fields to be 1. For this choice, the fields $v^{(1 ik)}_{1 1}$, $v^{(1 ik)}_{1 Q}$, $w^{(1k;p)}_{\alpha 1}$ and $w^{(1k;p)}_{\alpha Q}$ may be taken as  independent fields, while the other fields   $v^{(1 ik)}_{1 \beta}$  and $w^{(1k;p)}_{\alpha \beta}$, $\beta\neq 1,Q$, are expressed through $v^{(1 ik)}_{1 Q}$ and $w^{(1k;p)}_{\alpha Q}$
owing to the eqs.
(\ref{2_UBRv},\ref{2_UBRw}). However we will use the dependent variables   $u_{\alpha\beta}$ and $q^{(p)}_{\alpha\beta}$ defined by the eqs.(\ref{new_fields})  instead of $v^{(1 ik)}_{1 Q}$ and $w^{(1k;p)}_{\alpha Q}$ respectively in order to simplify the final nonlinear PDEs. This choice  allows us to remove algebraic relations (\ref{2_UBRv},\ref{2_UBRw}) from the final complete  system of nonlinear PDEs. 
Finally, we observe that, for our choice of the first superscript in the fields, parameters $A^{(j)}_1$, $j>2$, appear in the combinations $A^{(j)}_1 - {\cal{P}}^{(j1)}_\alpha$. For this reason  we may put $A^{(j)}_1=0$, $j>2$  without loss of generality. To simplify formulae, we assume that the matrix $A^{(2)}$ satisfies the following two   requirements:
\begin{eqnarray}
&&1. \;\;\; A^{(2)}_1=-1, \\\nonumber
&& 2. \;\;\;
A^{(2)}_i\neq A^{(2)}_j, \;\;i\neq j, \;\;i,j=1,\dots,Q. 
\end{eqnarray}  
All in all, applying 
$\displaystyle \sum_{\beta, k=1}^Q \cdot {\cal{B}}^{(2;1)}_\beta 
(\hat R^{-1})^{(k1)}_{\beta \tilde \beta}$ to the 
eq.(\ref{2_Nw_01}) with $\alpha=1$ and to the eq.(\ref{2_Nw_w2}) 
with $n=1$, putting $\alpha=1$ into the eq.(\ref{2_Nw_0}) and 
$\beta=1$ into the eq. (\ref{2_Nw_w}) 
we end up with 
the system of nonlinear PDEs
(\ref{Prop:sum_w}-\ref{Prop:sum_us}) for the fields (\ref{new_fields}) (we write the  symbolic set of equations   in the same order as they are written in the system (\ref{Prop:sum_w}-\ref{Prop:sum_us})):
\begin{eqnarray}
&&
E^{(56;1\beta p)}_{\alpha 1},\;\;\; \tilde E^{(57;1\beta p)}_{\alpha }\equiv \sum_{\gamma, k=1}^Q  E^{(57;1 k p)}_{\alpha\gamma }{\cal{B}}^{(2;1)}_\gamma 
(\hat R^{-1})^{(k1)}_{\gamma \beta},\\\nonumber
&&
E^{(52;1i\beta)}_{1 1},\;\;\; \tilde E^{(53;1i\beta)}_{\alpha }\equiv \sum_{\gamma, k=1}^Q  E^{(j3;1ik)}_{\alpha\gamma }{\cal{B}}^{(2;1)}_\gamma
(\hat R^{-1})^{(k1)}_{\gamma \beta},\\\nonumber
&&
 \tilde E^{(47;1\beta p)}_{\alpha }\equiv \sum_{\gamma, k=1}^Q  E^{(47;1kp)}_{\alpha\gamma }{\cal{B}}^{(2;1)}_\gamma 
(\hat R^{-1})^{(k1)}_{\gamma \beta},\\\nonumber
&&
E^{(42;1\alpha\beta)}_{1 1},\;\;\; \tilde E^{(j3;1\alpha \beta)}_{\alpha }\equiv \sum_{\beta, k=1}^Q  E^{(j3;1ik)}_{\alpha\gamma }{\cal{B}}^{(2;1)}_\gamma
(\hat R^{-1})^{(k1)}_{\gamma \beta},\;\;j=3,4,
\end{eqnarray}
where we have redefined constant coefficients in order to improve the structure of PDEs:
\begin{eqnarray}\label{coef0}
&&
s^{(j-1;w;m-1)}_{\alpha \beta}=s^{(j;w;m;\beta)}_{\alpha 1},\;\;m=2,3,4,j
,\;\;
 T^{(j-1;wwv)}_{\alpha \gamma\beta} =T^{(j;wv;\gamma\beta)}_{\alpha 1}
,\;\; \\\nonumber
 &&
\hspace{3cm}
 T^{(j-1;wqw;i_0)}_{\alpha \gamma \beta} =T^{(j;ww;\beta;i_0)}_{\alpha\gamma 1}
 ,\;\;j\ge 5\\\nonumber
 &&
 s^{(j-1;q;m-1)}_\alpha =\tilde s^{(j;w;m;1)}_{\alpha},\;\;m=2,3,j
 ,\;\;
 T^{(j-1;qwu)}_{\alpha \gamma} =\tilde T^{(j;wv;\gamma 1)}_{\alpha},\\\nonumber
 &&
 \hspace{3cm} 
  T^{(j-1;qqq;i_0)}_{\alpha \gamma}  =
  \tilde T^{(j;ww;1;i_0)}_{\alpha\gamma}
  ,\;\;j\ge 4,\\\nonumber
 &&
  s^{(j-1;v;m-1)}_{\alpha\beta} =s^{(j;v;m;\alpha\beta)}_{1},\;\;m=2,3,j,\;\;
  T^{(j-1;vvv)}_{\alpha\gamma\beta} =T^{(j;vv; \alpha\gamma\beta)}_{1} 
  ,\\\nonumber
  &&\hspace{3cm}
  T^{(j-1;vuw;i_0)}_{\alpha\gamma  \beta } =
  T^{(j;vw;\alpha\beta;i_0)}_{1\gamma}
  ,j\ge 4,\\\nonumber
 &&
  s^{(j-1;u;m-1)}_{\alpha} =\tilde s^{(j;v;m;\alpha)}_{1} ,\;\;m=2,j
  ,\;\;
  T^{(j-1;uvu)}_{\alpha\gamma}=\tilde  T^{(j;vv;\alpha\gamma )}_{1}
  ,\\\nonumber
  &&\hspace{3cm}
T^{(j-1;uuq;i_0)}_{\alpha\gamma}=
 \tilde T^{(j;vw;\alpha;i_0)}_{1\gamma},\;\;j\ge 3.
\end{eqnarray}
 This ends the proof of the item (c) of Proposition. The eq. (\ref{Prop:3_Nw_w2_red}) of the item (d.5) follows from the eq. (\ref{Prop:sum_q}) if one  puts $ v_{nk}=u_{nk}=0$.
Existence of the infinitely many commuting flows, as indicated in the item (d.1) of  Proposition, is assotiated with the arbitrary parameter $j$ in the system (\ref{2_Nw_0}, \ref{2_Nw_01}, \ref{2_Nw_w}, \ref{2_Nw_w2}). This remark ends the proof of  Proposition.

%%%%%%%%%%%%%%%%%%%%%%%%%
\subsection{Remark on the eq.(\ref{2_sp_0})}
\label{Remarks}

We will show, that the linear equation  
(\ref{2_sp_0_star}) is
 automatically satisfied due to 
the  eq.(\ref{2_UBR}).
In fact,
applying $\sum_{\alpha=1}^Q G_{\tilde \alpha \alpha}*$ to the eq.(\ref{2_sp_0_star})  and replacing  $\tilde \alpha$ by $\alpha$ in the result,   one gets the next expression for $ \hat F$:
\begin{eqnarray}
 \hat F^{(j;i_1 k)}_{\gamma\beta} = 
\sum_{n=1}^{Q}\sum_{\alpha=1}^Q ( \hat R^{-1})^{(i_1 n)}_{\gamma \alpha} R^{(nk)}_{\alpha\beta} B^{(j;n)}_\beta.
\end{eqnarray}
 The eq.(\ref{2_sp_0_star}) gets the next form:
\begin{eqnarray}\label{hatFU}
 U^{(nk)}_{\alpha\beta} =\sum_{i,\gamma,\gamma_1=1}^Q 
 U^{(n i)}_{\alpha\gamma} {\cal{B}}^{(2;n)}_\gamma ( \hat R^{-1})^{(i n)}_{\gamma\gamma_1} 
R^{(n k)}_{\gamma_1 \beta}  ,\;\;n\neq \beta
\end{eqnarray}
in view of the eq.(\ref{2_UBR}),
where we have canceled $ {\cal{B}}^{(j;n)}_\beta$ assuming that $n\neq \beta$.
Eq. (\ref{hatFU}) is the identity.
To show this, we apply
$\displaystyle\sum_{\beta,k=1}^Q \cdot {\cal{B}}^{(2;n)}_\beta (\hat R^{-1})^{(kl)}_{\beta \delta}$ to the eq.(\ref{hatFU}),  use the  eq.
(\ref{RinvR}) to simplify the RHS and use the eq.(\ref{2_UBR}) to simplify the
LHS of the resulting equation. 

%%%%%%%%%%%%%%%%%%%%%


\begin{thebibliography}{100}

\bibitem{GGKM}
C.S.Gardner, J.M.Green, M.D.Kruskal, R.M.Miura, Phys.Rev.Lett,
{\bf 19}, 1095 (1967)


\bibitem{ZSh1}
V.E.Zakharov and A.B.Shabat, Funct.Anal.Appl., {\bf 8}, 43 (1974) 


\bibitem{ZSh2}
V.E.Zakharov and A.B.Shabat, Funct.Anal.Appl., {\bf 13},  13 (1979) 


\bibitem{ZM}
V.E.Zakharov and S.V.Manakov, Funct.Anal.Appl., {\bf 19}, 11 (1985) 

\bibitem{BM}
L.V.Bogdanov and S.V.Manakov, J.Phys.A:Math.Gen., {\bf 21}, L537 (1988)

\bibitem{AKNS}
M.J.Ablowitz and H.Segur, {\it Solitons and Inverse Scattering Transform},
(SIAM, Philadelphia, 1981)


\bibitem{ZMNP}
V.E.Zakharov, S.V.Manakov, S.P.Novikov and L.P.Pitaevsky, 
{\it Theory of Solitons. The Inverse Problem Method}, 
Plenum Press (1984)

\bibitem{K}
B.Konopelchenko, {\it  Solitons in Multidimensions},
World Scientific, Singapore (1993)


\bibitem{Calogero}
F.Calogero in {\it What is integrability} by V.E.Zakharov,
Springer, 1 (1990)


\bibitem{C_int1}
F.Calogero and Ji.Xiaoda,
J.Math.Phys, {\bf 32}, 875 (1991)

\bibitem{C_int2}
F.Calogero and Ji.Xiaoda,
J.Math.Phys, {\bf 32}, 2703 (1991)

\bibitem{C_int3}
F.Calogero,
J.Math.Phys, {\bf 33}, 1257 (1992)

\bibitem{C_int4}
F.Calogero,
J.Math.Phys, {\bf 34}, 3197 (1993)

\bibitem{C_int5}
F.Calogero and Ji.Xiaoda,
J.Math.Phys, {\bf 34}, 5810 (1993)


\bibitem{Zak}
V. E. Zakharov, {\it On the dressing method}, in Inverse Methods in Action, 
edited by P. C. Sabatier, Springer-Verlag, Berlin, 602 (1990).

\bibitem{Zakharov1}
V. E. Zakharov,
{\it Integrable Systems in Multidimensional Spaces},
Lecture Notes in Physics, {\bf 153}, 
Springer-Verlag , Berlin, 190 (1982).

\bibitem{Zakharov2}
V. E. Zakharov,
{\it Multidimensional Integrable Systems}
in Proceedings of the International Congress of Mathematicians, 
PWN Warsaw, 1225 (1983).



\bibitem{BC}
R. Beals and R. R. Coifman, {\it
 Scattering, transformations spectrales et equations d'evolution 
nonlineare, I and II} at Seminaire Goulaoic-Meyer-Schwartz 1980-1981, 
exp. 22, and 1981-1982, exp. 21, Ecole Polytechnique, Palaiseau

\bibitem{ABF}
M. J. Ablowitz, D. BarYaacov and A. S. Fokas, Stud.Appl.Math., 
{\bf 69}, 135 (1983)





\bibitem{SAF} P.M.Santini, M.J.Ablowitz and A.S.Fokas,
J.Math.Phys., {\bf 25}, 2614  (1984).



\bibitem{Z}
 A.Zenchuk,
 J.Physics A:Math.Gen. {\bf 37}, nn 25,  6557 (2004).



\bibitem{ZS} A.I.Zenchuk, P.M.Santini, J. Phys. A: Math. Gen. {\bf 39}  5825 (2006)

\bibitem{Z2}  A.I.Zenchuk, math.AP/0603294 

\bibitem{ZS3}
 A.I.Zenchuk and P.M.Santini,
 J.Phys.A:Math.Theor, {\bf 40} 6147  (2007), 
  nlin.SI/0701031\\
  

\end{thebibliography}
\end{document}